%% file: main.tex
\documentclass[10pt,twocolumn,twoside]{IEEEtran}

\usepackage{amsmath,amsfonts}
\usepackage{algorithmic}
\usepackage{algorithm}
\usepackage{array}
\usepackage{textcomp}
\usepackage{stfloats}
\usepackage{url}
\usepackage{verbatim}
\usepackage{graphicx}
\usepackage{cite}
\usepackage{bbm}
\usepackage{bm}
 \usepackage{mdframed}
 \usepackage{amsfonts}
\usepackage{amsthm}
\usepackage{ bbold }
\usepackage{multirow}
\usepackage[table,xcdraw]{xcolor}
\usepackage{mathtools}
\usepackage{amssymb}
\usepackage{subcaption}
\usepackage{soul}

\usepackage{etoolbox,nicefrac}%
\usepackage{xpatch}
\usepackage{blindtext}
\usepackage{setspace}

\usepackage{tocloft}%
\newlength{\articlesectionshift}%
\let\LaTeXStandardSection\section
\let\LaTeXStandardTheSection\thesection
\let\LaTeXStandardTheSubSection\thesubsection
\let\LaTeXStandardTheSubSubSection\thesubsubsection
\let\LaTeXStandardTheParagraph\theparagraph
\makeatletter
\newcounter{titlecounter}

\xpretocmd{\maketitle}{\ifnumgreater{\value{titlecounter}}{1}}{\clearpage}{}{} 
\xpatchcmd{\maketitle}{\let\maketitle\relax\let\@maketitle\relax}{\refstepcounter{titlecounter}\begingroup
  \addtocontents{toc}{\begingroup\addtolength{\cftsecindent}{-\articlesectionshift}}%
  \addcontentsline{toc}{section}{\protect{\numberline{\thetitlecounter}{\@title}}}%
  \addtocontents{toc}{\endgroup}
}{%
  \typeout{Patching was successful}
}{%
  \typeout{patching failed}
}%

\def\@IEEEdestroythesectionargument#1{\LaTeXStandardSection{#1}}%

\xapptocmd{\maketitle}{%
\renewcommand{\thesection}{\LaTeXStandardTheSection}%
\renewcommand{\thesubsection}{\LaTeXStandardTheSubSection}%
\renewcommand{\thesubsubsection}{\LaTeXStandardTheSubSubSection}%
\renewcommand{\theparagraph}{\LaTeXStandardTheParagraph}%
}{}{}%

\newtheorem{Fact}{Fact}
\newtheorem{Lemma}{Lemma}

\newtheorem{Theorem}{Theorem}
\newtheorem{Def}{Definition}

\newtheorem{Asm}{Assumption}

\newtheorem{Remark}{Remark}

\definecolor{orange}{RGB}{255,107,0}

\DeclareMathOperator*{\minimize}{\textrm{minimize}}

\newcolumntype{M}[1]{>{\centering\arraybackslash}m{#1}}

\input{sym_marco.tex}

\title{Downlink MIMO Channel Estimation from Bits: Recoverability and Algorithm}
\author{Rajesh Shrestha, Mingjie Shao, Mingyi Hong, Wing-Kin Ma, and Xiao Fu%
\thanks{

R. Shrestha and M. Shao contributed equally.

R. Shrestha and X. Fu are with the School of EECS at Oregon State University, Corvallis, OR 97331, USA.
M. Shao was with the School of Information Science and Engineering, Shandong University, Qingdao 266237, china. He is now with the Key Laboratory of Systems and Control, Institute of Systems Science, Academy of Mathematics and Systems Science (AMSS), chinese Academy of Sciences (CAS), Beijing 1100149, China.
W.-K. Ma is with the Department of EE, The Chinese University of Hong Kong, Hong Kong SAR of China.  
M. Hong is with the Department of ECE at University of Minnesota, Minneapolis, MN 55455, USA.  
(Corresponding author: Xiao Fu)

The work of R. Shrestha and X. Fu was supported in part by the National Science Foundation (NSF) under project NSF CCF-2210004. The work of M. Hong was supported in part by the NSF CIF-2414372 and NSF ECCS-2426064. The work of W.-K. Ma was supported by a General Research Fund of Hong Kong Research Grant Council under Project ID 14203721. The work by M. Shao was supported in part by the National Natural Science Foundation of China under Grant 62401340
and the Natural Science Foundation of Shandong Province under Grant ZR2023QF103. 
}
}

\begin{document}
\maketitle

\begin{abstract}
In frequency division duplex (FDD) massive MIMO systems, a major challenge lies in acquiring the
downlink \emph{channel state information} (CSI) 
at the base station (BS) from limited feedback sent by the user equipment (UE).
To tackle this fundamental task, our contribution is twofold:
First, a simple feedback framework is proposed, where
a compression and Gaussian dithering-based quantization strategy is adopted at the UE side, and then a maximum likelihood estimator (MLE) is formulated at the BS side. 
\emph{Recoverability} of the MIMO channel under the widely used double directional model is established. Specifically, analyses are presented for two compression schemes---showing one being more overhead-economical and the other computationally lighter at the UE side.
Second, to realize the MLE, an \emph{alternating direction method of multipliers} (ADMM) algorithm is proposed. The algorithm is carefully designed to integrate a sophisticated \textit{harmonic retrieval} (HR) solver as subroutine, which turns out to be the key of effectively tackling this hard MLE problem.
Extensive numerical experiments are conducted to
validate the efficacy of our approach. 

\end{abstract}


\section{Introduction}

In \textit{frequency division duplex} (FDD) systems, downlink and uplink transmissions operate over different carrier frequencies, leading to oftentimes uncorrelated channels \cite{goldsmith2005wireless,love2008anoverview}.
In such systems, the downlink CSI at the {\it base station} (BS) is acquired through feedback sent from the {\it user equipment} (UE); see, e.g., \cite{jindal2006mimo,love2008anoverview}.
With the advent of 5G and beyond, the antenna array size in MIMO systems has grown significantly larger than before, involving tens or even hundreds of antennas at both the BS and the UE \cite{jindal2006mimo,ngo2013energy,gao2015spatially}.
However, the feedback signaling channel often has limited capacity.
Consequently, designing an accurate downlink CSI estimation scheme at the BS using limited feedback has become a major challenge in FDD MIMO systems \cite{lu2014overview,jiang2015achievable,hu2017channel}.

Vector quantization (VQ) codebook-based schemes \cite{jindal2006mimo,love2008anoverview} are arguably the most well-known methods for limited feedback, but naive VQ encounters scalability challenges in the era of massive MIMO.
More recent works such as \cite{li2021pushing} and \cite{li2023csi} considered  more advanced precoding matrix indicator (PMI) and channel quality indicator
(CQI) feedback, but similar challenges remain.
Many works use low-dimensional parameterizations of the large channel matrix (e.g., using stochastic models \cite{auyeung2007ontheperformance,choi2013noncoherent} or geometric models \cite{rao2014distributed,gao2015spatially,alevizos2018limited}), and estimate
the key parameters of the channel at the UE \cite{zhou2017sparse,qian2019algebraic,zhang2024integrated,kuo2012compressive,qian2018tensor}. This way, the UE can just feedback the key parameters other than the entire channel matrix, substantially reducing the overhead.
In particular, the works \cite{bajwa2010compressed,kuo2012compressive,gao2015spatially,qian2019algebraic,dai2018fdd,alevizos2018limited} employ the double directional parameterization, where the channel matrix is represented by $K$ paths with path losses, angle-of-departure (AoD) and angle-of-arrivals (AoA) being the parameters. The path losses, AoDs and  AoAs are then estimated via methods such as tensor decomposition \cite{zhou2017sparse,qian2018tensor,lin2020tensor} and compressive sensing \cite{kuo2012compressive,alkhateeb2014channel,zhou2017sparse,alevizos2018limited}.
A notable challenge in this line of work is that the UEs have limited computational resources (e.g., power, memory, and CPU capacity), yet channel parameter estimation is often a nontrivial task.
As a workaround, the works \cite{zhou2017sparse,gao2015spatially,alevizos2018limited} advocated schemes where the UE only compresses the channel information, while the major computations for CSI recovery are conducted at the BS. Two dictionary-based sparse algorithms were proposed in \cite{zhou2017sparse,alevizos2018limited}, using discretized spatial representations of AoD and AoA to construct the dictionary.

The latter genre is appealing in alleviating computational burdens of the UEs. However, some critical gaps remain.
First, the existing methods in this category primarily use sparse regression-based CSI recovery at the BS,
which involves a discretized spatial dictionary \cite{alkhateeb2014channel,zhou2017sparse,alevizos2018limited,hu2017channel}. The size of dictionaries grows quickly when the spatial resolution becomes finer---leading to serious scalability challenges.
Second, theoretical understanding of limited feedback-based downlink CSI estimation under the channel models in \cite{zhou2017sparse,alevizos2018limited} has been lacking---it has been unclear how many feedback bits are required to recover the CSI to a certain degree. Such understanding is essential for implementing these systems in a reliable, predictable way.

\smallskip

\noindent
{\bf Contributions.} This work aims to address these challenges. Our contributions are as follows:

\noindent
$\bullet$ {\bf Recoverabilty-Guaranteed CSI Acquisition from Bits}: We propose to compress the channel matrix at the UE and then use a Gaussian dithering based quantizer to convert the entries of the channel matrix to bits. 
This way, the original complex-valued channel matrix is reduced to a substantially smaller amount of binary values.
A \textit{maximum likelihood estimator} (MLE) is formulated at the BS accordingly. We propose two different compression schemes and a unified recovery method: The first compression scheme uses simple least squares operations to acquire the channel matrix at the UE and enjoys low feedback overheads for CSI recovery; the second uses fewer computation flops at the UE, with the tradeoff of costing more feedback overhead.
Notably, our approach does not use discretized spatial dictionaries, and thus circumvents the scalability issues encountered in previous works, e.g., \cite{alkhateeb2014channel,zhou2017sparse,alevizos2018limited,hu2017channel}.
More importantly, we show that recoverability of the downlink MIMO channel matrix under both schemes is guaranteed.

\noindent
$\bullet$ {\bf ADMM Algorithm via RELAX-based Subproblem Solving}: We propose to tackle
the formulated MLE problem under both feedback schemes using a unified optimization strategy.
Under the double directional channel model, the MLE amounts to a challenging quantized and compressed matrix recovery problem---due to antenna array manifold constraints.
Standard optimization strategies such as gradient descent were found to struggle to attain reasonable solutions (as shown later),
possibly because the complicated problem structure renders an undesired optimization landscape.
We propose an efficient {\it alternating direction method of multipliers} (ADMM) algorithm.
Our method ``decouples'' the complex recovery problem into manageable subproblems, enabling us to separately address the quantization effects and the constraints imposed by the array manifold.
In particular, it allows us to use sophisticated \textit{harmonic retrieval} (HR) algorithms, e.g., RELAX \cite{li1996efficient}, to efficiently handle the subproblem.

\smallskip

The preliminary version of this work appeared at Asilomar 2022~\cite{shao2022massive}.
The conference version includes one of the compression schemes and the ADMM algorithm design.
The journal version additionally includes (i) a new compression scheme, (ii) comprehensive recoverability analysis for both schemes, and (iii) more extensive experiments. 

\smallskip

\noindent
{\bf Notation.}
In this work, $\Rbb$ and $\Cbb$ denote the real and complex domains, respectively; $x$, $\bx$, $\bX$ and $\setX$ denote a scalar, a vector, a matrix, and a set, respectively;
$\bX^*$, $\bX^{\top}$, $\bX^{\rm{H}}$ and $\bX^\dagger$  denote the complex conjugate, the transpose, the hermitian transpose, and the pseudo-inverse of matrix $
\bX$, respectively;
$\Re$ and $\Im$ return the real part and imaginary part of its argument, respectively;
$\| \bx \|_{2}$ denote the $\ell_2$-norm of $\bm x$ and $\| \bX \|_2$ the spectral norm of $\bX$; $\| \bX \|_{\rm{F}}$ is the Frobenius norm; $|\setX|$ denotes the cardinality of the set $\setX$; $\mbox{Diag}$ puts the vector $\bx$ as the diagonal elements of $\bX$; $\otimes$ and $\odot$ denote the Kronecker product and the Khatri-Rao product, respectively; ${\cal CN}(\bm \mu, \bm \Sigma)$ and ${\cal N}(\bm \mu, \bm \Sigma)$ denote the complex Gaussian and real Gaussian distributions with mean $\bm\mu$ and covariance $\bm \Sigma$, respectively.

\section{Signal Model and Background} \label{sec:background}

\begin{figure}[!t]
    \centering
    \includegraphics[width=0.95\columnwidth]{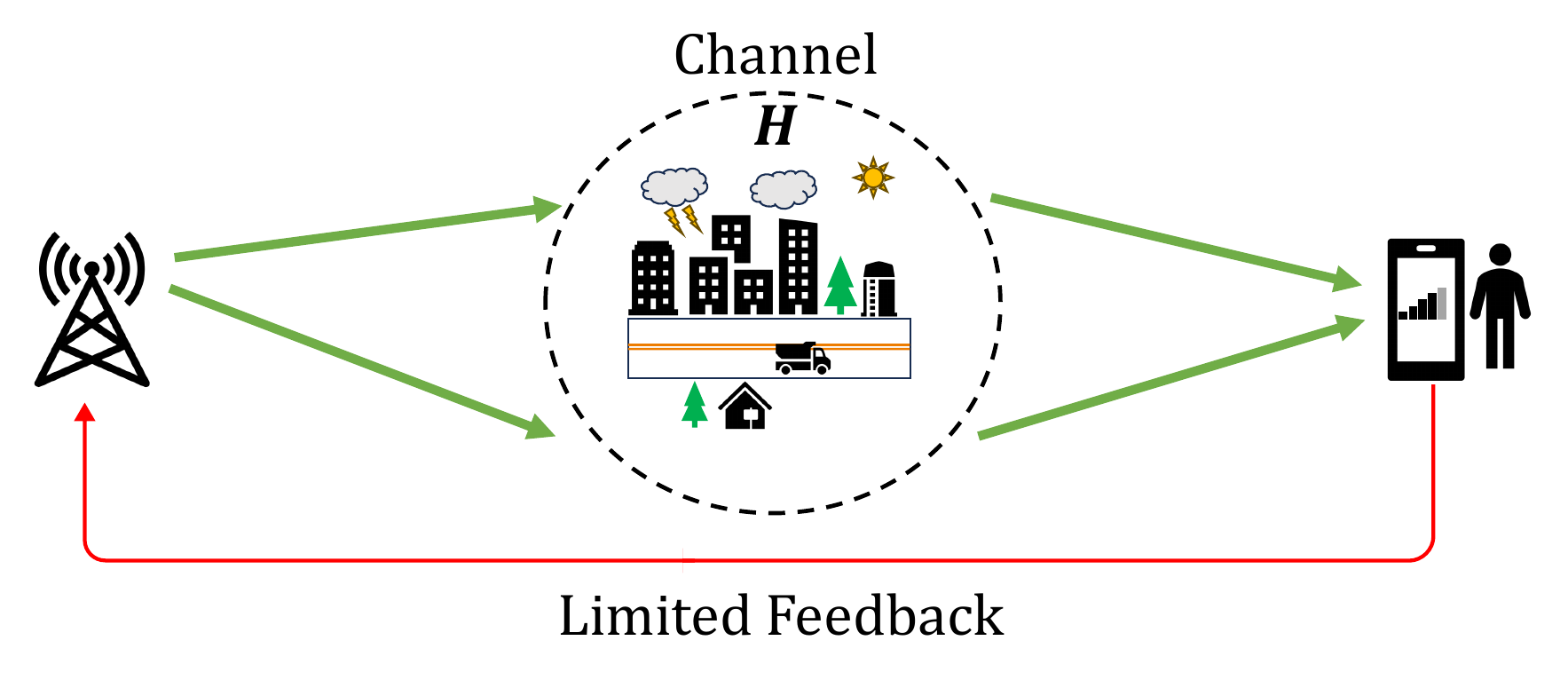}
    \caption{The limited feedback scenario considered in this work.}
    \label{fig: limited feedback framework}
\end{figure}

\subsection{Signal Model}
We consider the downlink channel  estimation problem in an FDD MIMO system as shown in Fig.~\ref{fig: limited feedback framework}.
The BS has $N$ transmit antennas, and the UE has $M$ receive antennas.
The downlink channel from the BS to the UE is assumed to remain static during the channel coherence time.
In FDD systems, the BS acquires the downlink CSI based on the UE's feedback.
To put this into the context, in the downlink training phase, the BS transmits pilot signals to the UE, and the received signal at the UE side can be written as
\begin{equation}\label{eq:downlink}
  \begin{split}
    \by_t =  \bH \bs_t + \bn_t, \quad t =1,\ldots, T,
  \end{split}
\end{equation}
where $t$ is the index of the pilot sequence, $\by_t \in \Cbb^{M}$ is the received signal at the UE side, $\bs_t \in \Cbb^N$ is the transmitted pilot signal, $\bn_t$ is the additive complex Gaussian noise, i.e., $\bn_t\sim{\cal CN}(\bm 0, \sigma_n^2 \bI)$, and
$\bH\in \Cbb^{M\times N}$ is the downlink channel.
The pilot signal $\bS:=[\bs_1,\ldots, \bs_T]$ is known by both the BS and the UE.
Suppose that both the BS and the UE are equipped with uniform linear arrays (ULA).
We consider the widely-adopted double directional channel model, specifically, the
finite scatterer channel model comprising of $K$ paths \cite{heath2016anoverview}:
\begin{equation}\label{eq:channel}
  \begin{split}
    \bH =  \sum_{k=1}^K  \beta_k \ba_{r}(\theta_k) \ba_{t}(\phi_k)^{\rm{H}},
  \end{split}
\end{equation}
where $\beta_k \in \Cbb$ is the channel path loss of the $k$th path;
$\theta_k, \phi_k \in [-\pi, \pi]$  are  AoA and AoD of the $k$th path, respectively;
$\ba_r(\theta_k)$ and $\ba_t(\phi_k)$  are the steering vectors associated with the AoA $\theta_k$ and AoD $\phi_k$, respectively, which are given by
\begin{equation}\label{eq:steer}
  \begin{split}
    \ba_r(\theta) = [1, e^{-\jj\frac{2\pi d}{\lambda}\sin(\theta)}, \ldots, e^{-\jj\frac{2\pi d}{\lambda}\sin(\theta)(M-1)}]^{\top},\\
    \ba_t(\phi) = [1, e^{-\jj\frac{2\pi d}{\lambda}\sin(\phi)}, \ldots, e^{-\jj\frac{2\pi d}{\lambda}\sin(\phi)(N-1)}]^{\top},
  \end{split}
\end{equation}
in which $d$ is inter-antenna spacing; and $\lambda$ is the carrier wavelength.
One can express the channel model in a matrix form:
\begin{equation}\label{eq:ch_com}
  \bH  = \bA_r(\bm \theta)\mbox{Diag}(\bm \beta) \bA_t (\bm \phi)^{\rm{H}},
\end{equation}
where the notation $\bm \theta = [\theta_1,\ldots, \theta_K]^{\top}$, $\bm \phi =[\phi_1,\ldots, \phi_K]^{\top}$, and $\bm \beta =[\beta_1,\ldots, \beta_K]^{\top}$.

In FDD systems, a naive feedback scheme is as follows:
the UE estimates the downlink channel $\bH$ from the received signal $\bY:= [\by_1,\ldots, \by_T]$ using the known pilot sequences $\bS$;  this often can be done with relatively light computations, e.g., via least squares.
Then, the UE sends the estimated channel $\widehat{\bH}$  to the BS via a feedback channel.
However, in the massive MIMO regime, the channel matrix $\bH$ has a large dimension.
In such circumstances, designing limited feedback-based high-accuracy CSI recovery at the BS is well-motivated.

\subsection{Limited Feedback-based CSI Estimation: Prior Art}
For large-scale $\bH$'s, quantization techniques such as VQ and codebook-based recovery at BS can be adopted to reduce the feedback overhead \cite{jindal2006mimo,love2008anoverview}.
VQ maps all the channel matrix realizations to several ``centroids'' and only feeds back the index of the centroid for each instance.
Beyond VQ, there are primarily two types of methods to feedback $\bm H$ using limited overheads.
The first type lets the UE estimate the channel parameters $\bm \theta$, $\bm \phi$ and $\bm \beta$ from $\widehat{\bH}$ \cite{kuo2012compressive,alevizos2018limited,qian2019algebraic,qian2018tensor,lin2020tensor}. 
Then, the UE only sends these parameters to the BS, which has a much lower dimension than that of $\bH$.
However, estimating these parameters often amounts to solving problems related to multi-dimensional harmonic retrieval, which are non-convex and challenging \cite{stoica2005spectral,liu2007multidimensional,sha2019harmonic,lin2020tensor,qian2019algebraic,qian2018tensor,lin2020tensor,nion2010tensor}.
Consequently, this type of approaches come at the cost of increased computational burden at the UE side. This is not ideal, as computational resources---e.g., power, CPU capacity, and memory---at the UE side are usually limited.
The second type of approaches shifts the computational burden to the BS side. This is arguably more desirable, as the BS has resources to perform computationally more intensive operations.
A line of work discretizes the angular space of $\bm \theta$ and $\bm \phi$, forms a large angular dictionary, and transforms the parameter estimation problem into a compressive sensing (CS) problem \cite{alkhateeb2014channel,zhou2017sparse,alevizos2018limited,kuo2012compressive}. 
The CS problem is then tackled at the BS side.
Nevertheless, in order to accurately estimate the channel parameters, it requires the angular discretization to be sufficiently dense, which in turn could significantly increase the problem dimension and computational/memory complexity.

Besides the difficulties in methodology design,
another notable challenge lies in theoretical understanding. 
For example, under the double directional channel model (and its variants), it has been unclear how many bits are sufficient to recover the downlink CSI at the BS to a target accuracy. However, establishing such understanding is essential for building reliable CSI recovery algorithms.

\section{Proposed Approach}
\label{sec:CQF}
In this section, we propose a compression-and-quantize strategy at the UE and an MLE-based CSI recovery criterion at the BS. We develop two schemes under this framework.

\subsection{Scheme 1: Estimation, Compression, and Quantization }
\label{sec:schm1}

\subsubsection{Operations at UE}
In this scheme, we assume that the UE first estimates $\bm H$ via relatively simple operations, e.g., least squares.

In practice, one can choose the pilot signal $\bS$ to be row orthogonal, i.e.,
$\bS\bS^{\rm{H}} = \bI.$
As a result, the LS estimation of $\bH$ under \eqref{eq:downlink} can be efficiently computed via
\begin{equation}\label{eq:est2}
\widehat{\bH} = \bY \bS^{\rm{H}}.
\end{equation}
Note that estimating $\bm H$ is relatively easy compared to estimating the key parameters $\bm \theta$, $\bm \beta$, and $\bm \phi$ \cite{sha2019harmonic,qian2019algebraic}.
To proceed, we express the channel matrix as a real-valued vector
\begin{equation}\label{eq:vec}
     \hat{\bh} = \mbox{vec}(\overline{\bH}), \quad \overline{\bH}= \begin{bmatrix}
      \Re(\widehat{\bH}) \\
      \Im(\widehat{\bH})
    \end{bmatrix},
\end{equation}
where $\Re(\widehat{\bH})$ and $\Im(\widehat{\bH})$ represent the real and imaginary parts of $ \widehat{\bH}$, respectively.
We denote $\hat{\bh} \in \Rbb^{J}$ as the vectorized $\widehat{\bH}$ with $J=2MN$.
We compress the channel $\hat{\bh}$ by a random matrix  $\bA \in \Rbb^{R\times J}$ with $R\ll J$ and
\[
    a_{i,j} \sim {\cal N}\left(0, \frac{1}{R} \right), \quad i=1,\ldots, R, ~ j = 1,\ldots, J.
\]
The compression matrix $\bA$ is known to both the BS and the UE. In practice, $\bA$ can be agreed upon the BS and UE via protocols; it can be generated using the same pseudo random number generator and seed at both sides before each transmission.

The compressed channel is given by
\begin{equation}\label{eq:compress}
  \bx  = \bA \hat{\bh}.
\end{equation}
Further, we quantize the compressed channel $\bx$ through the following:
\begin{equation}\label{eq:quant}
  \br = {\cal Q}(\bx + \bv),
\end{equation}
where  $\cal Q$ is the uniform quantizer with $Q$-level output $\setU:= \{ q_1, q_2,\ldots, q_Q \}$, i.e.,
        \[
            {\cal Q}(x ) = q \Leftrightarrow x \in [\underline{b}[q],  \bar{b}[q]),~ q\in \setU,
        \]
        where $\underline{b}[q]$ and $\bar{b}[q]$ are the quantization boundaries associated with the quantized output $q$, and $\bm v$ is a dithering noise with
$\bv\sim {\cal N}(\bm 0, \sigma_{v}^2 \bI)$.
Dithering is widely used in quantization, which improves the recovery performance via making the quantization error less correlated with the input signal and more uniformly distributed \cite{lipshitz1992quantization,wannamaker2000atheory,timilsina2023quantized}.
The UE sends the compressed and quantized counterpart $\br$, instead of the full channel matrix $\widehat{\bH}$, to the BS.
Obviously, the feedback of $\br$ requires much less communication overhead than that of $\widehat{\bH}$.

\subsubsection{MLE Recovery at BS}
The BS receives the compressed and quantized CSI $\br$ through a designated signaling channel.
Subsequently, the BS estimates the full channel matrix $\bH$ from $\br$.
Based on the compression and quantization process in \eqref{eq:est2}-\eqref{eq:quant},
the likelihood function $p(\br| \bh ;  \bA)$ is given by
\begin{equation}\label{eq:LL}
\begin{split}
  p(\br| \bh;  \bA) = &~ \prod_{i=1}^{R} p(r_i| \bh; \ba_i),\\
  p(r_i|\bh;  \ba_i) = &~\Phi \left( \frac{\bar{b}[r_i] - \ba_{i}^{\top}\bh}{\sigma_v} \right)-\Phi \left( \frac{\underline{b}[r_i] - \ba_{i}^{\top}\bh}{\sigma_v} \right),
  \end{split}
\end{equation}
where $\Phi(x) = \int_{-\infty}^{x} \frac{1}{\sqrt{2\pi}} e^{-t^2/2} \ dt$ is the cumulative distribution function of standard Gaussian distribution. This formulation arises from the dithering process, which induces a distribution $ x_i \sim {\cal N}(\ba_i^T \bh, \sigma_v^2)$. Consequently, the probability of the quantized value $r_i$ corresponds to the probability mass of this Gaussian distribution within the boundaries $(\underline{b}[r_i], \bar{b}[r_i])$, as expressed in \eqref{eq:LL}.

As a result, the MLE of $\bh$ can be expressed as follows:
\begin{equation}\label{eq:ML}
  \begin{split}
    \minimize_{\bz, \bH}&~ -\log p(\br| \bh ;  \bA)\\
    \mbox{subject~to}&~  {\cal G}(\bz)  =\bH,\\
  \end{split}
\end{equation}
where we stack the channel parameters $\bz  = (\bm \theta, \bm \phi, \bm \beta)$, and the operator $\cal G$ denotes the channel model \eqref{eq:ch_com} that maps the channel parameters $\bz$ to the channel $\bH$.

\subsection{Scheme 2: Direct Compression and Quantization}
Scheme 1 performs $\bm H$ estimation and feedback in a sequential manner. Acquiring $\bH$ using \eqref{eq:est2} requires $T\geq N$ (i.e., that the pilot length is equal to or greater than $N$), but $N$ could be large in massive MIMO settings. As a result, the $\bm H$-estimation process at UE inevitably introduces delays in the feedback, which is undesired in fast varying wireless environments.
In this subsection, we propose to compress and quantize the received signal at UE, and send them back to BS without estimating $\bm H$ first.
This scheme reduces the ``preparation time'' before feeding back the channel; see  \cite{alevizos2018limited,li2021pushing,li2023csi}.

\subsubsection{Operations at UE}
At the UE, we compress the received signal vector $\by_t=\bm H\bm s_t+\bm n_t$ using the following
\begin{equation}\label{eq:ins_comp}
  x_t = \ba_t^{\top} \tilde{\by}_t,
\end{equation}
where $\tilde{\by}_t = [
                         \Re(\by_t)^\top,
                         \Im(\by_t)^\top]^\top
 \in \Rbb^{2M}$ is the real-valued expression of $\by_t$;
$\ba_t \in \Rbb^{2M}$ is the compression vector, which is assumed to follow $\ba_t \sim {\cal N}(\bm 0, 1/T^2\bm I)$.
In Scheme 2, we choose the random pilot signal $\bs_t \sim {\cal CN}(\bm 0, \bI)$.
Then, the compressed signal $x_t$ is further quantized to a few-bit scalar for feedback via
\begin{equation}\label{eq:ins_quant}
    r_t = {\cal Q}(x_t + v_t),
\end{equation}
where $v_t\sim {\cal N}(0, \sigma^2_v)$.
Then,  the UE sends $r_t$ for $t=1,\ldots,T$.
 Compared to Scheme 1, Scheme 2 bypasses estimating $\bm H$ at the UE side. 
Instead, it directly compresses the received signal and sends it back to the BS. 
This approach reduces computational overhead and saves time/energy at the BS.

\subsubsection{MLE Recovery at BS}
 The compression strategy of Scheme 2 leads to a channel estimation formulation at the BS side that is different from \eqref{eq:ML}. 
Note that $\tilde{\by}_t$ in \eqref{eq:ins_comp} can be written as
\[
    \tilde{\by}_t =  \tilde{\bH} \tilde{\bs}_t + \tilde{\bn}_t,
\]
where
\begin{equation}\label{eq:toreal}
    \tilde{\bH} = \begin{bmatrix}
                    \Re(\bH) & - \Im(\bH) \\
                    \Im(\bH) & \Re(\bH)
                  \end{bmatrix}, ~
    \tilde{\bs}_t = \begin{bmatrix}
                      \Re(\bs_t)  \\
                      \Im(\bs_t)
                    \end{bmatrix},~
                        \tilde{\bn}_t = \begin{bmatrix}
                      \Re(\bn_t)  \\
                      \Im(\bn_t)
                    \end{bmatrix}.
\end{equation}
Therefore, $\bx$ can be expressed as
\begin{equation}\label{eq:eqv}
  \begin{split}
    \bx = &~\begin{bmatrix}
          \ba_1^{\top} \tilde{\bH} \tilde{\bs}_1    \\
          \vdots\\
            \ba_{T}^{\top} \tilde{\bH} \tilde{\bs}_{T}
          \end{bmatrix}+
          \begin{bmatrix}\ba_1^{\top}\tilde{\bn}_1\\
          \vdots\\
          \ba_{T}^{\top}\tilde{\bn}_{T}\end{bmatrix}= \bD \check{\bh} + \check{\bn},
  \end{split}
\end{equation}
where
\begin{equation*}
  \begin{split}
         \bD = \begin{bmatrix}
         \tilde{\bs}_1^{\top} \otimes \ba_1^{\top}    \\
          \vdots\\
           \tilde{\bs}_{T}^{\top} \otimes \ba_{T}^{\top}
          \end{bmatrix}, \quad
          \check{\bh} =\mbox{vec}(\tilde{\bH}), ~
         \check{\bn}=\begin{bmatrix}\ba_1^{\top}\tilde{\bn}_1\\
          \vdots\\
          \ba_{T}^{\top}\tilde{\bn}_{T}\end{bmatrix},
  \end{split}
\end{equation*}
and we have used
$
    \ba_{i}^{\top} \tilde{\bH} \tilde{\bs}_i  =  (\tilde{\bs}_i^{\top} \otimes \ba_i^{\top}) \mbox{vec}(\tilde{\bH})
$
 \cite{horn1991topics}.
Then, the feedback signal model can be rewritten as
$$\br = {\cal Q}(\bD \check{\bh} + \check{\bv}),$$
where $\check{\bv} = \check{\bn} + \bv$ and it can be shown that 
$\check{\bv}\sim {\cal N}(\bm 0, { \mbox{Diag}(\check{\bm \sigma}_v^2)})$
 for $[\check{\bm \sigma}_v^2]_i = \sigma_n^2  \sum_{j=1}^{M} a_{i,j}^2 +  { \sigma}_v^2 $ given $\bA$ known as a prior at the BS side.
Similar to Scheme 1, the BS estimates $\check{\bh}$ through the following ML estimation problem
\begin{equation}\label{eq:ML_ins}
  \begin{split}
    \minimize_{\bz,  \bH}&~ - \log p(\br| \check{\bh} ; { \bD})\\
    \mbox{subject~to} &~ {\cal G}(\bz) =  \bH,\\
  \end{split}
\end{equation}
where the likelihood function is given by
\begin{equation*}
  \begin{split}
    p(\br| \check{\bh}; { \bD}) =&~  \prod_{i=1}^{{ T}} p(r_i|\check{\bh}; { \bd_i})\\
    p(r_i|\check{\bh}; { \bd_i}) = &~\Phi \left( \frac{\bar{b}[r_i] - \bd_{i}^{\top}\check{\bh}}{\check{\sigma_v}} \right)-\Phi \left( \frac{\underline{b}[r_i] - \bd_{i}^{\top}\check{\bh}}{\check{\sigma_v}} \right),
  \end{split}
\end{equation*}
and again
the mapping ${\cal G}$ represents the generation process from the channel parameter $\bz$ to the channel  $\bH$; the correspondence between $\bH$ and $\check{\bh}$ is shown in equations \eqref{eq:toreal}  and \eqref{eq:eqv}.
It is seen that Problems \eqref{eq:ML_ins} and \eqref{eq:ML} take the same form.

\begin{Remark} \label{remark_compar}
In general, Scheme 1 has more operations performed by the UE as it involves estimating $\bH$, compression, and quantization.
These give rise to $O(MNT+MNR)$ flops.
Scheme 2 saves time at UE as it directly compresses the $\bH\bm S$ without extracting $\bH$ first, which only uses $O(MT)$ flops.
Note that the flop reduction at the UE by Scheme 2 can be substantial relative to Scheme 1, as $N$ could be (much) larger than $M$. 
{  Such reduction is considered useful for time and energy savings.}
Nonetheless, as one will see in the next section, Scheme 2 gains in efficiency but loses in sample complexity---that is, Scheme 2 requires slightly more feedback bits to attain the same CSI recovery accuracy of Scheme 1. 
\end{Remark}

\section{Recoverability Analysis}
\label{sec:rec_ana}

This section analyzes the recoverability of the proposed channel feedback schemes.

\subsection{Preliminaries}
\label{sec:pre}

In the channel model \eqref{eq:channel}, it is reasonable to assume that the channel path loss $\beta_k$ is bounded. We denote the maximum amplitude of $\beta_k$ as $\kappa$, i.e., $|\beta_k|\leq \kappa$ for all ${ k}=1, \ldots, K$.
The parameter space of $\bz$, denoted by $\setZ$, is then described by
\begin{align}\label{eq:setZ}
  \setZ =&~ \setZ_{\theta}\times \setZ_{\phi} \times \setZ_{\beta},\\
   \setZ_{\theta} =&~ [-\pi,\pi]^K,~\setZ_{\phi} = [-\pi, \pi]^K,~\setZ_{\beta} = \{\bm \beta \mid \| \bm\beta \|_{\infty}\leq \kappa \},\nonumber
\end{align}
where $\setZ_{\theta}$, $ \setZ_{\phi}$ and $\setZ_{\beta}$ represent the parameter space of $\bm \theta$, $\bm \phi$ and $\bm \beta$, respectively.
It is seen that the set $\setZ$ is compact.
In addition, the parameter space of $\bh$ and $\check{\bh}$, generated by $\bH = {\cal G}(\bz)$ for $\bz\in \setZ$, are denoted by
\begin{equation}\label{eq:setGam}
  \Gamma : =\{ \bh \mid \bh = g ({\cal G}(\bz)),~ \bz\in \setZ \}
\end{equation}
\begin{equation}\label{eq:setGamcheck}
  \check{\Gamma} : =\{ \check{\bh} \mid \check{\bh} = \check{g} ({\cal G}(\bz)),~ \bz\in \setZ \}
\end{equation}
where $g$ and $\check{g}$ represents the mapping operation from $\bH$ to $\bh$ i.e. \eqref{eq:vec}, and to $\check{\bh}$ i.e. \eqref{eq:toreal}  followed by \eqref{eq:eqv}, respectively.
Our analysis will utilize the covering number of the above sets, which is defined as follows:
\begin{Def}[Covering Number \cite{shalev2014understanding}] Let $\mathcal{X} \subset \Rbb^{d}$ be a set of vectors. Then, a set $\Bar{\mathcal{X}}$ is said to be an $\epsilon$-net of $\mathcal{X}$ with respect to the Euclidean  distance if for any $\bx \in \mathcal{X}$ there exists an $\bx' \in \Bar{\mathcal{X}}$ such that $\|\bx - \bx'\|_2 \le \epsilon$.
The smallest cardinality of the $\epsilon$-net among all the possible $\epsilon$-nets of $\mathcal{X}$ is called the covering number of the $\mathcal{X}$ and is represented as $\setC(\mathcal{X}, \epsilon)$.
\end{Def}

In addition, we have the following lemma:
\begin{Fact}\label{fact:lip}
There exists a finite $L_{\cal G}<\infty$ such that
\begin{align}\label{eq:LG}
\| {\cal G}(\bz) - {\cal G}(\bz') \|_{\rm{F}} \leq L_{\cal G} \|  \bz - \bz' \|_2;
\end{align}  
that is, the function ${\cal G}$ is {Lipschitz} continuous.
\end{Fact}
Note that $L_{\cal G}<\infty$ exists because the array manifolds and the double directional channel model do not give rise to erratic changes under continuous inputs $\bm z$; see Appendix~\ref{app:proof_lip} in the supplementary material of this paper. 

It is not hard to show the following facts:
\begin{Fact}\label{fact:cover}
   The covering numbers of the sets $\cal Z$ and $\Gamma$ are upper bounded by
  \begin{equation*}
    \begin{split}
      {\setC}(\setZ, \epsilon) \leq \left( \frac{8\pi\kappa}{\epsilon^2}   \right)^{2K},~~~~~
       \setC(\Gamma, \epsilon) \leq  \left( \frac{8\pi\kappa L_{\cal G}^2}{\epsilon^2}   \right)^{2K}.
    \end{split}
  \end{equation*}
\end{Fact}
The proof is in Appendix~\ref{app:proof_cover}.
\begin{Fact} \label{fact:norm boundedness of h}
  Assume that the channel path losses satisfy $|\beta_k|\leq \kappa$ for $k=1,\ldots, K$, where $\kappa>0$. Then, we have
$
    \| \bH \|_{\rm{F}} \leq K\sqrt{MN} \kappa.
$
\end{Fact}
\noindent{\it Proof}: We have
\begin{equation*}
  \begin{split}
    \|\bH\|_{\rm{F}} =&~ \|\bA_{r}(\bm \theta) \mbox{Diag}(\bm\beta) \bA_t(\bm \phi)^{\rm{H}}\|_{\rm{F}}\\
     \le& \sum_{k=1}^K |\beta_k| \left\|\ba_r(\theta_k) \ba_t(\phi_k)^{\rm{H}} \right\|_{\rm{F}} = \sum_{k=1}^K |\beta_k| \sqrt{MN} \\
     ~\le& K\sqrt{MN}\|\bm\beta\|_{\infty},
  \end{split}
\end{equation*}
where the first inequality is by applying basic properties of norms (i.e., the triangle inequality and homogeneity). \hfill $\blacksquare$

Our analysis requires the characterization of the likelihood function.
Consider the model in \eqref{eq:quant}--\eqref{eq:LL}.
Denote
\begin{equation}\label{eq:like_fun}
    f_{r}(x) = \Phi \left( \frac{\bar{b}[r] - x}{\sigma_v} \right)-\Phi \left( \frac{\underline{b}[r] - x}{ \sigma_v } \right).
\end{equation}

{{ Note that when $x$ is bounded (technically, as $x$ is Gaussian distributed, it does not have a deterministic bounded support, yet the realizations can be assumed to be bounded with a probability of at least $1-\vartheta$ for some $ \vartheta \geq 0$)}}, the negative log-likelihood function satisfies
\begin{equation}\label{eq:U_f}
    \sup_{x} \max_{r\in \setU} - \log f_r(x) \leq U_f
\end{equation}
for a certain constant $U_f>0$ { with probability at least $1-\vartheta$}. 
In addition, as the gradient of $f_r(x)$ is continuous, by the extreme value theorem we have, { with probability $1-\vartheta$, that}
\begin{equation}\label{eq:L_f}
     \sup_{x} \max_{r\in \setU} \left|\frac{f_{r}'(x)}{f_r(x)} \right| \leq L_f
\end{equation}
for some constant $L_f>0$, where $L_f$ being small implies that the function $f_r(x)$ does not change too rapidly. 
Similarly, we have
\begin{equation}\label{eq:F_f}
 \inf_{x} \max_{r\in\setU}  \frac{(f_{r}'(x))^2}{f_r(x)} \geq F_f
\end{equation}
for some constant $F_f>0$ { with probability $1-\vartheta$}. 
The constant $F_f$ being bounded away from zero means that the function $f_{r}(x)$ does not change too slowly.

In the rest of this section, for clarity, we will denote $\bH^{\natural} $ as the ground-truth channel matrix.
We make the following assumption.
\begin{Asm}\label{asm:err}
  There exists a constant $\nu\in [0,+\infty)$ such that
  \begin{equation}\label{eq:err}
    \min_{\bz \in \setZ} \| {\cal G}(\bz) - \bH^{\natural} \|_{\rm{F}}\leq \nu.
  \end{equation}
  Also, denote the optimal solution to the above problem as
  \[
    \appxgt{\bz} \in \arg\min_{\bz \in \setZ} \| {\cal G}(\bz) - \bH^{\natural} \|_{\rm{F}}.
  \]
\end{Asm}
This assumption accounts for the possible effects of modeling error.

 \subsection{Recoverability Analysis of Scheme 1}
\label{sec:Rec1}

Let $\bh^{\natural}$, $\bh^{\star}$ and $\appxgt{\bh}$ be the ground-truth channel, the optimal solution of the MLE in \eqref{eq:ML}, and the closest point to the ground truth in $\Gamma$, respectively.
Using these notations, we show:
\begin{Theorem}\label{thm:rec_1}
  Assume that the entries of $\bA\in \Rbb^{R \times J}$ satisfy $a_{i,j} \sim {\cal N}(0, 1/R )$ for all $i,j$, { and that $x_i$ for all $i\in[R]$ are bounded with probability at least $1-\vartheta$ for $\vartheta \geq 0$.}
  In addition, assume that
  \begin{equation}\label{eq:R_order}
    R = \Omega \left( K \log\left(\sqrt{\kappa} L_{\cal G}K \right)   \right).
  \end{equation}
  Then, we have
  \begin{equation*}
\begin{split}
  &~ \frac{ \|\bh^{\natural} - \bh^{\star}  \|_2^2}{MN} \\ \leq &~\frac{64}{\sqrt{R}}\left(4 \frac{L_f}{F_f} (\tau + \frac{3}{4} \nu )  + 9\frac{U_f}{F_f} \sqrt{ { 2} \log(2/\eta)} \right)+ \frac{32 \psi^2}{R}\\
  \end{split}
  \end{equation*}
  with probability at least $1- 3\eta-{ \vartheta -}e^{-\Omega(R)}$, where $\tau = 4\left(2+ \sqrt{\nicefrac{J}{R}}\right) + 12  \sqrt{{4K}\log \left(2\sqrt{2\pi\kappa} L_{\cal G}   \right) }$. and $ \psi = \nicefrac{1}{K} + \nicefrac{\nu}{2} + 2\sqrt{2} + \sqrt{\nicefrac{2J}{R}}$.
\end{Theorem}

The proof of Theorem~\ref{thm:rec_1} is relegated to Appendix~\ref{app:proof_thm1}.
Theorem~\ref{thm:rec_1} shows that it requires $   R = \Omega \left( K \log(K)   \right)$ 
quantized measurements to ensure the {\it mean squared error} (MSE) of the estimated channel $\bm h^\star$ to reach the level of $O(1/\sqrt{R})$. Note that the result scales  with the inherent propagation path number $K$, not with the transmit and receive dimensions $M$ and $N$.

\begin{figure}[t!]
	\centering
	\centering
		\includegraphics[width=0.8\linewidth]{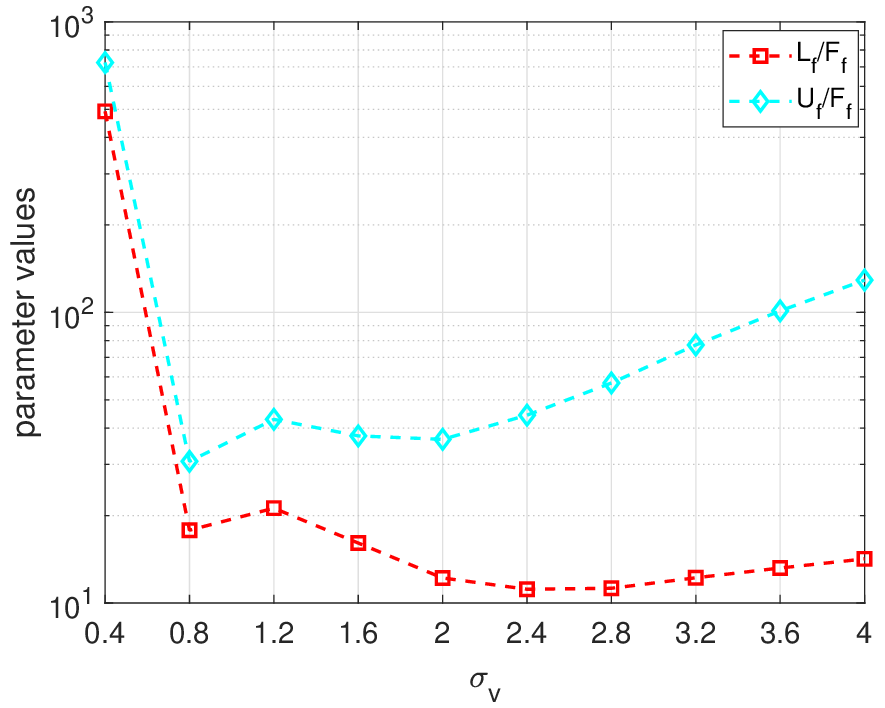}
	\caption{Illustration of ratios $L_f/F_f$ and $U_f/F_f$ for different dithering levels $\sigma_v$; $Q=4$.}
	\label{Fig:para_dither}
\end{figure}

\begin{Remark}
Some remarks regarding the dithering level $\sigma$ and quantization levels are as follows.
A large $\sigma$ could overwhelm the original signal and incur degraded recovery performance; on the other hand, an overly small $\sigma_v$ does not bring any dithering effect.
Theorem~\ref{thm:rec_1} attests to this insight, which shows that a large $L_f$  will make the MSE worse---note that a large $L_f$ is caused by small $\sigma_v$.
We further validate this intuition by examining the ratios $U_f/F_f$ and $L_f/F_f$ in Theorem~\ref{thm:rec_1}.
Fig.~\ref{Fig:para_dither}
shows these ratios under different dithering levels. 
The ratios first decrease and then increase, which also means that the dithering noise should be neither too small nor overly large.

\end{Remark}

\subsection{Recoverability Analysis of Scheme 2}
\label{sec:Rec2}

The recoverability analysis of Scheme 2 is more challenging because   the matrix $\bD$ has a more special structure whose properties are not immediately clear.
Nonetheless, by aborative derivations, we are also able to show that Scheme 2 also admits recoverability of $\bm H^\natural$:
\begin{Theorem}\label{thm:rec_2}
Assume that $\bD$ has a size of $T\times J$ with $J=4MN$, where
$\bD  = \nicefrac{1}{T} (\bS^{\top} \odot \bA^{\top})^{\top},$
{ that $x_i$ for all $i\in[T]$ are bounded with probability at least $1-\vartheta$ for $\vartheta\geq 0$,}
and that the entries of $\bm S$ and $\bm A$ are independently drawn from the standard normal distribution.
Assume that the following holds:
  \begin{equation}\label{eq:T_order}
    T = \Omega \left( K^2 \left(\log (\sqrt{\kappa} L_{\cal G} K) \right)^2  \right).
  \end{equation}
  Then, we have
  \begin{equation*}
\begin{split}
  &~ \frac{ \|\bh^{\natural} - \bh^{\star}  \|_2^2}{MN} \\ \leq &~\frac{64}{\sqrt{T}}\left(4 \frac{L_f}{F_f} (\tau + \frac{3}{4} \nu )  + 9\frac{U_f}{F_f} \sqrt{ { 2} \log(2/\eta)} \right)+ \frac{32\psi^2}{ T}
  \end{split}
  \end{equation*}
  with probability at least $1- 3\eta-{ \vartheta -}e^{-\mathbf{\Omega}(\sqrt{T})}$, where $\tau = 4\left(2+ \sqrt{\nicefrac{J}{T}}\right) + 12  \sqrt{{4K}\log \left(2\sqrt{2\pi\kappa} L_{\cal G}   \right) }$ and $ \psi = \nicefrac{1}{K} + \nicefrac{\nu}{2} + 2\sqrt{2} + \sqrt{\nicefrac{2J}{T}}$.
\end{Theorem}

The proof of Theorem~\ref{thm:rec_2} is relegated to  Appendix~\ref{app:proof_thm2} in the
supplementary material of this paper.

\begin{Remark} 
Compared with Scheme 1 (cf. Theorem~\ref{thm:rec_1}), Scheme 2 offers weaker guarantees. 
More specifically, given $T=R=G$, the two schemes have the same upper bound of the recovery MSE. 
However, comparing \eqref{eq:R_order} and \eqref{eq:T_order}, this upper bound holds for Scheme 2 under fewer channel paths relative to Scheme~1 (i.e., $K=O(\sqrt{G})$ v.s. $K=O(G)$). Also, the probabilities of recovery of both schemes are $1-3\eta - 2\exp(-\Omega(G))$ and $1-3\eta - 2\exp(-\Omega(\sqrt{G}))$, respectively.
Combining with Remark~\ref{remark_compar}, one can see that Scheme 1 has stronger recoverability guarantees in general, while Scheme 2 uses less computational resources at the UE side.
\end{Remark}

\begin{Remark}
Despite its importance, the recoverability of array manifold-structured matrices from quantized and compressed measurements had not previously been established.
The topic is related to low-rank matrix/tensor completion from quantized measurements (see, e.g.,
\cite{davenport2014onebit,ghadermarzy2018learning,lee2020tensor,timilsina2023quantized}).
The works in \cite{zhang2020spectrum,shrestha2022deep,timilsina2023quantized} established recoverability over nonlinear manifolds represented by neural networks and multilinear structures, which is closely related to our setting.
Nonetheless, some key analytical tools (e.g., Hoeffding's inequality) used under completion settings cannot be applied to our Gaussian compression case.
The latter was only studied in quantized compressive sensing for sparse signals \cite{plan2013robust,boufounos2015quantization,xu2019quantized,saab2018quantization}. 
However, their proofs are based on $\ell_1$ minimization, and thus are not suitable for analyzing complex array manifold-structured, non-sparse signals.
Our analysis builds upon a \textit{set-restricted eigenvalue condition}
(S-REC) \cite{bora2017compressed} tailored for the double directional channel structure and a link between Gaussian-dithering-induced MLE and the Kullback-Leibler (KL) divergence (see \cite{timilsina2023quantized,lee2020tensor}).
This new proof pipeline also provides a viable path for establishing guarantee in problems with other complex structural constraints under Gaussian compression.

\end{Remark}

\section{Proposed Channel Estimation Algorithm}
\label{sec:alg}
This section presents an algorithm for solving \eqref{eq:ML} and \eqref{eq:ML_ins}.
We describe the algorithm based on \eqref{eq:ML} and will briefly comment on the latter that can be handled by the same algorithm.
A natural thought is to use the gradient descent (GD) algorithm to tackle \eqref{eq:ML},
as the constraint can be ``absorbed'' into the objective, resulting in differentiable form shown in \eqref{obj: gradient descent}:
\begin{equation} \label{obj: gradient descent}
    \minimize_{\bz} ~ -\log p(\br| {\cal G}(\bz))
\end{equation}
However, gradient descent performs unsatisfactorily (see Fig.~\ref{fig:illustration of direct GD failure})---possibly because the Vandermonde manifold-induced landscape has too many local minima.
In this work, we propose an ADMM algorithm \cite{boyd2011foundations} to tackle problem \eqref{eq:ML} that leverages sophisticated harmonic retrieval (HR) solvers to handle the array manifold.

To begin with, the augmented Lagrangian function associated with problem \eqref{eq:ML} is expressed as follows:
\begin{multline}
        {\cal L}(\bz, \bH, \bm \Lambda ) =  -\log p (\br | \bh) \\
        + \langle \bm \Lambda, {\cal G}(\bz)-\bH \rangle + \frac{\rho}{2} \|{\cal G}(\bz)-\bH\|_{\rm{F}}^2,
\end{multline}
where $\bm \Lambda$ is the Lagrangian dual variable and $\rho > 0$ is a hyperparameter.
In each iteration, the ADMM algorithm performs the following operations:
\begin{mdframed}
{\bf ADMM for Solving \eqref{eq:ML}:}
\begin{subequations} \label{eq:ADMM}
\begin{align}
    &\bz^{\ell} = \underset{\bz}{\arg \min} \left\|{\cal G}(\bz) - \left( \bH^{\ell-1} - \frac{\bm \Lambda^{\ell-1}}{\rho} \right) \right\|_{\rm{F}}^2, \label{eq:z_sub}\\
    &\bH^{\ell} = \underset{\bH}{\arg \min} - \log p(\br|\bh) + \frac{\rho}{2} \left\| \bH - {\cal G}\left(\bz^{\ell}\right) - \frac{\bm \Lambda^{\ell-1}}{\rho} \right\|_{\rm{F}}^2, \label{eq:H_sub}\\
    &\bm \Lambda^{\ell} = \bm \Lambda^{\ell-1} + \rho \left( {\cal G}( \bz^{\ell} )  - \bH^{\ell}\right) \label{eq:lambda_ADMM},
\end{align}
\end{subequations}
\end{mdframed}
where $\ell$ denotes the iteration number.
As mentioned, the MLE problem \eqref{eq:ML} is challenging because of the manifold structure (represented by ${\cal G}(\bm z)$). Under the ADMM update rule, estimating $\bm z$ from ${\cal G}(\bm z)$ is now only present in the subproblem \eqref{eq:z_sub}, having a relatively ``clean'' least squares-based denoising form.
Next, we describe how to solve the subproblems \eqref{eq:z_sub} and \eqref{eq:H_sub}, respectively.

\subsection{ Modified RELAX for Subproblem \eqref{eq:z_sub}}
The subproblem \eqref{eq:z_sub} can be viewed as an HR problem, if one regards ($\bH^{\ell-1}-\bm \Lambda^{\ell-1}/\rho$) as a noisy multi-dimensional harmonic signal \cite{stoica2005spectral,nion2010tensor}. 
This is because, similar to HR, Eq.~\eqref{eq:z_sub} also aims to recover the frequency components { $(\bm \theta, \bm \phi)$}  and amplitudes $\bm \beta$ from the noisy two-dimensional signal $(\bH^{\ell-1}-\bm \Lambda^{\ell-1}/\rho)$.
Noisy multi-dimensional HR poses a challenging nonconvex optimization problem.
Nonetheless, as a classical task in array processing, many sophisticated HR algorithms exist.
We propose to employ the celebrated RELAX method \cite{li1996efficient} to tackle \eqref{eq:z_sub}. Note that the original RELAX algorithm essentially discretizes the spatial domain (i.e., $[-\pi,\pi]$) and estimates $\theta_i$ and $\phi_i$ over discrete spatial grids. This is not desired in CSI recovery. Hence, we also propose a simple modification step to the RELAX iterations as remedy.

To explain the modified RELAX algorithm,  note that the subproblem in \eqref{eq:z_sub} can be written as
\begin{equation}\label{eq:relax_opt}
    \bz = \underset{(\bm \beta, \bm \phi, \bm \theta)}{\arg\min} \left\|\bm \zeta - \sum_{k=1}^K \ba_t^*(\phi_k) \otimes \ba_r(\theta_k) \beta_k \right\|_2^2,
\end{equation}
where $\bm \zeta = \mbox{vec}( \bH - \bm \Lambda/\rho)$.
The RELAX algorithm's hierarchy consists of two levels: The outer level estimates $\{\beta_i, \phi_i, \theta_i\}_{i=1}^{\eta}$ and varies $\eta$ from $1$ to $K$ incrementally.
Within each outer step $\eta$, the inner loop uses \textit{block coordinate descent} (BCD) to estimate $( \beta_1, \phi_1, \theta_1),\ldots,( \beta_\eta, \phi_\eta, \theta_\eta)$ together. This way, RELAX gradually finds the most primary harmonics and was numerically found to be robust against convergence to poor local minima.
To be more specific, the outer iteration $\eta$ (where $\eta\in\{1,\ldots,K\}$) deals with the following:
\begin{equation}\label{eq:relax_opt_inter}
     \minimize_{\{\beta_i, \phi_i, \theta_i\}_{i=1}^{\eta}} \left \|\bm \zeta - \sum_{i=1}^\eta \ba_t^*(\phi_i) \otimes \ba_r(\theta_i) \beta_i \right \|_2^2.
\end{equation}
The inner iterations update
$(\beta_i, \phi_i, \theta_i)$ for $i \in \{1, \ldots, \eta \}$ in an alternating optimization manner by solving
\begin{equation}\label{eq:BCD}
    (\beta_i, \phi_i, \theta_i) = \underset{(\beta_i, \phi_i, \theta_i)}{\arg \min} \left \| \bm \zeta_i -  \ba_t^*(\phi_i) \otimes \ba_r(\theta_i) \beta_i \right \|_2^2,
\end{equation}
where
$    \bm \zeta_i = \bm \zeta - \sum_{j=1, j \neq i}^{\eta} \ba_t^*(\phi_j) \otimes \ba_r(\theta_j) \beta_j.
$
Problem \eqref{eq:BCD} 
is shown to be equivalent to the following:
\begin{align}
    (\phi_i, \theta_i) 
    & = \arg \max_{\phi_i, \theta_i} \left|\ba_r(\theta_i)^{\rm{H}} \mat(\bm \zeta_i) \ba_t(\phi_i)) \right|^2  \label{eqn:relaxformulation},
\end{align}
where we have $  \beta_i = \nicefrac{(\ba_t^{ *}(\phi_i) \otimes \ba_r(\theta_i))^{\rm{H}} \bm \zeta_i}{MN} $ and the matricization function ``${\rm mat}(\cdot)$'' is the inverse operation of ``${\rm vec}(\cdot)$''. The original RELAX algorithm approximates the solution of \eqref{eqn:relaxformulation} using 2D-FFT.
RELAX is widely recognized as an effective method for handling 2D HR \cite{stoica2005spectral}.
Nonetheless, the 2D-FFT step works over a discretized array manifold, which means that the accuracy of RELAX is limited by the discretization resolution (i.e., the size of the DFT matrix used).
To pull the RELAX-found solutions back to the continuous space, we further refine the solution by running gradient ascent for solving \eqref{eqn:relaxformulation}; that is,
\begin{align}\label{eq:gradientascent}
    (\phi_i,\theta_i) \leftarrow   (\phi_i,\theta_i)  + \alpha \nabla f_{i,\eta}^{\rm HR}(\phi_i,\theta_i)
\end{align}
where $ f_{i,\eta}^{\rm HR}(\phi_i,\theta_i)=|\ba_r(\theta_i)^{\rm{H}} \mat(\bm \zeta_i) \ba_t(\phi_i) |^2$.
The gradient updates in \eqref{eq:gradientascent} follows the 2D-FFT step for solving \eqref{eqn:relaxformulation}.  The modified RELAX (\texttt{Mod-RELAX}) algorithm is summarized in Appendix~\ref{app: mod-relax}.

\subsection{EM Algorithm for Subproblem \eqref{eq:H_sub}}

Many algorithms can handle the subproblem in \eqref{eq:H_sub}, e.g., the proximal gradient algorithm proposed in the context of quantized compressive sensing \cite{zymnis2009compressed}. 
We propose to employ an \textit{expectation-maximization} (EM) approach that showed empirical success in handling similar objective functions as in \eqref{eq:H_sub} \cite{mezghani2010multiple,shao2024accelerated}. { The EM algorithm is based on optimization of the following variational upper bound:
\begin{equation*}
\begin{split}
       &~ -\log \left[ \Phi \left( \frac{\bar{b}[r_i] - \ba_{i}^{\top}\bh^j}{\sigma_v} \right)-\Phi \left( \frac{\underline{b}[r_i] - \ba_{i}^{\top}\bh^j}{\sigma_v} \right) \right] \\
    \le    &~ \frac{1}{2\sigma_v^2} | g_i - \ba_i^T \bh |^2 +c,
        \end{split}
\end{equation*}
where
\begin{equation}
     g_i = ~ \ba_i^{\top} \bh^j -\sigma \frac{ \phi\left( \frac{\bar{b}[r_i] - \ba_{i}^{\top}\bar{\bh}^j}{\sigma_v} \right) - \phi \left( \frac{\underline{b}[r_i] - \ba_{i}^{\top}\bar{\bh}^j}{\sigma_v} \right) }{\Phi \left( \frac{\bar{b}[r_i] - \ba_{i}^{\top}\bar{\bh}^j}{\sigma_v} \right)-\Phi \left( \frac{\underline{b}[r_i] - \ba_{i}^{\top}\bar{\bh}^j}{\sigma_v} \right) }
\end{equation}
for any $\bar{\bh}$, and $c$ is a  constant independent of $\bh$.} The equality holds when $\bar{\bh}=\bh$.

The EM algorithm consists of the following two steps: { updating the upperbound followed by maximization with respect to this upperbound.}
\begin{align}
     g_i^j = &~ \ba_i^{\top} \bh^j -\sigma \frac{ \phi\left( \frac{\bar{b}[r_i] - \ba_{i}^{\top}\bh^j}{\sigma_v} \right) - \phi \left( \frac{\underline{b}[r_i] - \ba_{i}^{\top}\bh^j}{\sigma_v} \right) }{\Phi \left( \frac{\bar{b}[r_i] - \ba_{i}^{\top}\bh^j}{\sigma_v} \right)-\Phi \left( \frac{\underline{b}[r_i] - \ba_{i}^{\top}\bh^j}{\sigma_v} \right) },~\forall i\in[R],\label{eq:E_step}\\
    \bh^{j+1} &= \underset{\bh}{\arg \min} \frac{1}{2\sigma^2}   \|\bA \bh - \bg^j\|_2^2 + \frac{\rho}{2} \| \bh - \bu^{\ell}\|_2^2,\nonumber\\
    &=   \left( \frac{1}{2\sigma^2} \bA^T \bA + \frac{\rho}{2} \bI \right)^{-1} \left( \frac{1}{\sigma^2} \bA^T \bg^j + \rho \bu^{\ell} \right), \label{eq:M_step}
\end{align}
where $j$ denotes  the iteration index of the EM algorithm, \eqref{eq:E_step} and \eqref{eq:M_step} are the E-step and M-step, respectively,
and
\begin{equation}\label{eq:mu}
  \bu^{\ell}  = \mbox{vec}\left({\cal G}\left(\bz^{\ell}\right) + \frac{\bm \Lambda^{\ell-1}}{\rho} \right).
\end{equation}
It is seen that both E-step and M-step have closed-form solution; also the matrix inverse in the M-step of updating $\bh$ can be computed in advance and is not required to compute in each EM iteration.
Thus, the per-iteration complexity of the EM algorithm in \eqref{eq:E_step}-\eqref{eq:M_step} stays at a low level.

\smallskip

\begin{figure}[!t]
    \centering
    \includegraphics[width=.85\linewidth]{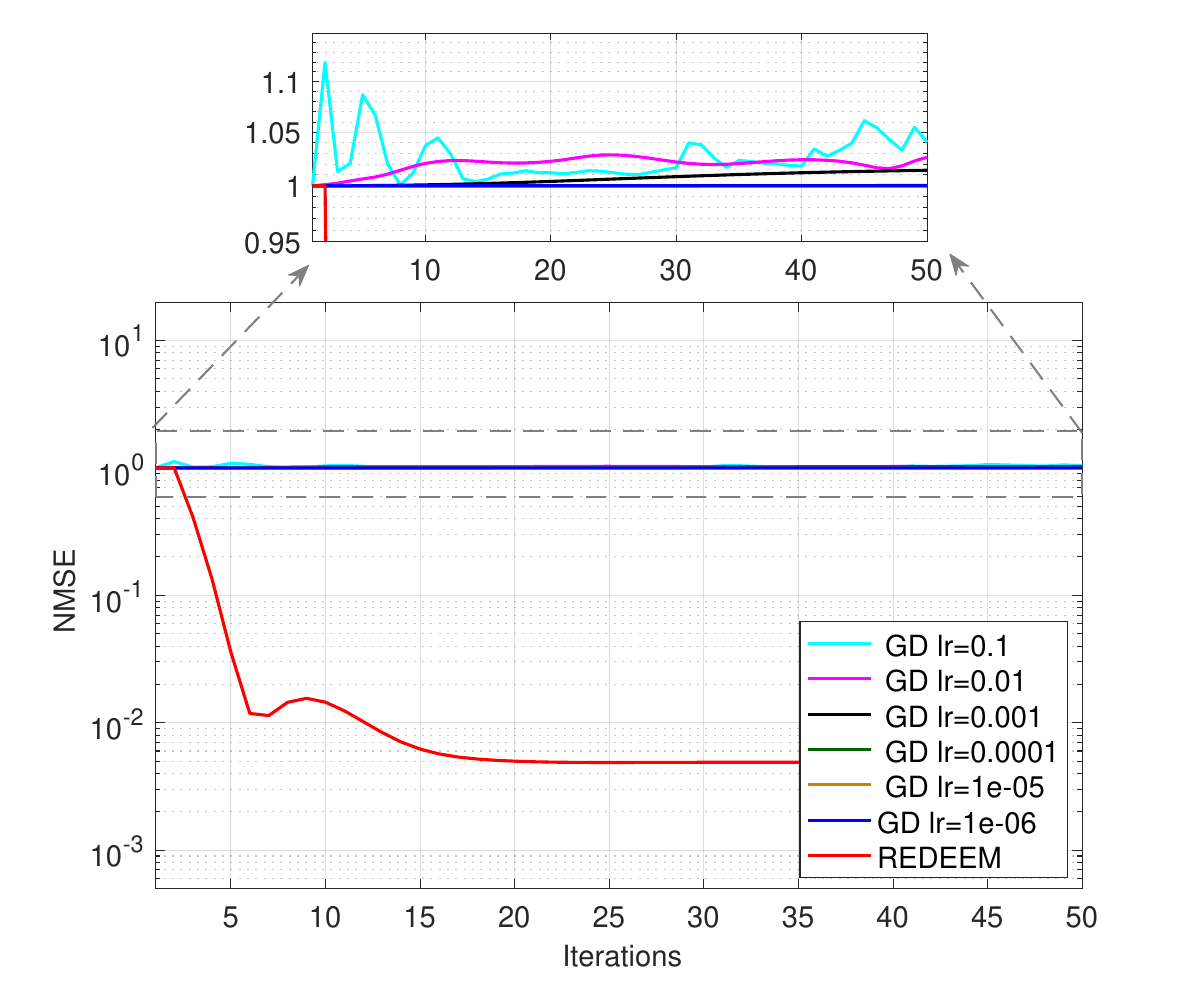}
    \caption{NMSEs (see definition in \eqref{eq:nmse}) of the estimated $\widehat{\bm H}$ by the proposed ADMM algorithm and plain-vanilla \texttt{GD} against iterations.  $(M,N)=(16,32)$, $Q=3$, $R=300$, $\sigma = 0.03$ and $K=6$.}
    \label{fig:illustration of direct GD failure}
\end{figure}

To summarize, the $\bm z$-update and $\bm H$-update in our ADMM algorithm in \eqref{eq:ADMM} are carried out by \texttt{Mod-RELAX} and \texttt{EM}, respectively.
The algorithm in \eqref{eq:ADMM} was for solving Problem~\eqref{eq:ML}, i.e., the MLE under Scheme 1.
To apply the ADMM algorithm to solving Problem \eqref{eq:ML_ins} under Scheme 2, one can simply replace $\br$, $\bA$,  $\bh$ and $\sigma_v$ by $\br$, $\bD$, $\check{\bh}$ and $\check{\sigma}$ in \eqref{eq:ML_ins}, respectively;
also note that the correspondence between $\bh$ and $\bH$ needs to be replaced by the correspondence between $\check{\bh}$ and $\bH$.
We refer to our algorithm in \eqref{eq:ADMM} as \underline{R}ELAX and \underline{E}M-assisted alternating \underline{d}ir\underline{e}ction m\underline{e}thod of \underline{m}ultipliers (\texttt{REDEEM}).
Fig.~\ref{fig:illustration of direct GD failure} shows the sharp contrast between the convergence characteristics of the \texttt{REDEEM} algorithm and directly applying gradient descent (\texttt{GD}) to Problem \eqref{eq:ML}.
{ We also present detailed per-iteration complexity analysis for \texttt{REDEEM} in Appendix \ref{sec:per-iteration complexity at BS}.}

\section{Numerical Results} \label{sec:sim}

{ In this section, we will first present basic simulations using the signal model \eqref{eq:channel}. Then, we will test the methods under more challenging scenarios generated by the DeepMIMO toolbox \cite{Alkhateeb2019deepmimo}.}

\subsection{{Basic Simulations}}
\label{sec:channel model simulation}

\subsubsection{Data Generation}

In the simulations, the BS and the user are assumed to be equipped with ULAs, with the inter-antenna spacing $d = \lambda/2$.
The channel is generated according to the double directional channel model in \eqref{eq:ch_com}.
The AoD and AoA angles $\{\phi_i\}_i, \{ \theta_i \}_i$ are randomly sampled from the uniform distribution on $(-\frac{\pi}{2}, \frac{\pi}{2})$. 
Each complex path loss $\beta_i$ is generated  with its amplitude $|\beta_i|$ and angle $\angle \beta_i$ uniformly sampled from $(0.5,1)$ and $(0,2 \pi)$, respectively.
The quantizer has uniform quantization intervals. 
{ 
The min and max values of 1000 data samples, treated as historic CSI samples, are used as the min and max quantization boundaries in our experiments. The dithering noise level $\sigma_v$ (and $\check{\sigma}_v$) is set to $25\%$ of the maximal signal amplitude.}
Our results are averaged over 1,000  trials with randomly generated data in each trial. { Unless otherwise noted, all the experiments in Sec.~\ref{sec:channel model simulation} use above settings.}

\subsubsection{Performance Metric}
Following the literature \cite{alevizos2018limited, sun2018limited, mo2014channel, mo2018channel}, we use the averaged \textit{normalized mean squared error} (NMSE), { and \emph{average beamforming gain} (BGain)} to measure the CSI reconstruction performance of all algorithms.
Given the ground-truth channel matrix $\bH^{\natural}$ and the estimate $\widehat{\bH}$, the NMSE and BGain is defined as follows:
\begin{equation}\label{eq:nmse}
    \text{NMSE} = \Exp \left[ \frac{\|\bH^{\natural} - \widehat{\bH} \|_{\rm{F}}^2}{\|\bH^{\natural} \|_{\rm{F}}^2} \right],
\end{equation}
\begin{equation}\label{eq:bgain}
     \text{BGain} = \Exp \left[ \tr(\widehat{\bH}^H \bH)^2/\| \widehat{\bH}  \|_F^2\right].
\end{equation}

\subsubsection{Baselines}

We compare our proposed methods with a diverse set of baselines. First, we consider approaches that use limited feedback and relegate most computational burden of the channel estimation to the BS side. This includes 
the sparse regression based
MLE (\texttt{SparseMLE}) \cite{alevizos2018limited}, and \texttt{EMMP} \cite{mo2014channel}. 
Note that the original \texttt{SparseMLE} method uses its own compression strategy (where the compression matrix is essentially a Kronecker product of $\bm S^\top$ and a Gaussian compression matrix).
We also implement a version of \texttt{SparseMLE} that uses the estimation-and-then-compression strategy of Scheme 1. We refer to this variant as \texttt{SparseMLE (Scheme 1)}. { We apply the same dithering technique and noise level for all the benchmarking methods.}
Second, as a reference, we also present the results of methods that carry out heavy computations at the { UE} for estimating the key parameters from $\bH$. This includes \texttt{OMP-SQ} \cite{alevizos2018limited} and \texttt{UE-HR} that performs HR on $\bm H$ at the UE side; in our experiments, \texttt{UE-HR} employs RELAX \cite{li1996efficient} for HR.
These results are used to provide insights into the gap between the accuracy of full-precision measurement-based estimations and those relying on quantized measurements.

\subsubsection{Algorithm Settings}
For methods requiring channel estimation at UE, we utilize orthogonal variable spreading factor (OVSF) code as the pilot signals; this includes the proposed \texttt{REDEEM (Scheme 1)},  \texttt{SparseMLE (Scheme 1)}, \texttt{EMMP} and \texttt{UE-HR}. 
For other methods, we use Gaussian random pilots.  The length of pilot sequences for these schemes  is set as $4N$. 
The SNR of the received pilot signals is set as 25dB for all the methods unless specified otherwise.
To implement our ADMM algorithm, the step size $\rho$ is set to 1 for all the experiments.
All the optimization variables are initialized to $\bm {0}$.
For the $\bm z$-subproblem, the inner loop of \texttt{Mod-RELAX} (see lines 5-9 in Algorithm~\ref{alg:RELAX}) is stopped after 60 BCD iterations or the relative change of $(\beta_j, \phi_j, \theta_j)_{j=1}^{\eta}$ is smaller than $10^{-3}$.
The GA step in our subproblem solver \texttt{Mod-RELAX} runs for at most 20 iterations. 
For the $\bm H$-subproblem in our ADMM framework, the maximum iteration for the \texttt{EM} method is set to 60.

\subsubsection{Results}
\label{sec:sim_perform}

\begin{figure}[t!]
	\centering
	\begin{subfigure}[b]{.49\linewidth}
	\centering
		\includegraphics[width=1.1\linewidth]{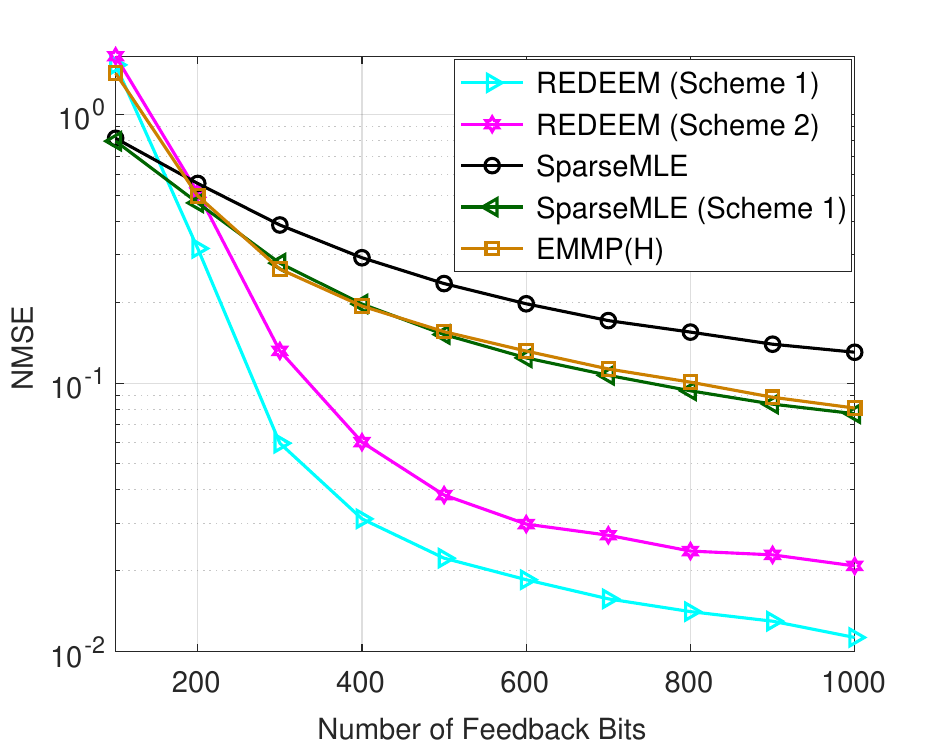}
		\caption{$(M,K,Q)=(16,6,2)$.}\label{fig:quant_1bit}
	\end{subfigure}
    \hfill
	\begin{subfigure}[b]{.49\linewidth}
	\centering
		\includegraphics[width=1.1\linewidth]{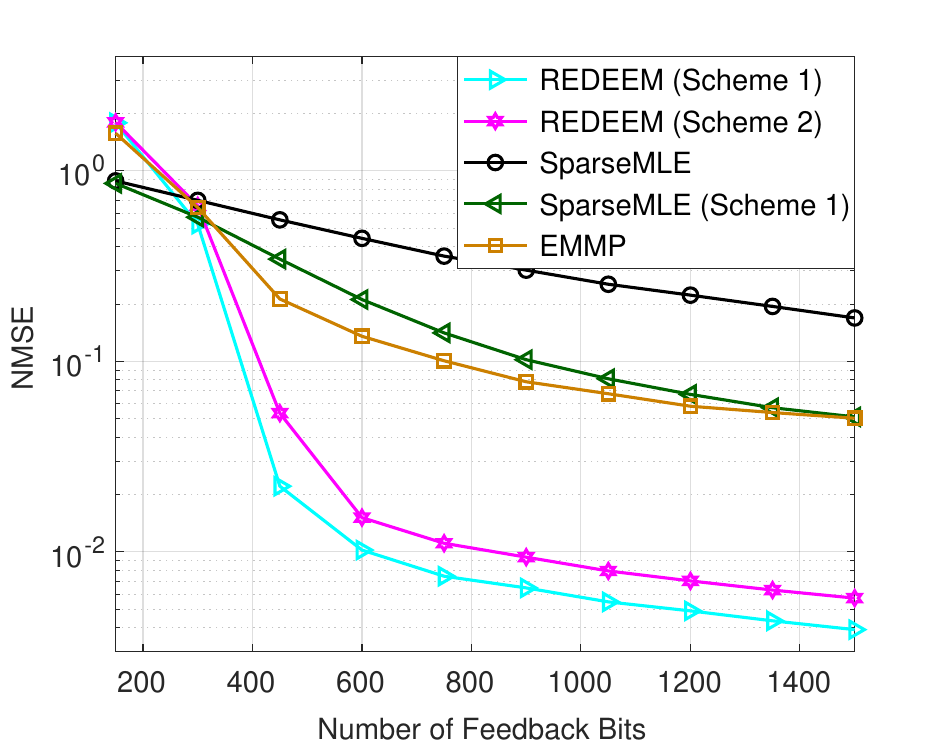}
		\caption{$(M,K,Q)=(24,8,3)$.}\label{fig:quant_2bit}
	\end{subfigure}
	\caption{NMSE Performance under different $Q$'s when $N=32$.}
	\label{Fig:feedback_bit}
\end{figure}

\begin{figure}[t!]
	\centering
	\begin{subfigure}[b]{.49\linewidth}
	\centering
		\includegraphics[width=1.1\linewidth]{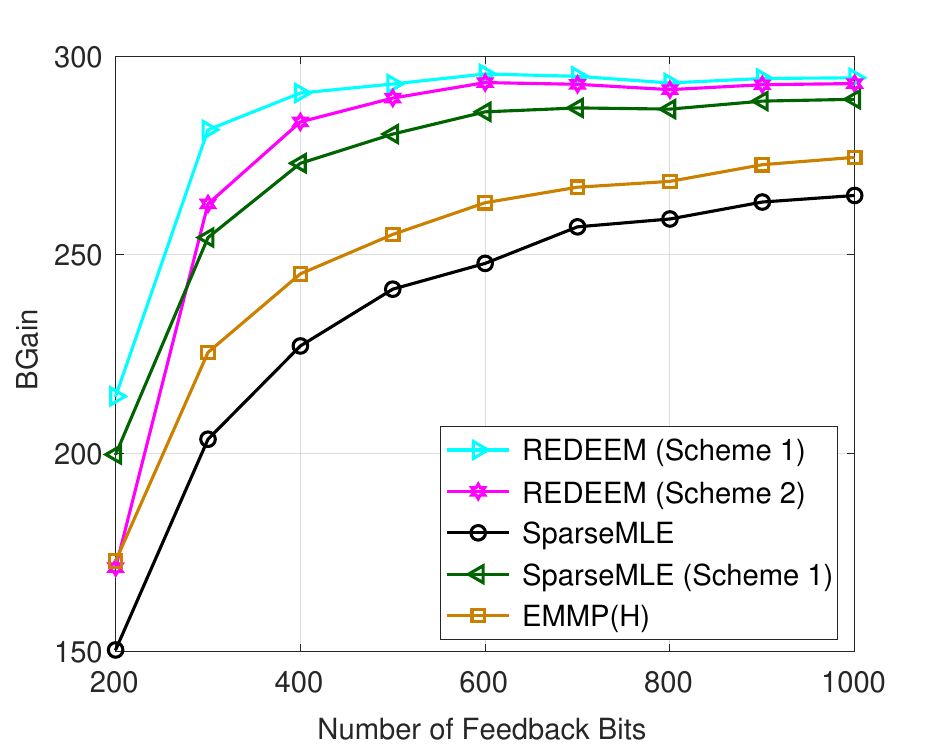}
		\caption{ $(M,K,Q)=(16,6,2)$.}\label{fig:quant_2bit bgain}
	\end{subfigure}
    \hfill
	\begin{subfigure}[b]{.49\linewidth}
	\centering
		\includegraphics[width=1.1\linewidth]{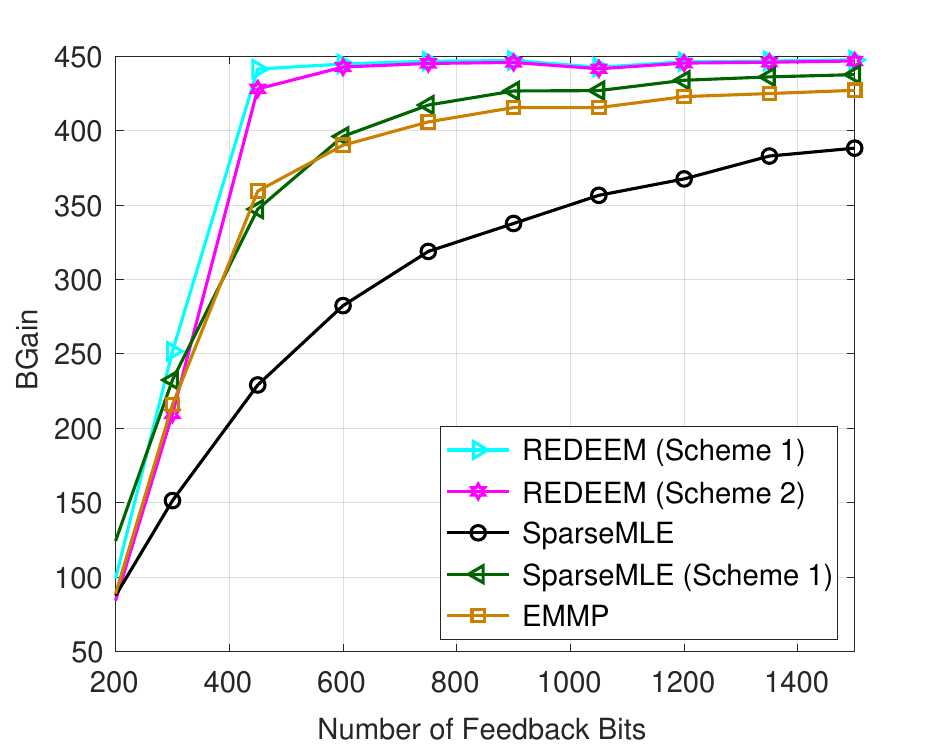}
		\caption{ $(M,K,Q)=(24,8,3)$.}\label{fig:quant_3bit bgain}
	\end{subfigure}
	\caption{ BGain performance under different $Q$'s when $N=32$.}
	\label{Fig:feedback_bit on bgain}
\end{figure}

{ Fig.~\ref{Fig:feedback_bit} and Fig.~\ref{Fig:feedback_bit on bgain} show the  NMSE and BGain performance, respectively, under various numbers of feedback bits. Fig.~\ref{Fig:feedback_bit} (a) and (b) present the NMSE results under $Q=2$ bits and $Q=3$ bits, respectively.}
It is seen that both of the proposed schemes achieve promising channel estimation accuracy.
In particular, Fig.~\ref{Fig:feedback_bit} (a) shows that when 500 bits are used, Schemes 1 and 2 reach NMSEs of around $2\times 10^{-2}$ and $4\times 10^{-2}$, respectively, while the baselines all admit NMSEs larger than $10^{-1}$. { Fig.~\ref{Fig:feedback_bit on bgain} (b) shows that when $Q=3$ and 500 bits are used for feedback, our methods are able to attain a BGain value larger than $430$, while other methods are below around $360$. } Note that if one feeds back the whole $\bm H$ with full precision (i.e., using 32 bits to represent a real value), 32,768 bits would be needed. Using 500 bits costs 1.5\% of the overhead. 

\begin{figure}[t!]
    \centering
    \includegraphics[width=0.8\linewidth]{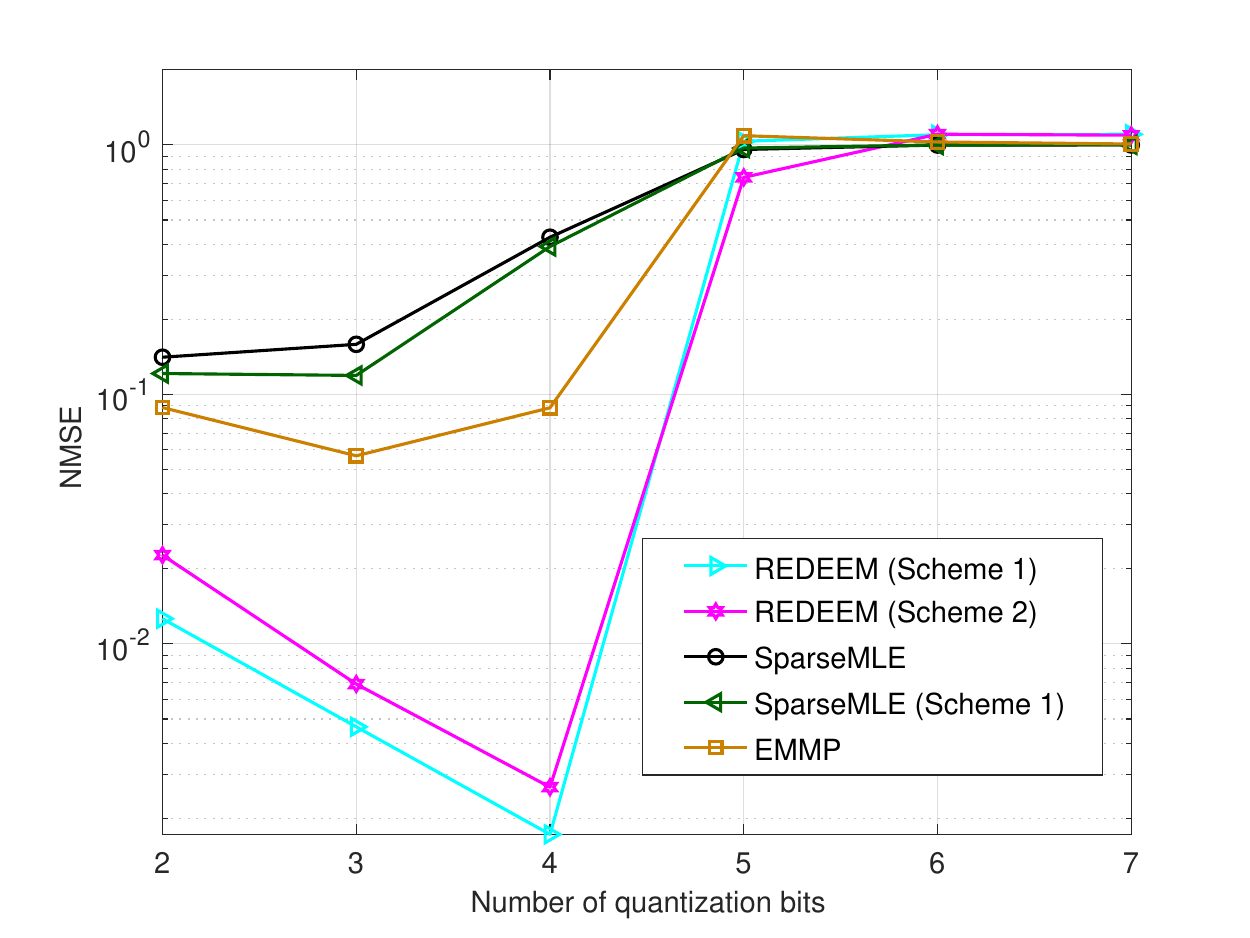}
    \caption{ NMSE performance under different $Q$'s with a fixed 900 bits feedback budget; $N=32$, $M=16$ $K=6$.}
    \label{fig:tradeoff between quantization bits and measurements R or T}
\end{figure}

{ Fig. ~\ref{fig:tradeoff between quantization bits and measurements R or T} shows the NMSE performance under a $900$-bit feedback budget as $Q$ changes. 
Note that both of our schemes enjoy growing accuracy when $Q$ increases from 2 to 4 and
attain their best performance at $4$-bit quantization.
This is because increased $Q$ makes brings more fine-grained quantization interval.
However, when $Q>5$, allocating more bits to the quantizer leaves too few measurements ($R$ for Scheme $1$ and $T$ for Scheme $2$) to for feedback, hurting the overall performance.
This presents an interesting tradeoff for the system designers to balance with.
}

\begin{figure}
	\centering
    \includegraphics[width=0.9\linewidth]{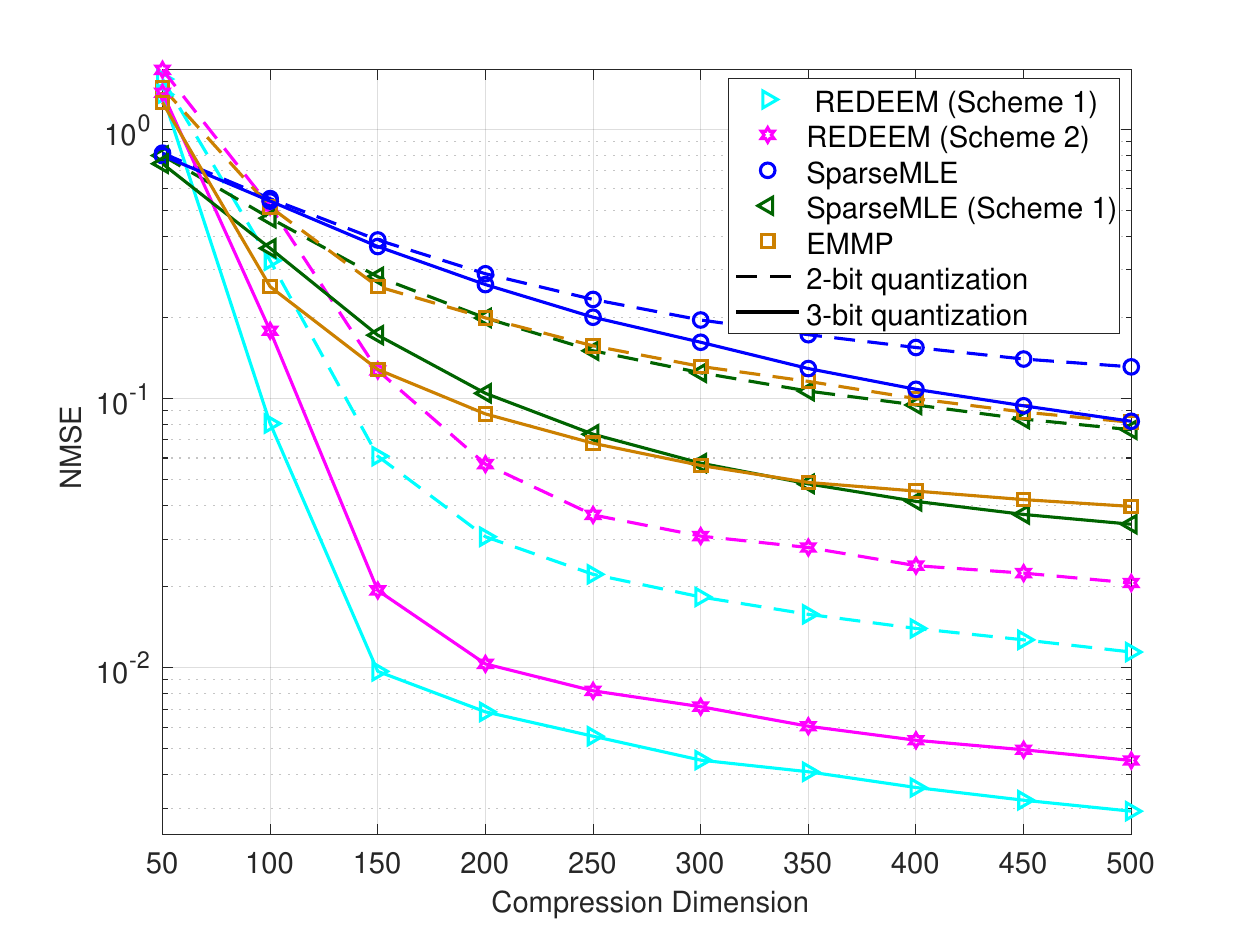}
	\caption{NMSE performance under different number of measurements ($R$ and $T$ for Scheme 1 and 2); $N=32$, $M=16$, $K=6$; Solid line: $Q=3$; Dash line: $Q=2$.}  
	\label{Fig:quant}
\end{figure}

Fig.~\ref{Fig:quant} shows the NMSE performance under different  feedback dimensions (i.e., $R$ and $T$ in Schemes 1 and 2, respectively). It is seen that with increasing quantization levels, the performance of all the methods improves. 
In particular, with  2 to 3 bits, our proposed schemes can achieve promising estimation accuracy, and perform much better than the other methods. 
Interestingly,  with the same number of  600  feedback bits, the case of 3-bit quantization and compression dimension $R=200$ performs better than the case of 2-bit quantization and compression dimension $R=300$,  showing that the quantization precision may play a more important role in our recovery algorithm.

\begin{figure}[t!]
    \centering
\begin{subfigure}[b]{.49\linewidth}
	\centering 
       \includegraphics[width=1.1\linewidth]{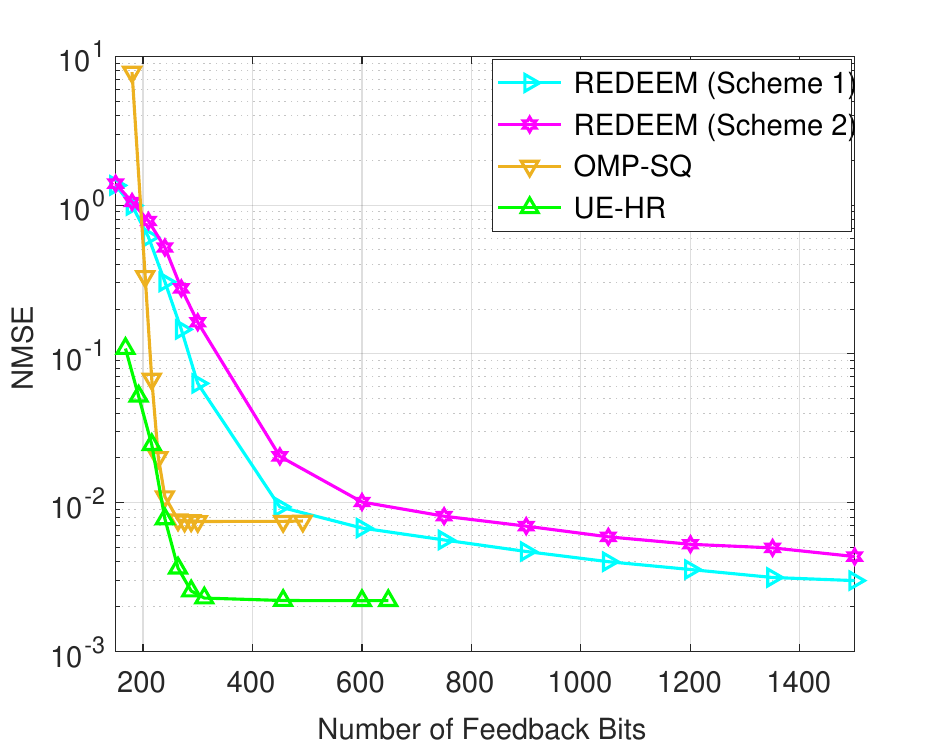}
    \caption{ NMSE of the methods.}
    \label{fig:PE1-bscompute}
	\end{subfigure}
    \hfill
	\begin{subfigure}[b]{.49\linewidth}
	\centering
    \includegraphics[width=1.1\linewidth]{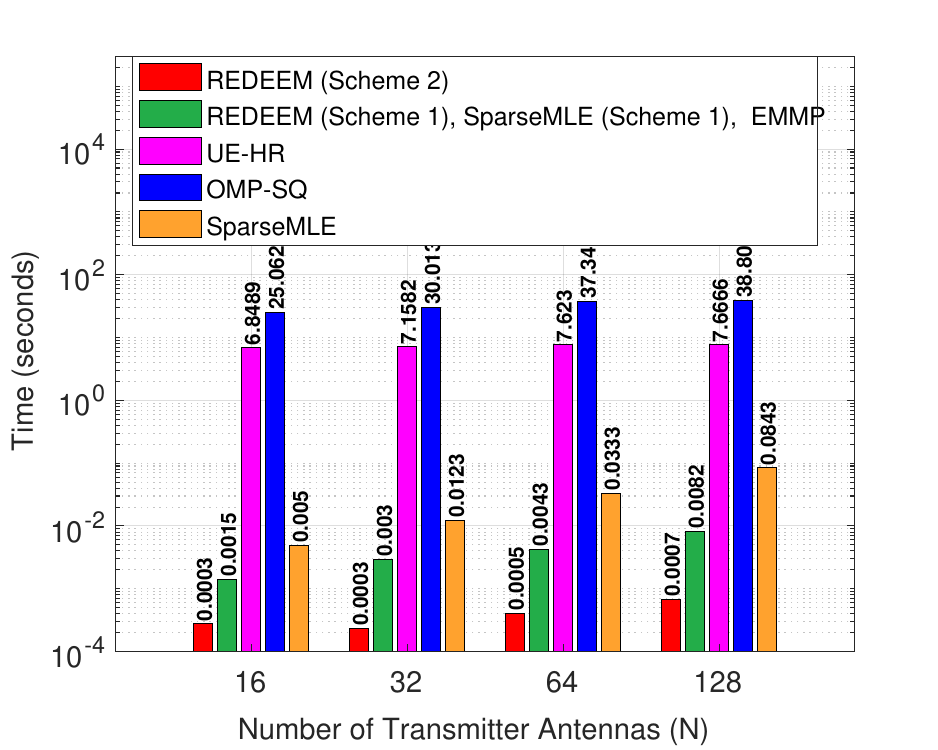}
    \caption{PT at UE; 900 feedback bits. }
    \label{fig:feedback preparation time}
    \end{subfigure}
    \caption{Performance of the proposed methods and the UE-based methods; $N=32$, $M=16$, $K=6$, $Q=3$.}
    \label{fig:BScomputecomparison}
\end{figure}

Fig.~\ref{fig:BScomputecomparison} (a) and Fig.~\ref{fig:BScomputecomparison} (b) compare the proposed schemes with baselines that estimate the channel parameters at the UE, namely, \texttt{OMP-SQ} and \texttt{UE-HR}. 
The former presents the NMSE performance and the latter shows the ``preparation time { (PT)}'' performance, which is the processing time needed at UE before sending the feedback bits. The preparation time reflects the complexity of UE operations, e.g., Gaussian compression in \eqref{eq:compress}, dithered quantization in \eqref{eq:quant}, or the rank-1 projection of our Scheme 2 in \eqref{eq:ins_comp}. 
{ For methods that estimate $(\bm \theta,\bm \beta,\bm\phi)$ at the UE, the preparation time also includes the time for executing the channel estimation algorithms (e.g., least squares).}
In general, one does not wish to spend long preparation time at the UE side due to the fast changing nature of wireless channels.
It is seen that both \texttt{OMP-SQ} and  \texttt{UE-HR} yield favorable NMSE performance. Nonetheless, they attain low NMSEs at the cost of sharply increased computational burden and the preparation time at the UE side.
 On the other hand, our proposed method only requires light process at the UE side---that is, when $Q\times R=600$
our Schemes 1 and 2 only cost 0.0419\% and  0.00419\% of UE-HR's preparation time to reach a reasonable NMSE of $ 1 \times 10^{-2}$, respectively. In particular, to achieve the same channel estimation accuracy as Scheme 1, Scheme 2 reduces the computational complexity at the UE side without significantly increasing the feedback overhead. This may help reduce the overall delay at the BS in estimating the downlink channel.

\begin{figure}[t!]
    \centering
    \begin{subfigure}[t]{0.49\linewidth}
     \includegraphics[width=1.1\linewidth]{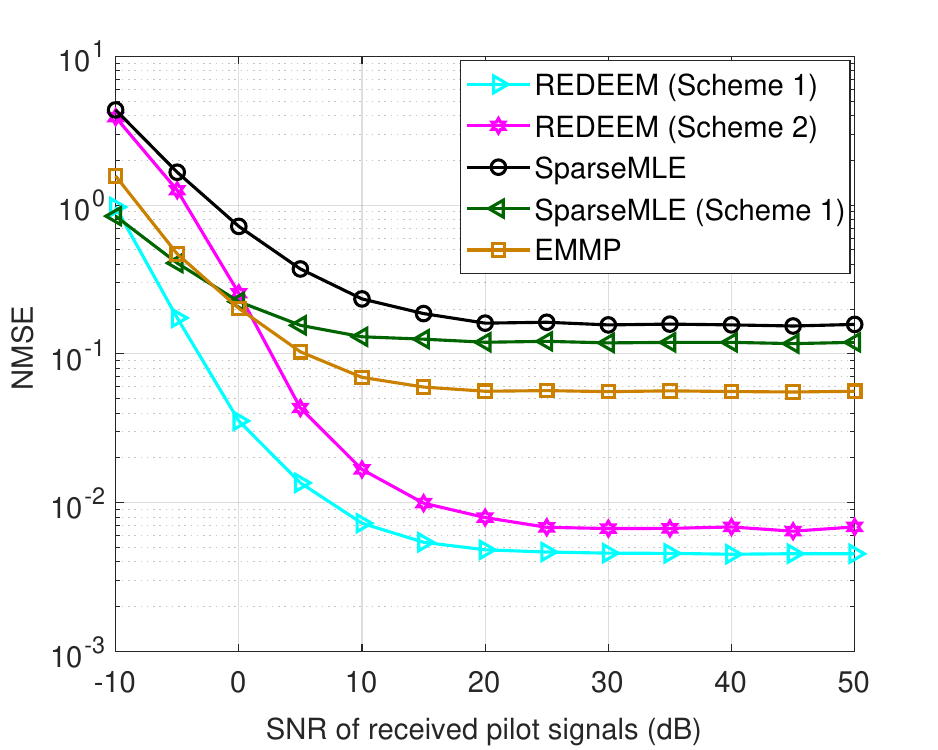}
    \caption{ NMSE vs SNR; $N=32$, $M=16$, $K=6$, $Q=3$, 900 feedback bits. }
    \label{fig:PE3}
    \end{subfigure}
    \hfill
    \begin{subfigure}[t]{0.49\linewidth}
        \includegraphics[width=1.1\linewidth]{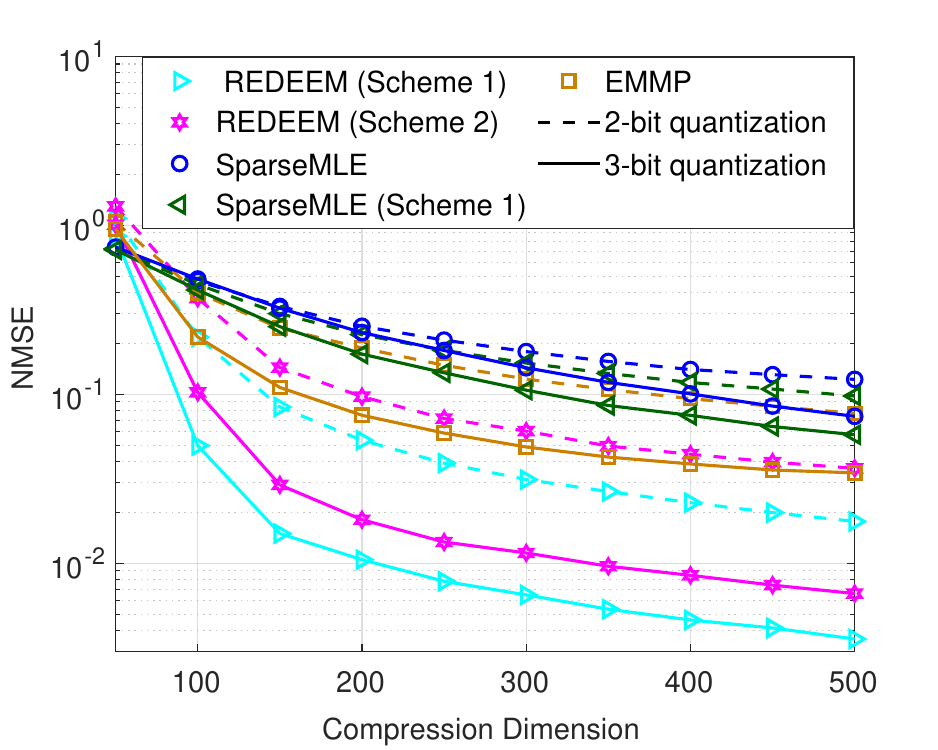}
        \caption{ NMSE vs $R$ and $T$ when $|\beta_i|$'s are sampled from  [0.01,1].}
        \label{fig:highdynamic range of pathloss}
    \end{subfigure}
    \caption{ Robustness evaluation with respect to SNR of received signals and high dynamic range of path loss}
    \label{fig:robustness evaluation}
\end{figure}

{ Fig.~\ref{fig:robustness evaluation} (a) and Fig.~\ref{fig:robustness evaluation} (b) show the sensitivity of the methods under a couple of environmental parameters.
Specifically,  Fig.~\ref{fig:robustness evaluation} (a) shows the CSI recovery performance under different SNRs, where we fix the compression dimension $R$ and $T$ to be $300$} and use $Q=3$.
{ The SNR refers to the received pilot signal to noise ratio at the UE.}
It is seen that all the considered algorithms get better as SNR increases. Our proposed schemes show a significant performance margin than existing designs---the NMSEs are at least 40\% lower than those of the baselines when SNR$\geq 5$dB. { Fig.~\ref{fig:robustness evaluation} (b) shows the NMSE under a wide range path-loss amplitudes (i.e., $\{ \beta_i \}$).
In this simulation, the $|\beta_i|$'s are drawn uniformly from a dynamic range of $[0.01, 1]$ (instead of $[0.5,1]$ as in previous simulations). One can see that our proposed schemes still outperform other baselines. }

\begin{figure}[t!]
    \centering
\begin{subfigure}[t]{.49\linewidth}
    \includegraphics[width=1.1\linewidth]{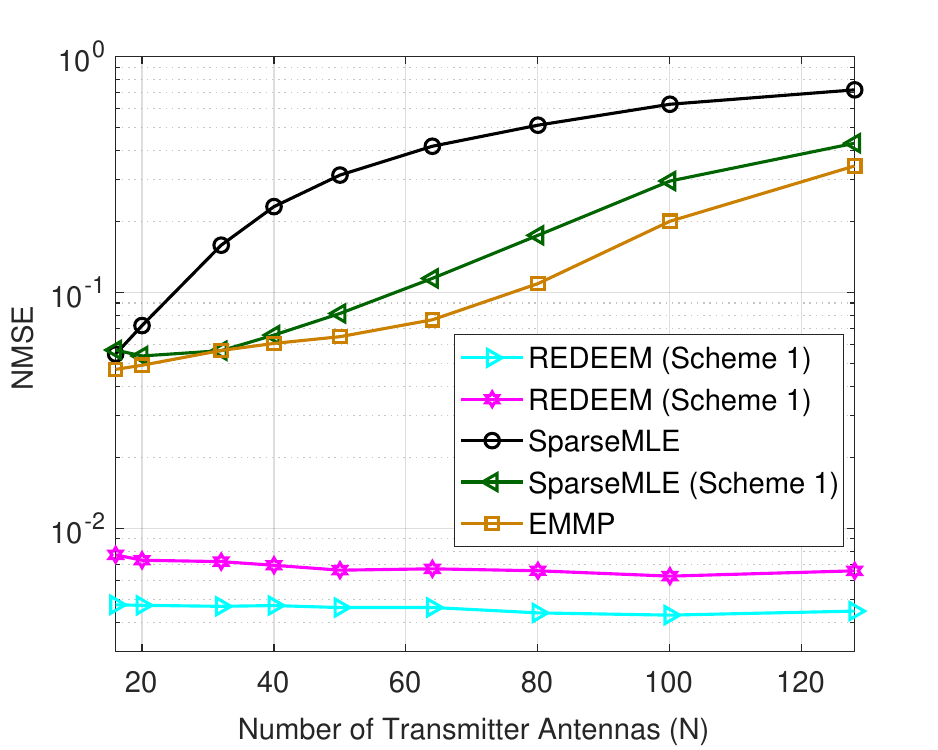}
    \caption{NMSE v.s. $N$; $M=16$, $K=6$, $Q=3$, 900 feedback bits.}
    \label{fig:PE5}
\end{subfigure}
\hfill
   \begin{subfigure}[t]{.49\linewidth}
        \centering
        \includegraphics[width=1.17\linewidth]{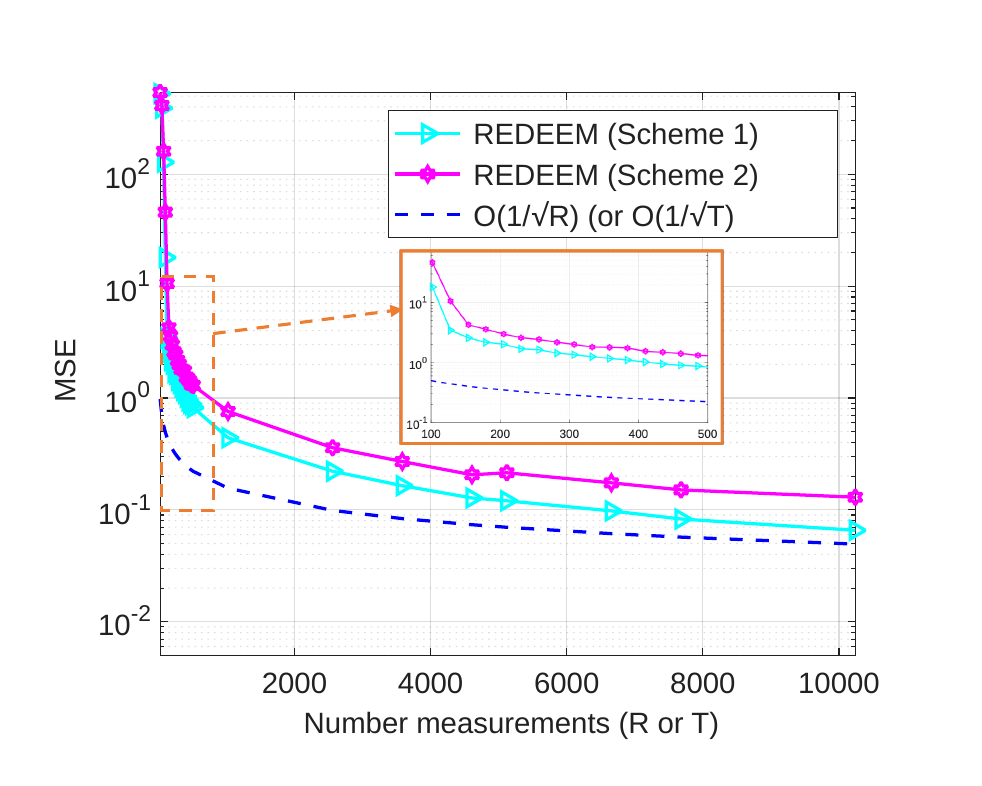}
        \caption{ MSE vs $R$ and $T$.}
        \label{fig:tightness of mse bounds}
    \end{subfigure}
    \caption{ Theorem validation: insensitivity to $N$ and MSE decaying rate when $R$ and $T$ grow.}\label{fig:validation} 
\end{figure}

{ 
Fig.~\ref{fig:validation} validates the our recoverability theorems.
In particular,
}
Fig.~\ref{fig:validation} (a) shows the impact of increasing the number of $N$, i.e., the number of transmitter antennas.
The NMSEs of the proposed schemes are almost unaffected under this test setting.
In contrast, the performance of other methods degrades faster as the antenna size increases.  
{ 
The insensitivity of our methods to the growing $N$ is consistent with our theorems: The performance of our method is not affected by $M$ or $N$ when $R$ is larger than $\Omega(K)$.}
{ 
Fig. \ref{fig:validation} (b) shows that the algorithm-obtained MSE values indeed decays with a rate roughly at the predicted order, namely, $O(1/\sqrt{R})$ and $O(1/\sqrt{T})$ for Schemes 1 and 2, respectively.
}

\begin{figure}[t!]
    \centering
    \includegraphics[width=0.8\linewidth]{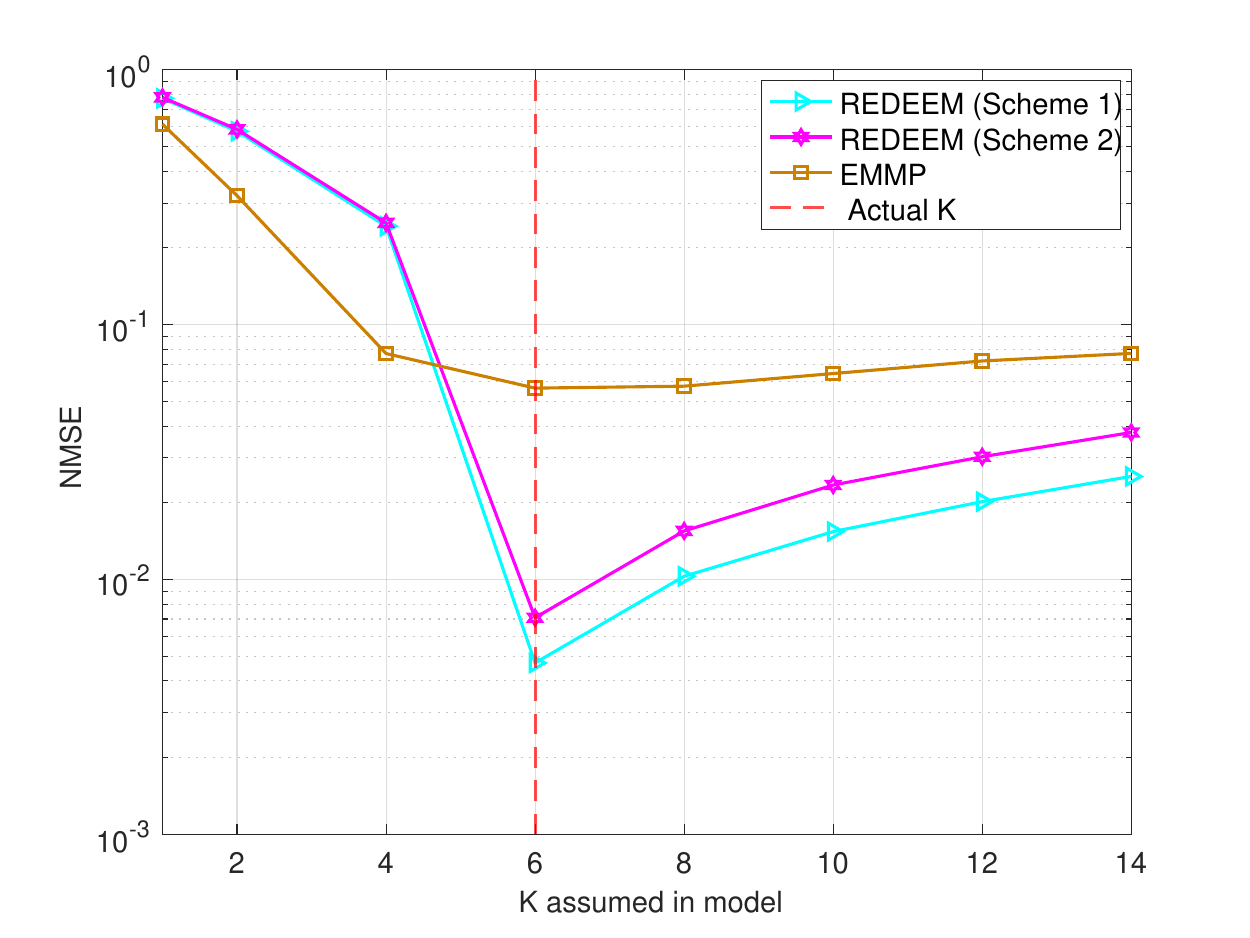}
    \caption{Sensitivity to the deviation of $K$ from the actual value; $M=32$, $N=16$, Actual $K=6$, $Q=3$, 900 feedback bits.}
    \label{fig:PE4}
\end{figure}

Fig.~\ref{fig:PE4} tests the algorithm robustness against mis-specified $K$, i.e., the number of paths.
The ground-truth $K$ is set to $6$.
It is seen that our methods perform best when the value  $K$  exactly matches the actual $K$ value. When $K$ is over-estimated, the NMSE only degrades mildly. However, when $K$ is under-estimated, the CSI recovery sees significant deterioration. This is understandable, as over-estimation of the rank of a matrix does not lose information, yet under-estimation is more detrimental.

\subsection{DeepMIMO-based Simulations}
\subsubsection{Data Generation} 
We generate the channel matrix by using the DeepMIMO-v2 tool \cite{Alkhateeb2019deepmimo, Remcom} that uses ray-tracing to simulate the real-world channels. 
We present our results in three different scenarios provided by DeepMIMO: \emph{Outdoor 1 Blockage Scenario (O1)}, \emph{Boston5G Scenario (B5)}, and \emph{Indoor 3 Scenario (I3)}.

For each scenario, DeepMIMO uses several predefined BS locations, identified using labels such as ``BS1'' and ``BS2''. 
For the \emph{O1} scenario, we use BS3, and for both \emph{B5} and \emph{I3}, we select BS1.  The exact positions of these BS are shown in their corresponding scenario figures; see \cite{Alkhateeb2019deepmimo} and the illustration in \url{https://www.deepmimo.net/scenarios/}. 
The UEs are placed into the environment space uniformly at random. For both the BS and UE, we use the same antenna configuration and interelement spacing as in Section \ref{sec:channel model simulation}, along with identical quantization intervals and dithering noise levels. All experiments here employ  $N=32$ transmit antennas, $M=16$ receive antennas, $K=6$ propagation paths, and $3$-bit quantization. 

Compared to our basic simulations, DeepMIMO creates more complex and challenging scenarios.
DeepMIMO labels each channel with its Line-of-Sight (LOS) status, denoting whether an unobstructed direct path exists between the BS and UE exists. We refer to those with a clear direct path as With-Line-of-Sight (WLOS) channels, and  those without any direct path as Non-Line-of-Sight (NLOS) channels. 
The DeepMIMO uses the ray-tracing model to generates channel whose combined path loss statistics are listed in Table.~\ref{tab:combined_pathloss_stats}; ``combined'' here means an overall characterization of all path losses.
It is seen that the path losses generated by DeepMIMO have a wide dynamic range, presenting challenging channel estimation scenarios.

\begin{table}[!t]

\centering
\caption{ Combined path loss characteristics for the DeepMIMO scenarios; provided \cite{Alkhateeb2019deepmimo}.}
\label{tab:combined_pathloss_stats}
\small
\setlength{\tabcolsep}{4pt}
\begin{tabular}{|l|c|c|c|}
\hline
\textbf{Scenario} & \textbf{Max (dB)} & \textbf{Min (dB)} & \textbf{Mean \(\pm\) Std (dB)} \\
\hline
\emph{O1} & \(151.41\) & \(58.93\) & \(86.51 \pm 8.88\) \\
\emph{B5} & \(133.03\) & \(60.61\) & \(93.02 \pm 13.77\) \\
\emph{I3} &  \(79.25\) & \(45.18\) & \(53.89 \pm 3.21\)  \\
\hline
\end{tabular}
\end{table}

\subsubsection{Results}
\begin{figure}[!t]
    \centering
    \begin{subfigure}[b]{0.49\linewidth}
    	\centering
        \includegraphics[width=1.1\linewidth]{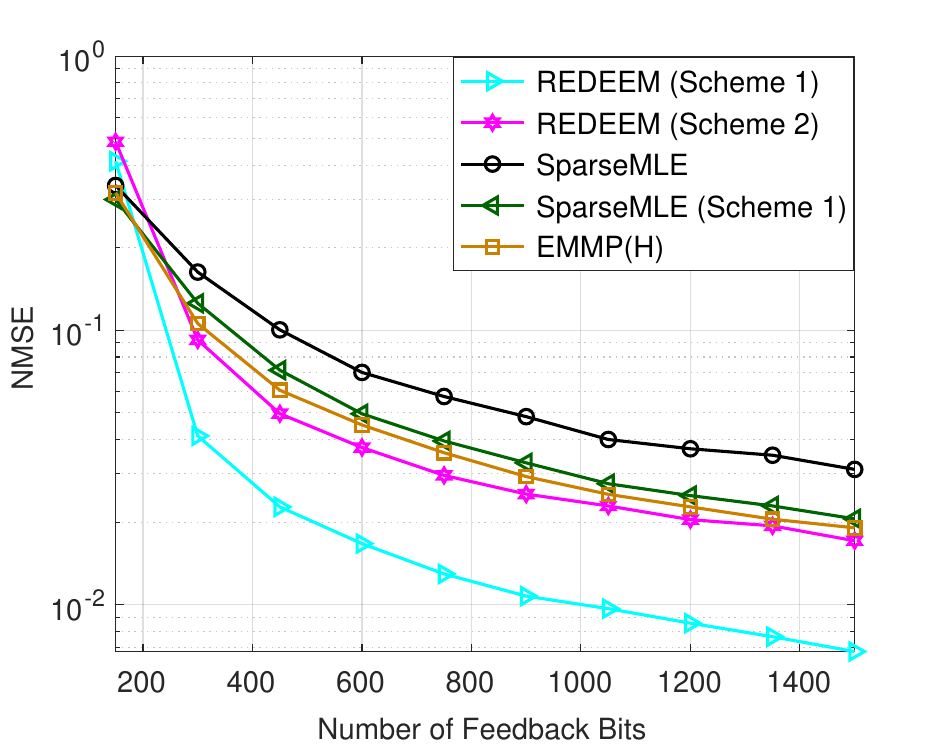}
        \caption{ WLOS}
        \label{fig: LOS for O1}
    \end{subfigure}
    \begin{subfigure}[b]{0.49\linewidth}
    	\centering
        \includegraphics[width=1.1\linewidth]{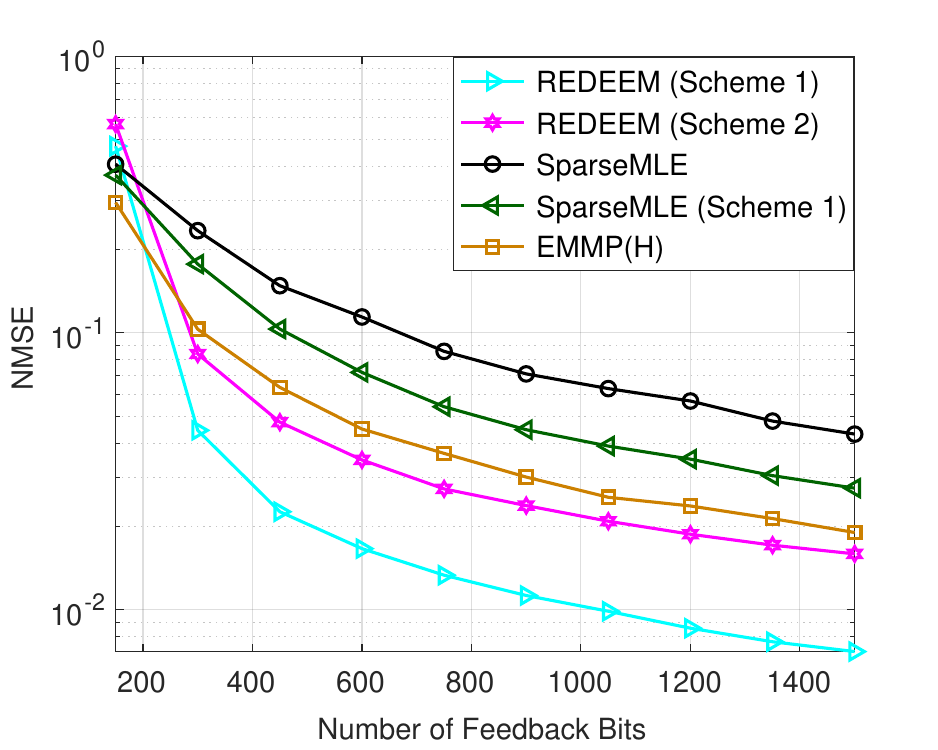}
        \caption{ NLOS}
        \label{fig: NLOS for O1}
    \end{subfigure}
    \caption{ Performance under various numbers of measurements under the \emph{O1} scenario; $N=32$, $M=16$, $K=6$, $Q=3$. }
    \label{fig:performance in O1}
\end{figure}

\begin{figure}[!t]
    \centering
    \begin{subfigure}[b]{0.49\linewidth}
    	\centering
        \includegraphics[width=1.1\linewidth]{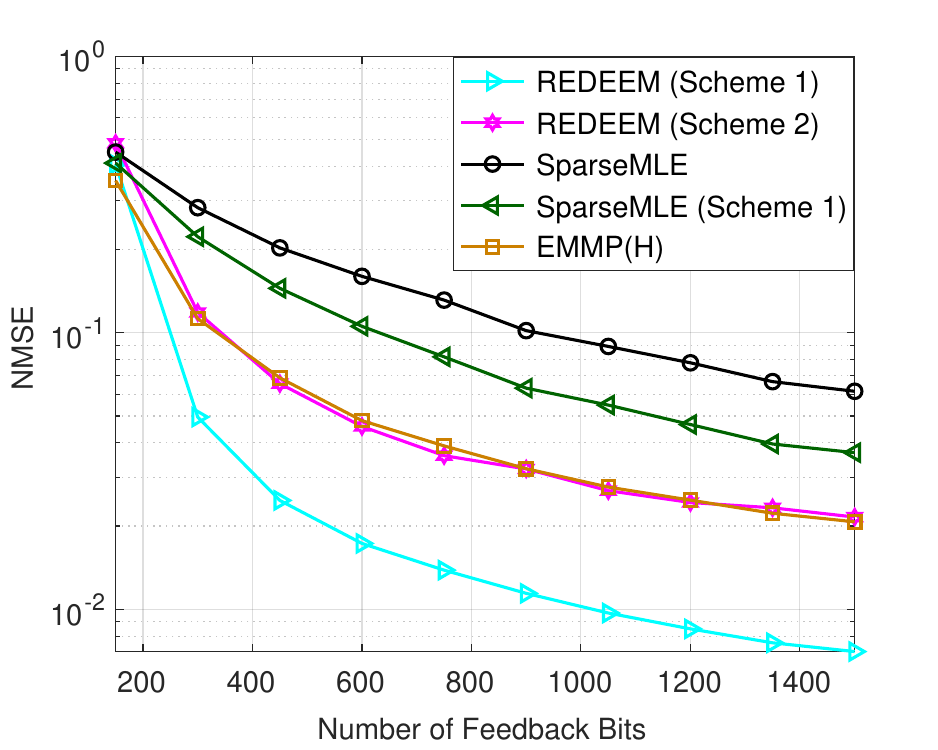}
        \caption{ WLOS}
        \label{fig: LOS for B5}
    \end{subfigure}
    \begin{subfigure}[b]{0.49\linewidth}
    	\centering
        \includegraphics[width=1.1\linewidth]{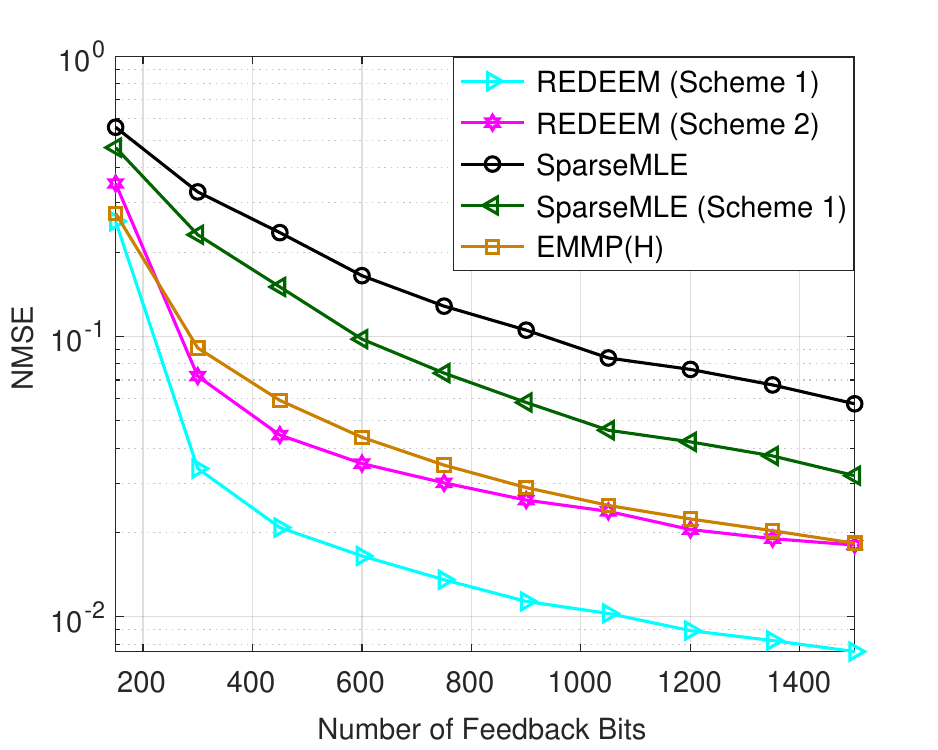}
        \caption{ NLOS}
        \label{fig: NLOS for B5}
    \end{subfigure}
    \caption{ Performance under various numbers of measurements under the \emph{B5} scenario; $N=32$, $M=16$, $K=6$, $Q=3$. }
    \label{fig:performance in B5}
\end{figure}

\begin{figure}[!t]
    \centering
    \begin{subfigure}[b]{0.49\linewidth}
    	\centering
        \includegraphics[width=1.1\linewidth]{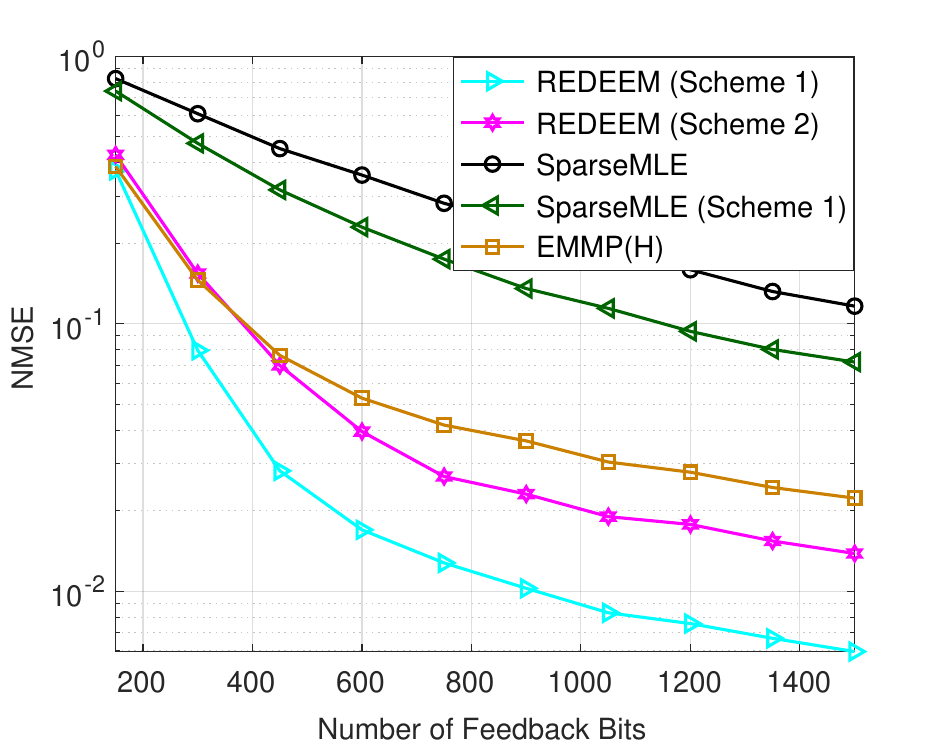}
        \caption{ WLOS}
        \label{fig: LOS for I3}
    \end{subfigure}
    \begin{subfigure}[b]{0.49\linewidth}
    	\centering
        \includegraphics[width=1.1\linewidth]{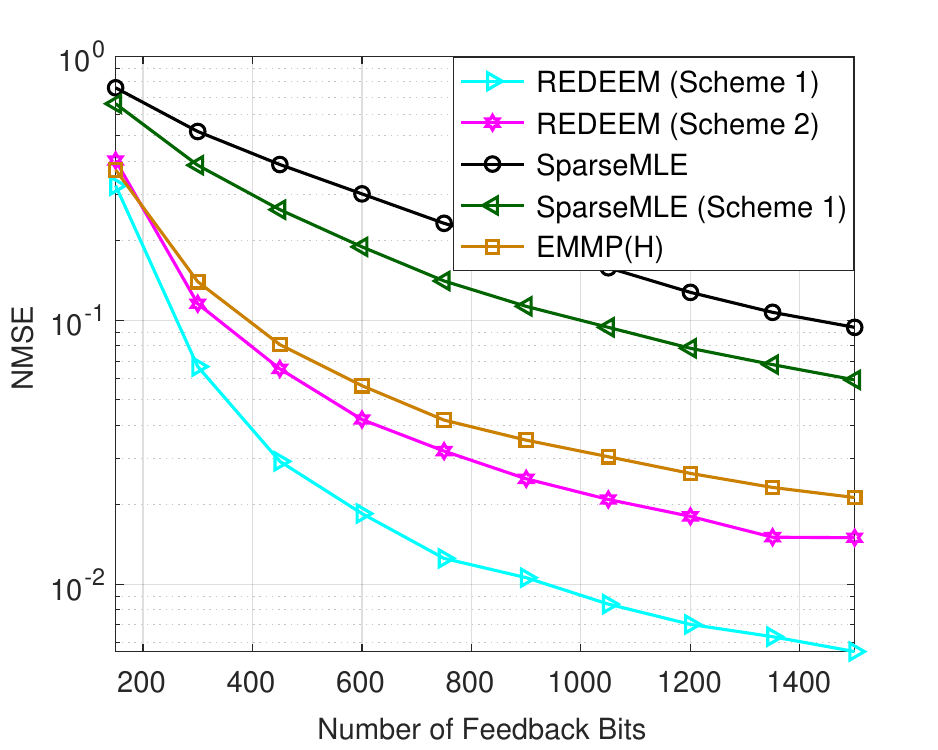}
        \caption{ NLOS}
        \label{fig: NLOS for I3}
    \end{subfigure}
    \caption{ Recovery performance under different feedback dimension for \emph{I3} scenario; $N=32$, $M=16$, $K=6$, $3$-bit quantization. }
    \label{fig:performance in I3}
\end{figure}

Fig. \ref{fig:performance in O1} compares the NMSE performance of the methods for the \emph{O1} scenario. 
Fig. \ref{fig:performance in O1} (a) shows the performance for WLOS paths, whereas \ref{fig:performance in O1} (b) shows for NLOS paths. 
It is seen that our methods have promising channel estimation results compared to the baselines---consistent with the results observed in the basic simulations.
Similarly, Figs. \ref{fig:performance in B5} and \ref{fig:performance in I3} showthat our proposed methods work well under channel configurations under the \emph{B5} and \emph{I3} scenarios, respectively.

\color{black}

\section{Conclusion}\label{sec:conc}

In this paper, we revisited the limited feedback-based CSI estimation problem in FDD MIMO systems. 
We proposed a random compression and noise dithering-based quantization framework for CSI feedback. 
Our approach features two different compression schemes that demonstrate a trade off between UE preparation time and feedback overhead.
Under our framework, most of the heavy computations are carried out by the BS.
We showed that, through MLE-based recovery at the BS, the proposed schemes can provably recover the downlink CSI under reasonable conditions, providing valuable insights into sample complexity of recovering array manifold-constrained channels.
To handle the challenging optimization problem associated with the MLEs,
we proposed an efficient ADMM algorithm, namely, \texttt{REDEEM}, that judiciously employs a sophisticated, classic HR algorithm to tackle its hard subproblem.
We conducted extensive numerical simulations to test the performance of our method, showcasing promising channel estimation accuracy.

{ 

There are several promising directions for future exploration. First, one could investigate \textit{more signal-adaptive designs} for the compression matrix. In this work, we used random compression matrices primarily to preserve properties from the JL Lemma, which was central to our recoverability analysis. However, leveraging structure in the channel or data to design tailored compression matrices could further reduce feedback overhead---an appealing and practical direction.
Second, on the theoretical front, \textit{more fine-grained characterizations} could be developed---for instance, clarifying the relationships between $U_f$, $L_f$, and $F_f$, and how they depend on the quantization level.
Third, from an implementation standpoint, \textit{integrating online optimization} or \textit{learning-based techniques} could significantly enhance computational efficiency and scalability.

}

\appendices

\section{Mod-RELAX Method} \label{app: mod-relax}

\begin{algorithm}
\caption{The Mod-RELAX algorithm \eqref{eq:z_sub}} \label{alg:RELAX}
\footnotesize
\begin{algorithmic}[1]
    \STATE \textbf{Input:} $\bH$, $\bm \Lambda$;
    \STATE Initialize $\bm\beta =\bm 0$;
  
    \FOR{$\eta = 1,\ldots, K$}
\REPEAT
    \FOR{$i = 1,2,\ldots, \eta-1$}
       
        \STATE Compute $\bm \zeta_i$ as in \eqref{eq:BCD};
        \STATE Use 2D-FFT to find a discretized solution $(\phi_i, \theta_i,\beta_i)$;
        \STATE Run gradient ascent \eqref{eq:gradientascent} multiple steps to refine the solution $(\phi_i, \theta_i,\beta_i)$;
        
    \ENDFOR
\UNTIL{$(\beta_j, \phi_j, \theta_j)_{j=1}^\eta$ converges}
    \ENDFOR
    \RETURN $(\beta_k, \phi_k, \theta_k)_{k=1}^K$
\end{algorithmic}
\end{algorithm}

\section{Proof of Fact~\ref{fact:cover}}
\label{app:proof_cover}

The parameter domains of AoA and AoD are given by \(\bm \theta \in [-\pi, \pi]^{K}, \quad \bm \phi\in [-\pi, \pi]^K\), repectively.
Then, the covering number for the $\epsilon_{1}$-net of $\setZ_{\theta}$ and $\setZ_{\phi}$ are given by
$$
\setC(\setZ_{\theta} \times \setZ_{\phi}, \epsilon_1) \leq \left( \nicefrac{\pi}{\epsilon_1} +1 \right)^{2K}\leq \left( \nicefrac{2\pi}{\epsilon_1}   \right)^{2K}
$$
for small $\epsilon_1$.
Also, for $\| \bm\beta \|_{\infty} \leq \kappa$, which can be guaranteed if $\|\Re(\bm \beta); \Im(\bm \beta)  \|_{\infty}\leq \kappa\sqrt{2}/2$, the covering number of the $\epsilon_2$-net for this set is given by
\[
    \setC(\setZ_{\beta}, \epsilon_2) \leq \left( \nicefrac{\sqrt{2}\kappa}{2\epsilon_2} +1 \right)^{2K}\leq \left( \nicefrac{2\kappa}{\epsilon_2}   \right)^{2K}.
\]
Thus, the covering number of the $\epsilon$-net of $\setZ $ for $\epsilon =  \sqrt{\epsilon_1^2+ \epsilon^2_2}$ is given by
\begin{equation*}
  \begin{split}
    \setC(\setZ, \epsilon) \leq &~\left( \nicefrac{2\pi}{\epsilon_1}   \right)^{2K} \left( \nicefrac{2\kappa}{\epsilon_2}   \right)^{2K} 
    = \left( \nicefrac{4\pi \kappa}{\epsilon_1\epsilon_2}   \right)^{2K}.
  \end{split}
\end{equation*}
When $\epsilon_1 = \epsilon_2 =\epsilon/ \sqrt{2}$, the right hand side expression attains its minimum, and
$ \setC(\setZ, \epsilon) \leq ~\left( \nicefrac{8\pi\kappa}{\epsilon^2}   \right)^{2K}.$
Moreover, since the mapping $\cal G$ is $L_{\cal G}$-Lipschitz according to Fact~\ref{fact:lip}, the covering number for $\Gamma$ is given by 
$ \setC(\Gamma, \epsilon) \leq  \left( \nicefrac{8\pi\kappa L_{\cal G}^2}{\epsilon^2}   \right)^{2K}.$
This completes the proof.

\section{Proof of Theorem~\ref{thm:rec_1}}
\label{app:proof_thm1}

We rewrite the ML estimation problem as
\begin{equation}\label{eq:ML_ref}
  \min_{\bx \in \setS} {\cal L} (\bx): = -\frac{1}{R} \sum_{i=1}^{R} \sum_{q\in \setU} \mathbb{1}(r_i = q) \log (f_q(x_i)),
\end{equation}
where 
$\setS : = \{ \bx = \bA \bh,~\bh \in\Gamma \},$
$\setU$ is the discrete quantization range set, and $f_q$ is the likelihood function defined in \eqref{eq:like_fun}.
We denote $\bx^{\natural} = \bA \bh^{\natural}$,  $\bx^{\star} = \bA \bh^{\star}$ and $\appxgt{\bx} = \bA \appxgt{\bh}$ as the compressed quantities of the ground-truth  $\bh^{\natural}$, the MLE solution $\bh^{\star}$ and the closest point $\appxgt{\bh}$ to the ground truth in $\Gamma$, respectively.
Due to the optimality of $\bh^{\star}$, we have
$
    {\cal L}(\bx^{\star}) \leq {\cal L}(\appxgt{\bx}).
$
Therefore:
\begin{equation}\label{eq:gen_bound1}
  \begin{split}
     &{\cal L}(\bx^{\star}) \leq {\cal L}(\appxgt{\bx})\\
     \Rightarrow & {\cal L}(\bx^{\star}) - \Exp [ {\cal L}(\bx^{\star})] +\Exp [ {\cal L}(\bx^{\star})] - \Exp[{\cal L}(\bx^{\natural})] + \\
      &\Exp[{\cal L}(\bx^{\natural})] - {\cal L}(\bx^{\natural}) +  {\cal L}(\bx^{\natural}) -{\cal L}(\appxgt{\bx}) \leq 0\\
      \Rightarrow & \Exp[ {\cal L}(\bx^{\star})] - \Exp[{\cal L}(\bx^{\natural})]\leq   \Exp [ {\cal L}(\bx^{\star})] - {\cal L}(\bx^{\star}) \\
      & + |{\cal L}(\bx^{\natural})- \Exp[{\cal L}(\bx^{\natural})] | + |{\cal L}(\appxgt{\bx}) -{\cal L}(\bx^{\natural})|,
  \end{split}
\end{equation}
where the expectation is over $\bm r$; the randomness is introduced from $\bv$ and $\bA$.  We can write
\begin{equation*}
  \begin{split}
  & \Exp[{\cal L}(\bx^{\natural})] -  \Exp[ {\cal L}(\bx^{\star})] \\
  = & \Exp\left[  -\frac{1}{R} \sum_{i=1}^{R} \sum_{q\in \setU} \mathbb{1}(r_i = q) \log \left(\frac{f_q(x^{\natural}_i)}{f_q(x_i^{\star})}\right) \right] \\
  = &  -\frac{1}{R} \sum_{i=1}^{R} \sum_{q\in \setU} \Exp\left[\mathbb{1}(r_i = q) \right] \log \left(\frac{f_q(x^{\natural}_i)}{f_q(x_i^{\star})}\right)\\
  = & -\frac{1}{R} \sum_{i=1}^{R} \sum_{q\in \setU} f_{q}(x^{\natural}_i) \log \left(\frac{f_q(x^{\natural}_i)}{f_q(x_i^{\star})}\right)\\
  = & - {\rm KL}(\bx^{\natural}|| \bx^{\star}),
  \end{split}
\end{equation*}
where ${\rm KL}(\bx^{\natural}|| \bx^{\star}) =  \frac{1}{R}\sum_{i=1}^{R} {\rm KL}(x_i^{\natural}|| x_i^{\star})$ with
\(
    {\rm KL}(x_i^{\natural}|| x_i^{\star}) = \sum_{q\in \setU} f_{q}(x^{\natural}_i) \log \left(\frac{f_q(x^{\natural}_i)}{f_q(x_i^{\star})}\right)
\)
denoting the KL divergence between the two distributions induced by $x^{\natural}_i$ and $x_i^{\star}$, respectively.
As a result, the inequality \eqref{eq:gen_bound1} can be written as
\begin{equation*}
\begin{split}
  &  {\rm KL}(\bx^{\natural}|| \bx^{\star}) \leq \\
  &   \underbrace{\Exp [ {\cal L}(\bx^{\star})] - {\cal L}(\bx^{\star})}_{\rm term (a)}
       + \underbrace{|{\cal L}(\bx^{\natural})- \Exp[{\cal L}(\bx^{\natural})] |}_{\rm term (b)} + \underbrace{|{\cal L}(\appxgt{\bx}) -{\cal L}(\bx^{\natural})|}_{\rm term (c)}.
      \end{split}
\end{equation*}
Next, we analyze the terms (a), (b), and (c).

Consider (b) first. Note that $W_i =-\sum_{q\in \setU} \mathbb{1}(r_i = q) \log (f_q(x_i^{\natural}))$ for all $i$ are independent, where $W_i \in [0,U_f]$ holds.
Then, for any $0<\eta<1$, with probability at least $1-\eta{  -\vartheta}$, we have:
\[
 |{\cal L}(\bx^{\natural})- \Exp[{\cal L}(\bx^{\natural})] | \leq U_f \sqrt{ \frac{1}{2R}\log(2/\eta)},
\]
where we have used the Hoeffding's inequality.
The term (c) can be bounded as follows:
\begin{equation*}
  \begin{split}
    &|{\cal L}(\appxgt{\bx}) -{\cal L}(\bx^{\natural})|\\
    = & \frac{1}{R} \left| \sum_{i=1}^{R} \sum_{q\in \setU} \mathbb{1}(r_i = q) \log \left(\frac{f_q(\appxgt{x}_i)}{f_q(x_i^{\natural})}\right) \right| \\
    \leq & \frac{1}{R}\sum_{i=1}^{R}  \sum_{q\in \setU}  \mathbb{1}(r_i = q)  | \log(f_q(\appxgt{x}_i)) - \log(f_q(x_i^{\natural}) ) | \\
    \leq & \frac{1}{R} \sum_{i=1}^{R}  \sum_{q\in \setU}  \mathbb{1}(r_i = q) L_f  | \appxgt{x}_i - x_i^{\natural} |\\
    = & \frac{ L_f}{R}  \| \appxgt{\bx} - \bx^{\natural} \|_1 \leq  \frac{ L_f}{\sqrt{R}}  \| \appxgt{\bx} - \bx^{\natural} \|_2,
  \end{split}
\end{equation*}
where the first inequality is due to triangle inequality, the second inequality is due to the fact that the function $\log(f_q(x))$ is $L_f$-Lipschitz continuous\footnote{It follows that $\log(f_q(x_1)) - \log(f_q(x_2))\leq \max_{x}|\frac{f'_q(x)}{f_q(x)}|\cdot|x_1-x_2|$. By noting \eqref{eq:L_f}, it leads to $\max_{x} \left|\frac{f'_q(x)}{f_q(x)} \right|\leq L_f$ and the $L_f$-Lipschitz continuity of function $\log(f_q(x))$.}  
and the last inequality is due to the norm inequality $\|  \bx \|_1\leq \sqrt{R} \| \bx \|_2$ for $\bx \in \Rbb^{R}$.
By the Johnson-Lindenstrauss Lemma \cite{Baraniuk2008}, with probability as least $1-2e^{-\Omega(\alpha^2 R)}$, we have
\(
         \| \appxgt{\bx} - \bx^{\natural} \|_2 \leq (1+ \alpha) \| \appxgt{\bh} - \bh^{\natural} \|_2 
         \leq (1+ \alpha) \nu,
\)
where the second inequality is due to Assumption~\ref{asm:err}.

Term (a) amounts to the difference between sample average ${\cal L}(\bx^{\star})$ and the expected value $\Exp[{\cal L}(\bx^{\star})]$.
{ In addition, note that { all elements of } $\bx$ is bounded with probability  at least $1-\vartheta$.}
Hence, we can bound term (a) by the Rademacher complexity bounds \cite[Theorem 26.5]{shalev2014understanding}.
With probability of at least $1- \delta { - \vartheta}$,  we have
\begin{equation}\label{eq:terma}
  \begin{split}
    \Exp [ {\cal L}(\bx^{\star})] - {\cal L}(\bx^{\star}) \leq 2 \widehat{\setR}( {\cal W} ) + 4 U_f \sqrt{\frac{2\log(4/\delta)}{R}},
  \end{split}
\end{equation}
where the function class $\cal W$ is defined as
$  {\cal W} =\{ \br \rightarrow {\cal L}(\br, \bx) \mid \bx = \bA \bh,~ \bH \in \Gamma \}$
and the empirical Rademacher complexity of a function class $\setW$ is defined as
\begin{equation}
    \widehat{\setR}(\setW) :=  \Exp_{\bm \sigma}\left[ \underset{w \in \setW }{\sup} \frac{1}{R} \sum_{i=1}^R \sigma_i w(r_i) \right]
\end{equation}
where, $\sigma_1,\ldots,\sigma_R$ are independent Rademacher random variables uniformly sampled from $\{-1, 1 \}$.

By invoking the Talagrand's contraction Lemma \cite[Lemma 26.9]{shalev2014understanding} and the bound, we get
\begin{equation}\label{eq:rad_bound}
\widehat{\setR}( {\cal W} ) \leq L_f \widehat{\setR} (\setS),
\end{equation}
where $\setS$  denotes the parameter space of $\bx$ in \eqref{eq:compress} as
 \begin{equation}\label{eq:setS}
    \setS := \{ \bx\mid \bx = \bA \bh,~ \bH \in \Gamma \}.
  \end{equation}
By the Dudley's entropy integral \cite{bartlett2017spectrally}, the empirical Rademacher complexity can be bounded by the covering number
\begin{equation}\label{eq:cover_bound}
    \widehat{\setR} (\setS) \leq \xi (\setS),
\end{equation}
where
\(
    \xi(\setS) = \inf_{\mu>0} \left( \frac{4\mu}{\sqrt{R}} + \frac{12}{R} \int_{\mu}^{\sqrt{R}} \sqrt{\log \setC(\setS, \epsilon)}  \ d \epsilon \right).
\)

Since $\sqrt{\log \setC(\setS, \mu)}$ increases as $\mu$ decreases, the integral can be upper bounded by
$
\int_{\mu}^{\sqrt{R}} \sqrt{\log \setC(\setS, \epsilon)}  \ d \epsilon  \leq \sqrt{R}  \sqrt{\log \setC(\setS, \mu)}.$
As a result, we can bound
\(
    \xi(\setS) \leq \inf_{\mu>0} \left( \frac{4\mu}{\sqrt{R}} + \frac{12}{R} \sqrt{R}  \sqrt{\log \setC(\setS, \mu)} \right).
\)

It remains to characterize the covering number of the parameter space $\setS$.
Let $\bx$ be a point in $\Gamma$ and $\bz$ be  the nearest point to $\bx$ in the  $\mu/\| \bA \|_2$-net of $\Gamma$. 
As a result, $\bA \bx$ is a point in $\setS$  based on \eqref{eq:setS}.
Then, we have 
\[
\| \bA \bx - \bA \bz \|_2 \leq \| \bA \|_2 \cdot \| \bx - \bz \|_2 \leq \mu.
\]
This implies that the set $\{\bA \bz\mid \bz \in \mu/\| 
 \bA\|_2\mbox{-net of } \Gamma\}$  must include a $\mu$-net of $\setS$.
As a result, we get
\(
    \setC(\setS, \mu) \leq \setC(\Gamma, \mu/\|\bA\|_2).
\)

From \cite[Corollary 5.35]{vershynin2011introduction}, we have
\begin{align}\label{eq:A2}
 \| \bA \|_{2} \leq 2+ \sqrt{\nicefrac{J}{R}}    
\end{align}
with probability at least $1-2e^{-R/2}$.
We can choose $\mu = 2+ \sqrt{\nicefrac{J}{R}}$. Then, we arrive at
\(
    \xi(\setS) \leq \frac{4{ (2+\sqrt{J/R})}}{\sqrt{R}} + \frac{12}{\sqrt{R}  } \sqrt{{4K}\log \left(2\sqrt{2\pi\kappa} L_{\cal G}   \right) }.
\)

This result, combined with \eqref{eq:terma}, \eqref{eq:rad_bound} and \eqref{eq:cover_bound}, yields
\(
    \Exp [ {\cal L}(\bx^{\star})] - {\cal L}(\bx^{\star}) \leq  2\frac{L_f}{\sqrt{R}  } \tau   + 4 U_f \sqrt{\frac{2\log(4/\delta)}{R}} 
\)
with probability at least   $1- \delta { - \vartheta}-2e^{-R/2}$, where
$
\tau = 4\left(2+ \sqrt{\nicefrac{J}{R}}\right) + 12  \sqrt{{4K}\log \left(2\sqrt{2\pi\kappa} L_{\cal G}   \right) }.$

By choosing $\delta =2\eta$ and $\alpha =1/2$, and summing up the bounds on terms (a)-(c), we have
\begin{equation*}
  \begin{split}
    &{\rm KL}(\bx^{\natural}|| \bx^{\star}) \\
    \leq & 2\frac{L_f}{\sqrt{R}  } \tau   + 4 U_f \sqrt{\frac{2\log(2/\eta)}{R}} +\frac{3}{2} \nu  \frac{ L_f}{\sqrt{R}} +U_f \sqrt{ \frac{1}{2R}\log\frac{2}{\eta}}\\
    = &2\frac{L_f}{\sqrt{R}  } (\tau + \frac{3}{4} \nu )  + \frac{9}{{\sqrt{2}}} \frac{U_f}{\sqrt{R}} \sqrt{\log(2/\eta)}\\
    = & \frac{1}{\sqrt{R}}\left(2 L_f (\tau + \frac{3}{4} \nu )  + \frac{9}{{ \sqrt{2}}}U_f \sqrt{ \log(2/\eta)} \right)
  \end{split}
\end{equation*}
with probability at least $1- 3\eta{ - \vartheta}-e^{-\Omega( R)} $.

Next, we relate the KL divergence with $\| \bA (\bh^{\natural} - \bh^{\star}) \|_2$.
From \cite{davenport2014onebit}, it is known that the KL divergence is lower bounded by the Hellinger distance
\(
    d_{H}^{2}(\bx^{\natural}, \bx^{\star}) \leq {\rm KL}(\bx^{\natural}|| \bx^{\star}),
\)
where the Hellinger distance is defined by
\begin{equation*}
  \begin{split}
    &d_{H}^{2}(\bx^{\natural}, \bx^{\star})  =\frac{1}{R}\sum_{i=1}^{R}  d_{H}^{2}(x_i^{\natural}, x_i^{\star}) \\
 & d_{H}^{2}(x, y)  = \Big(\sqrt{f_q(x)}-\sqrt{f_q(y)}\Big)^2 +\\
 &\Big(\sqrt{1-f_q(x)}-\sqrt{1-f_q(y)}\Big)^2.
  \end{split}
\end{equation*}
Also, from \cite{cao2015categorical}, it is shown that the Hellinger distance can be further lower bounded by
\(
    \frac{F_f}{4} \frac{\| \bA (\bh^{\natural} - \bh^{\star}) \|_2^2}{R}  \leq  d_{H}^{2}(\bx^{\natural}, \bx^{\star}).
\)

As a result, we have
\begin{equation}\label{eq:KL_rec}
 \frac{\| \bA (\bh^{\natural} - \bh^{\star}) \|_2^2}{R}\ \leq \frac{4}{F_f}{\rm KL}(\bx^{\natural}|| \bx^{\star}).
\end{equation}
Consider the following Lemma.
\begin{Lemma}\label{lem:s_rec}
(S-REC) Let the compressing matrix $\bA\in \Rbb^{R \times J}$  and    $a_{i,j} \sim {\cal N}(0, 1/R )$, $ i=1,\ldots, R$, $j = 1,\ldots, J$.
  Let
  \[
   R = \Omega \left( K \log\frac{2\sqrt{2}\sqrt{\pi\kappa} L_{\cal G}}{\epsilon}   \right).
  \]
  Then, the matrix $\bA$ satisfies the S-REC property in the following sense:
  \[
    \| \bA (\bh - \bh') \|_2 \geq \gamma \|\bh - \bh'  \|_2 - \epsilon, \forall \bh, \bh' \in \Gamma
  \]
  with $0<\gamma<1$ and $\epsilon>0$,   with probability $1- e^{- \Omega((1-\gamma)^2R)}$.
\end{Lemma}
\begin{IEEEproof}
The proof is in the same vein of \cite[Lemma  4.1]{bora2017compressed}. 
The difference is as follows: \cite[Lemma  4.1] {bora2017compressed} is for the case where the parameter space is an $\ell_2$-norm ball, while in our case, the parameter space is the bounded manifold $\Gamma$, which is compact. 
We replace the {  covering number} for the $\ell_2$-norm ball in \cite[Lemma  4.1]{bora2017compressed} by the covering number for the manifold $\Gamma$, and the remaining proof remains the same. 
This leads to the desired result in Lemma~\ref{lem:s_rec}.
\end{IEEEproof}

Applying Lemma~\ref{lem:s_rec}, we have
\begin{align}
     \| \bA \dot{\bh} - \bA \bh^*\|_2  &\le  \| \bA \dot{\bh} - \bA \bh^\natural\|_2 + \| \bA \bh^* - \bA \bh^\natural\|_2 \nonumber\\
    &\le  \left(2+\sqrt{\nicefrac{J}{R}}\right) \nu + \| \bA \bh^* - \bA \bh^\natural\|_2, \label{eqn: bound compressed close and compressed optimal}
\end{align}
which holds with probability at least $1-2e^{-R/2}$.
The first inequality is obtained using the triangle inequality, and the second uses \eqref{eq:A2}.

Moving forward, we have
\begin{align}
   \frac{\| \bh^{\natural} - \bh^{\star} \|_2^2}{MN} \underset{(a)}{\leq} &~ \frac{(\| \bh^{\natural} - \appxgt{\bh}  \|_2 + \| \appxgt{\bh} - \bh^{\star} \|_2)^2}{MN} \nonumber\\
    \underset{(b)}{\leq }&~ \frac{(\|\bA( \appxgt{\bh} - \bh^{\star}) \|_2 + \epsilon + \gamma \nu)^2}{\gamma^2 MN} \nonumber\\
    \underset{(c)}{\leq} &~ \frac{4(2+\sqrt{\nicefrac{J}{R}})^2 \nu^2 + 2 \| \bA \bh^* -\bA \bh^\natural \|_2^2 + 2(\epsilon+\gamma \nu)^2}{\gamma^2 MN} \nonumber\\
    \le &~ \frac{2 \|\bA \bh^* - \bA \bh^\natural\|_2^2 + 2 \epsilon'^2}{\gamma^2MN}, \label{eq:Ahbound}
\end{align}
where $(a)$ is due to triangle inequality, $(b)$  is due to Lemma \ref{lem:s_rec}, $(c)$ is due to \eqref{eqn: bound compressed close and compressed optimal},  and $\epsilon'=\epsilon+\gamma\nu+2\sqrt{2}+\sqrt{\nicefrac{2J}{R}}$.

By substituting \eqref{eq:KL_rec} into \eqref{eq:Ahbound} and using $R\leq 4MN$, one can obtain
\(
    \nicefrac{\|\bA(\bh^{\natural} - \bh^{\star} ) \|_2^2}{MN} \leq \nicefrac{8}{\sqrt{R}}(4 \nicefrac{L_f}{F_f} (\tau + \nicefrac{3}{4} \nu )  + 9\nicefrac{U_f}{F_f} \sqrt{{ 2} \log(2/\eta)} ),
\) and \( \nicefrac{\epsilon'^2}{MN} \leq  \nicefrac{4\epsilon'^2}{R}\).
Choosing $\gamma =1/2$ and $\epsilon =1/K$, we obtain the result in Theorem~\ref{thm:rec_1},
with probability at least $1- 3\eta{ - \vartheta}-e^{-\Omega(R)}$.

\bigskip


\bibliography{refs}
\bibliographystyle{IEEEtran}

\clearpage
\setcounter{page}{1}

{\bf Supplementary Material of {``Downlink Channel Estimation from Compressed and Quantized Feedback''}}

{

\section{Per-iteration complexity of \texttt{REDEEM} at BS} \label{sec:per-iteration complexity at BS}
The per-iteration complexity of our proposed Scheme 1 algorithm is analyzed below:
\begin{enumerate}
    \item \emph{Dual variable update}: This step involves only matrix additions. The time complexity is  $O(MN)$.
 
    \item \emph{EM} update: Note that the expensive inverse operation in \eqref{eq:M_step} doesn't depend on the optimization variables, and it can be just computed once in the beginning. The time complexities of E step and the M step are $O(RMN)$ and $O(MNR)$, respectively.
    \item \emph{Mod-RELAX} update: This step uses 2D-DFT for solving \eqref{eqn:relaxformulation} whose performance is reliant on the discretization resolution of the DFT matrix. Our implementation has a time complexity of \(O(\bar{I}_{GA} \bar{I}_{RLX}\bar{I}_{RLX} K^3 MN + \bar{I}_{RLX} K^2 MN \log MN)\).
    \item Total per-iteration complexity is \(O(\bar{I}_{\text{EM}} MNR + \bar{I}_{GA} \bar{I}_{RLX}  K^3 MN + \bar{I}_{RLX} K^2 MN \log MN)\).
\end{enumerate}
In the above, $\bar{I}_{\text{GA}}, \bar{I}_{\text{EM}}$ and $\bar{I}_{\text{RLX}}$  denote the maximum number of iterations for gradient ascent, EM, and Mod-RELAX, respectively.
Empirically, the algorithm converges in a few iterations. Note that Scheme 2 has the same per-iteration complexity as that of Scheme 1, with the measurement dimension $R$ simply replaced by $T$.
}
\section{Proof of Fact~\ref{fact:lip}}\label{app:proof_lip}
To prove that the mapping $\cal G$ is Lipschitz continuous, it suffices to show the Jacobian of $\cal G$ is bounded; see, e.g., \cite[Theorem 9.7]{rockafellar2009variational}.
Let us denote the Jacobian matrix w.r.t. $\theta_k$ as
$\nabla_{\theta_k} \bm H \in\mathbb{R}^{M\times N}$, where
$ [\nabla_{\theta_k} \bm H]_{m,n} = \nicefrac{ \partial [\bm H]_{m,n}  }{\partial \theta_k}$.
Then, it is readily seen that
\begin{align*}
    \left \| \nabla_{\theta_k} \bH \right \|_{\rm{F}} &\le |\beta_k| \frac{2\pi d}{\lambda} M\sqrt{MN}
\end{align*}
for all $k=1,\ldots,K$.
Similarly, we have
\[    \left \| \nabla_{\phi_k} \bH \right \|_{\rm{F}} \le |\beta_k| \frac{2\pi d}{\lambda} N\sqrt{MN},~
    \left \| \nabla_{\beta_k} \bH \right \|_{\rm{F}} \le \sqrt{MN}.\]

Then, it can be seen that there exists a finite $L_{\cal G}>0$ such that
\begin{align*}
    L_{\cal G} &\le \sup_{z\in {\cal Z}} \sqrt{\sum_{k=1}^K \left(  \left \| \nabla_{\theta_k} \bH \right \|_{\rm{F}}^2 + \left \| \nabla_{\phi_k} \bH \right \|_{\rm{F}}^2 + \left \| \nabla_{\beta_k} \bH \right \|_{\rm{F}}^2  \right)}\\
    &\le \sqrt{KMN} \sqrt{ 1+ \left(\frac{2\pi d \kappa}{\lambda}\right)^2(M^2+N^2)},
\end{align*}
 where $\kappa$ is the upper bound of $\| \bm \beta \|_\infty $ as defined in \eqref{eq:setZ}.
This proves the existence of a finite Lipschitz constant of the function ${\cal G}(\cdot)$.

\section{Proof of Theorem \ref{thm:rec_2}}
\label{app:proof_thm2}

The proof of Theorem~\ref{thm:rec_2} 
follow the same rationale as Theorem~\ref{thm:rec_1}, but is more challenging.
The crux is to verify whether the matrix $\bD$ satisfies the S-REC property.
Towards this goal, we show the following important result.

\begin{Lemma}\label{lem:s_rec2}
(S-REC of $\bD$) Let the compressing matrix $\bD\in \Rbb^{{ T} \times J}$  and
\[
    \bD  = \frac{1}{{ T}} (\bS^{\top} \odot \bA^{\top})^{\top},
\]
with
\begin{equation*}
  \begin{split}
   & s_{i,j} \sim {\cal N}(0, 1), \quad i= 1,\ldots, 2N,\quad j = 1,\ldots, { T},\\
    & a_{i,j} \sim {\cal N}(0,1), \quad i= 1,\ldots, 2M,\quad j = 1,\ldots, { T}
  \end{split}
\end{equation*}
and $J = 4MN$.
  Let
  \[
   { T} = \Omega \left( K^2 \left(\log\frac{4\sqrt{2}\sqrt{\pi\kappa} L_{\cal G}}{\epsilon} \right)^2  \right).
  \]
  Then, the matrix $\bD$ satisfies the S-REC property in the following sense:
  \[
    \| \bD (\check{\bh} - \check{\bh}') \|_2 \geq \gamma \|\check{\bh} - \check{\bh}'  \|_2 - \epsilon, \forall \check{\bh}, \check{\bh}' \in \check{\Gamma}
  \]
  with $0<\gamma<1$ and $\epsilon>0$,   with probability $1- e^{- \Omega((1-\gamma)^2{ T})}$.
\end{Lemma}

The proof for Lemma~\ref{lem:s_rec2} is shown in Section \ref{app:proof_lemmasrec2}.
With Lemma~\ref{lem:s_rec2}, the remaining proof is the same as those for Theorem~\ref{thm:rec_1}.

\section{Proof of Lemma~\ref{lem:s_rec2}}
\label{app:proof_lemmasrec2}

To show the   result in Lemma~\ref{lem:s_rec2}, we first show the matrix $\bD$ satisfies the distributional JL-Moment property, which is stated as follows in Lemma~\ref{lem:DJL}.
\begin{Lemma}\label{lem:DJL}
With
\[
 N_{S} \geq \max( c_1 \alpha^{-2} \log(1/\epsilon), c_2 \alpha^{-1} (\log(1/\epsilon))^2 ),
\]
and fixed $\alpha$
for any vector $\bx \in \Rbb^d$, we have
  \[
  \Pr \left( \left| \frac{ \| \bD \bx \|_2^2}{\|  \bx \|_2^2} -1 \right| \geq \alpha \right)\leq \epsilon.
  \]
Also, for any $\epsilon>0$ and fixed ${ T}$, if
\[
    \alpha \geq \max \left( \sqrt{\frac{c_1}{{ T}} \log \frac{1}{\epsilon}}, \frac{c_2^2}{{ T}}\left(\log\frac{1}{\epsilon}\right)^2 \right),
\]
 we have
   \[
  \Pr \left( \left| \frac{ \| \bD \bx \|_2^2}{\|  \bx \|_2^2} -1 \right| \geq \alpha \right)\leq \epsilon.
  \]
\end{Lemma}
The proof of Lemma~\ref{lem:DJL} will be shown later.
With the result in Lemma~\ref{lem:DJL}, we are able to show that the following result.
 \begin{Lemma}\label{lem:rec_lip}
Let the compressing matrix $\bD\in \Rbb^{{ T} \times J}$  and
\[
    \bD  = \frac{1}{{ T}} (\bS^{\top} \odot \bA^{\top})^{\top},
\]
with
\begin{equation*}
  \begin{split}
   & s_{i,j} \sim {\cal N}(0, 1), \quad i= 1,\ldots, 2N,\quad j = 1,\ldots, { T},\\
    & a_{i,j} \sim {\cal N}(0,1), \quad i= 1,\ldots, 2M,\quad j = 1,\ldots, { T}
  \end{split}
\end{equation*}
and $J = 4MN$.
  Let
  \[
   { T} = \Omega \left( K^2 \left(\log\frac{4\sqrt{2}\sqrt{\pi\kappa} L_{\cal G}}{\epsilon} \right)^2  \right).
  \]
  Let $M$ be a $\epsilon$-net of $\setZ$.
  Then for any $\check{\bh}\in \check{\Gamma}$, if
  \[
  \check{\bh}' = \arg\min_{\hat{\bh}\in {\cal G}(M)} \| \check{\bh} - \hat{\bh} \|_2,
  \]
  then we have $\| \bD (\check{\bh} - \check{\bh}') \|_2 = {O}(\epsilon)$ with probability at least $1-e^{- \Omega(\sqrt{{ T}})}$.
 \end{Lemma}
 The proof of Lemma~\ref{lem:rec_lip} will be shown later.
Now we are ready to show the main proof of Lemma~\ref{lem:s_rec2}.
 First, we define $\check{s}= \check{g}\circ {\cal G}$. We construct an $\epsilon/2L_{\cal G}$-net $M$ of $\setZ$.
Then $N = \check{s}(M)$ is an $\epsilon$-net of $\check{\Gamma}$.
Let $U$ denote the set of pairwise distance of $N$, i.e.,
\(
    U=\{ \check{s}(\bz_1) - \check{s} (\bz_2)  \mid \bz_1, \bz_2 \in M \}.
\)

Then, the cardinality $|U| = |N|^2$ and
\begin{equation}\label{eq:car_T}
  \log(|U|) \leq  8K \log \left(\frac{4\sqrt{2}\sqrt{\pi\kappa} L_{\cal G}}{\epsilon }\right).
\end{equation}
Let $\bz_1$ and $\bz_2$ be the two points in $\setZ$. Also, we let $\bz_1'$ and $\bz_2'$ be the nearest points in $M$ to $\bz_1$ and $\bz_2$, respectively.
This ensures
\begin{equation}\label{eq:bound1}
  \begin{split}
    \| \check{s} (\bz_1) - \check{s} (\bz_2) \|_2 \leq& \| \check{s} (\bz_1) - \check{s} (\bz_1') \|_2+ \|  \check{s} (\bz_1') - \check{s} (\bz_2')  \|_2 +\\
    &~\| \check{s} (\bz_2) - \check{s} (\bz_2') \|_2\\
    \leq &\|  \check{s} (\bz_1') - \check{s} (\bz_2')  \|_2  + 2\epsilon
  \end{split}
\end{equation}
where the first inequality is due to triangle inequality and the second inequality is due to the construction of $\epsilon$-net $N$.
Also, we have
\begin{equation}\label{eq:bound2}
  \begin{split}
   & \| \bD \check{s} (\bz_1') - \bD \check{s} (\bz_2') \|_2 \\
    \leq & \| \bD \check{s} (\bz_1') - \bD \check{s} (\bz_1) \|_2 + \| \bD \check{s} (\bz_1) - \bD \check{s}(\bz_2) \|_2\\
&~+ \| \bD \check{s} (\bz_2) - \bD \check{s} (\bz_2') \|_2\\
    \leq & \| \bD \check{s} (\bz_1) - \bD \check{s} (\bz_2) \|_2 + {O}(\epsilon).
  \end{split}
\end{equation}
where the first inequality is due to triangle inequality and the second inequality is due to Lemma~\ref{lem:rec_lip}.
By the result in Lemma~\ref{lem:DJL}, we have
\begin{equation}\label{eq:bound3}
  \begin{split}
    \| \bD \check{s} (\bz_1') - \bD \check{s} (\bz_2') \|_2 \geq (1-\alpha) \|  \check{s} (\bz_1') -   \check{s} (\bz_2') \|_2
  \end{split}
\end{equation}
with probability of at least $1-e^{-{\mathbf{\Omega}(\sqrt{{ T}})}}$.

Combining the above inequalities, it holds that
\begin{equation}\label{eq:bound4}
  \begin{split}
  (1-\alpha) \| \check{s} (\bz_1) - \check{s} (\bz_2) \|_2 \leq &~(1-\alpha) \|  \check{s} (\bz_1') -   \check{s} (\bz_2') \|_2 + O(\epsilon)\\
   \leq &~ \| \bD \check{s} (\bz_1') - \bD \check{s} (\bz_2') \|_2 + O(\epsilon)\\
   \leq &~ \| \bD \check{s} (\bz_1) - \bD \check{s} (\bz_2) \|_2 + O(\epsilon),
  \end{split}
\end{equation}
where the first inequality is due to \eqref{eq:bound1}, the second inequality is due to \eqref{eq:bound3}, and the final inequality is due to \eqref{eq:bound2}.
This implies that $\bD$ satisfies the S-REC property.


\section{Proof of Lemma~\ref{lem:DJL}}
\label{app:proof_lemma_DJL}
First, we introduce the definition of JL Moment property.
\begin{Def}
  We say that a matrix $\bA \in \Rbb^{m \times d}$ satisfies the $(\alpha, \epsilon,t)$-JL Moment property if for any parameter $\alpha, \epsilon \in [0,1]$, $t\geq 2$, and for every $\bx \in \Rbb^d$ with $\| \bx \|_2 =1$, we have
  \[
    \Exp[| \| \bA \bx \|_2^2 -1 |^t] \leq  \alpha^t \epsilon \mbox{ and } \Exp[\|  \bA \bx  \|_2^2] =1.
  \]
\end{Def}

Following the result in \cite[Theorem 42]{ahle2020oblivious} with the special case $c=2$, it can be verified that the $\bD$ satisfies the  $(\alpha, \epsilon,p )$-JL-moment property for any $p \ge 4$ if
\[
{ T}\geq\max(c_1 \alpha^{-2} \log 1/\epsilon, c_2 \alpha^{-1} (\log 1/\epsilon)^2)
\] for some constants $c_1,c_2>0$.
Then, by Markov inequality, it arrives that
  \begin{equation*}
  \begin{split}
  &\Pr \left( \left| \frac{ \| \bD \bx \|_2^2}{\|  \bx \|_2^2} -1 \right| \geq \alpha \right)= \Pr \left( \left| \frac{ \| \bD \bx \|_2^2}{\|  \bx \|_2^2} -1 \right|^p \geq \alpha^{p} \right)\\
  \leq &~ \frac{\alpha^p \epsilon}{\alpha^p} = \epsilon.
  \end{split}
  \end{equation*}


\section{Proof of Lemma \ref{lem:rec_lip}}\label{app:proof_lem_rec_lip}

From Lemma~\ref{lem:DJL}, we have
\(
  \Pr(\| \bD \bx \|_2 \geq (1+\alpha)\| \bx \|_2) \leq f
\)
for any $f>0$ if
\(
    \alpha  \geq \max\big\{ \sqrt{\frac{c_1}{{ T}} \log(1/f)},\frac{c_2^2}{{ T}} ( \log(1/f))^2 \big\}
\).

Let us construct a chain of epsilon nets of $\setZ$ with $M_0 \subseteq M_1 \subseteq M_2 \subseteq \ldots \subseteq M_l$ such that $M_i$ is a $\epsilon_i/2L_{\cal G}$-net with $\epsilon_i = \epsilon/2^i$. We know that there exists chain of nets such that
\begin{equation*}
  \begin{split}
    \log |M_i| \leq &~ 4K \log\left(\frac{4\sqrt{2}\sqrt{\pi\kappa} L_{\cal G}}{\epsilon_i}  \right)\\
    \leq &~4iK+  4K \log\left(\frac{4\sqrt{2}\sqrt{\pi\kappa} L_{\cal G}}{\epsilon}\right).
  \end{split}
\end{equation*}
Let $N_i = \check{g}({\cal G}(M_i))$. Then $N_i$ is a $\epsilon_i$-net of $\Gamma$, with $|N_i| = |M_i|$.
 For $i \in \{ 0,1,2,\ldots,l-1 \}$, let
    \begin{align}
        { U_i} = \{ \check{\bh}_{i+1}-\check{\bh}_i | \check{\bh}_{i+1} \in N_{i+1}, \check{\bh}_i \in N_i \} \nonumber
    \end{align}
    Then, the cardinality of each $ U_i$ can be bounded as follows
    \begin{equation}
    \begin{split}
        { \log |U_i|} \le & \log |N_{i+1}| + \log|N_{i}|\\
        \le & 12iK + 8K \log\left(\frac{4\sqrt{2}\sqrt{\pi\kappa} L_{\cal G}}{\epsilon }\right)\nonumber
    \end{split}
    \end{equation}
    With this, for particular setting of $\alpha_i$ and $f_i$ we have
    \begin{equation}
        \Pr(\|\bD \check{\bh}\|_2 > (1+ \alpha_i) \|\check{\bh}\|_2) \le f_i. \nonumber
    \end{equation}
    Thus by union bound, we have, for $i=1,\ldots, l$,
    \begin{align}
        \Pr(\|\bD \check{\bh}\|_2 \le (1 + \alpha_i)\|\check{\bh}\|_2, \forall \check{\bh} \in { U_i}) \ge 1- \sum_{i=0}^{l-1} |{ U_i}| f_i \nonumber
    \end{align}
This leads to
    \begin{align}
        \log (|{ U_i}| f_i) = \log(|{ U_i}|) + \log(f_i) \nonumber\\
        \le 12iK + 8K \log \left(\frac{4\sqrt{2}\sqrt{\pi\kappa} L_{\cal G}}{\epsilon }\right) + \log(f_i) \nonumber
    \end{align}
Now, let
    \({ T}^{1/2}= 12K \log \left(\frac{4\sqrt{2}\sqrt{\pi\kappa} L_{\cal G}}{\epsilon }\right),
    \) and
    \(
    \log f_i = -({ T}^{1/2}+16iK),
    \)
which leads to
    \begin{align*}
         \alpha_i &= \max \left(\frac{c_1^{1/2}}{{ T}^{1/2}}({ T}^{1/2}+16iK)^{1/2},\frac{c_2^2}{{ T}} ({ T}^{1/2}+16iK)^2 \right)
    \end{align*}
Then, we have
\begin{align}
    \log(|{ U_i}| f_i) &\le 12iK + \frac{2}{3} { T}^{1/2} -{ T}^{1/2} - 16iK \nonumber\\
    &= -4iK -\frac{\sqrt{{ T}}}{3} \nonumber
\end{align}
This implies that
\begin{align}
     &\Pr(\bD \check{\bh}_2 \le (1 + \alpha_i)\|\check{\bh}\|_2, \forall \check{\bh} \in { U_i}) \nonumber \\
\ge & 1- \sum_{i=0}^{l-1} e^{-4iK -\sqrt{{ T}}/3} \ge  1- e^{-\sqrt{{ T}}/3} \left( \frac{1}{1-e^{-1}} \right)\nonumber\\
\ge & 1- 2e^{-\sqrt{{ T}}/3} \nonumber
\end{align}
Then, for any $\check{\bh} \in \check{\Gamma}$, we can write
\begin{equation}\label{eq:tel}
  \begin{split}
    \check{\bh} = \check{\bh}_0 + \sum_{i=0}^{l-1} (\check{\bh}_{i+1} - \check{\bh}_{i}) + \check{\bh}^{f},
  \end{split}
\end{equation}
where $\check{\bh}_i \in \Gamma$ and $\check{\bh}^f = \check{\bh} -\check{\bh}_l$.

We have $\bh_{i+1}-\bh_i \in { U_i}$, and with probability of at least $1-2e^{-\sqrt{{ T}}/3}$, the following holds
\begin{align}
    \sum_{i=0}^{l-1} \|\bD (\check{\bh}_{i+1}-\check{\bh}_i)\|_2 &\le \sum_{i=0}^{l-1}(1+\alpha_i) \|\check{\bh}_{i+1}-\check{\bh}_i\|_2 \nonumber \\
    &\leq \epsilon  \sum_{i=0}^{l-1} \frac{1}{2^i} (1+\alpha_i) \nonumber
\end{align}
Since $\alpha_i$ decreases with value of ${ T}$ and  the term inside summation is decreasing, it can be upper bounded as
\begin{equation}
    \sum_{i=0}^{l-1} \|\bD (\check{\bh}_{i+1}-\check{\bh}_i)\|_2 \le O(\epsilon) \nonumber
\end{equation}
Then, for any $\check{\bh}, \check{\bh}' \in \check{\Gamma}$, with probability of at least $1-e^{-\mathbf{\Omega}(\sqrt{{ T}})}$, we have
\begin{align}
    \|\bD (\check{\bh} - \check{\bh}')\|_2
    &\le \sum_{i=0}^{l-1} \|\bD (\check{\bh}_{i+1} -\check{\bh}_i)\|_2 + \|\bD \check{\bh}_f\|_2 \nonumber\\
    &\le O(\epsilon) + \|\bD \check{\bh}_f\|_2 \nonumber
\end{align}
where $\check{\bh}^f = \check{\bh} -\check{\bh}_l$,
and  $\|\check{\bh}^f\|_2 = \|\check{\bh} -\check{\bh}_l \|_2 \le \epsilon_l = \frac{\epsilon}{2^l}$.
Then, $\|\bD \check{\bh}_f \|_2 \le \|\bD\|_2 \frac{\epsilon}{2^l}$.

We need the following result to proceed.
\begin{Lemma}\label{lem:KR_prod}
  Let
  \(
  \bD =   (\bA_1^{\top} \odot \bA_2^{\top})^{\top},
  \)
  where $\odot$ denotes the Khatri-Rao product, $[\bA_1]_{i,j} \sim {\cal N}(0,1/{ T})$ and $[\bA_2]_{i,j} \sim {\cal N}(0,1/{ T})$.
  Then, with probability at least $1- 4e^{-{ T}/2}$, it holds that
  \begin{equation}\label{eq:KR_pr}
    \|\bD\|_2 \leq 4+ 4 \sqrt{\frac{J}{{ T}}} + \frac{J}{{ T}}.
  \end{equation}
\end{Lemma}
The proof of Lemma~\ref{lem:KR_prod} will be shown soon.
Using result from Lemma \ref{lem:KR_prod},  with probability of at least $1-4e^{-{ T}/2}$, we have
\(
    \|\bD \check{\bh}^f\|_2 \le \left(4 + 4 \sqrt{\frac{{J}}{{ T}}} + \frac{{J}}{{ T}} \right) \frac{\epsilon}{2^l}.
\)

If we use $l = \log {J}$, then,
\(
    \|\bD \check{\bh}_f\|_2 = O({\epsilon}).
\)
Then, we have
\begin{equation}\label{eq:net}
\|\bD (\check{\bh} - \check{\bh}')\|_2  = O(\epsilon)
\end{equation}
 with probability of at least $1-e^{-{\mathbf{\Omega}(\sqrt{{ T}})}}$.



\section{Proof of Lemma~\ref{lem:KR_prod}}

  The proof idea follows the result in \cite{ibrahim2020recoverability}.  By the definition of spectral norm,
    \begin{subequations}\label{eq:kr_spec}
        \begin{align}
            \|\bA_1^{\top} \odot \bA_2^{\top}\|_2 &= \sigma_{\text{max}}(\bA_1^{\top} \odot \bA_2^{\top}) \nonumber\\
            &= \max_{\|\bx\|_2=1} \|(\bA_1^{\top} \odot \bA_2^{\top}) \bx\| \nonumber \\
            &= \max_{\| \bx\|_2=1} \|(\bA_1^{\top} \otimes \bA_2^{\top}) \mathbf{P} \bx \|_2 \label{eqn: addition of column selection matrix}\\
            & \le \max_{\|\bx\|_2=1}  \|\bA_1^{\top} \otimes \bA_2^{\top}\|_2 \|\mathbf{P}\|_2 \|\bx\|_2 \label{eqn: submultiplicative property of spectral norm}\\
            &=   \|\bA_1^{\top} \otimes \bA_2^{\top}\|_2 \max_{\|\bx\|_2=1} \|\bx\|_2 \label{eqn: spectral norm of P is 1} \\
            &=   \|\bA_1^{\top} \otimes \bA_2^{\top}\|_2 \nonumber = \|\bA_1\|_2 \|\bA_2\|_2 \nonumber
        \end{align}
    \end{subequations}
    where   $\mathbf{P}\in \{0,1\}^{T^2\times T}$ is a column selection matrix that  has used the fact that $\bA_1^{\top} \odot \bA_2^{\top}$ comprises of a subset of columns of $\bA_1^{\top} \otimes \bA_2^{\top}$ in equation \eqref{eqn: addition of column selection matrix}; \eqref{eqn: submultiplicative property of spectral norm} is due to the Cauchy-Schwarz inequality, and \eqref{eqn: spectral norm of P is 1} is due to the fact $\|\mathbf{P}\|_2 = 1$.

    From \cite[Corollary 5.35]{vershynin2011introduction}, for both $\bA_1$ and $\bA_2$, we have $\|\bA_i\|_2 \ge 2 + \sqrt{n_i/{ T}}$ with probability at most $2e^{-{ T}/2}$
    where, $n_1=2N$ and $n_2=2M$.
    Now, by using union bound we have
    \begin{align}
\Pr(\|\bA_1\|_2 \cdot \|\bA_2\|_2 \ge t^2) &\le  \Pr((\|\bA_1\|_2 \ge t) \cup (\|\bA_2\|_2 \ge t))\nonumber\\
       \le & \Pr({\|\bA_1\|_2 \ge t}) + \Pr({\|\bA_2\|_2 \ge t}) \nonumber
    \end{align}
    Then, combining above two results, we have
    \begin{align}
        &\Pr\left(\|\bA_1\|_2 \cdot \|\bA_2\|_2 \ge \left(2+\sqrt{\frac{J}{{ T}}}\right)^2\right) \nonumber \\
        &\le \Pr\left(\|\bA_1\|_2 \ge 2+\sqrt{\frac{J}{{ T}}}\right) + \Pr\left(\|\bA_2\|_2 \ge 2+\sqrt{\frac{J}{{ T}}}\right) \nonumber\\
        &\le \Pr\left(\|\bA_1\|_2 \ge 2+\sqrt{\frac{n_1}{{ T}}}\right) + \Pr\left(\|\bA_2\|_2 \ge 2+\sqrt{\frac{n_2}{{ T}}}\right) \nonumber
    \end{align}
    Therefore , we have
           \[
           \Pr\left( \|\bA_1\|_2 \cdot \|\bA_2\|_2 \ge 4+4\sqrt{\frac{J}{{ T}}}+\frac{J}{{ T}} \right) \le 4e^{-{ T}/2}
           \]
           This, together with \eqref{eq:kr_spec}, leads to the result in Lemma~\ref{lem:KR_prod}.
           
\end{document}

%% file: sym_marco.tex
\newcommand\bx{\ensuremath{{\bm x}}}
\newcommand\by{\ensuremath{{\bm y}}}

\newcommand\bh{\ensuremath{{\bm h}}}
\newcommand\bH{\ensuremath{{\bm H}}}

\newcommand\bz{\ensuremath{{\bm z}}}
\newcommand\bn{\ensuremath{{\bm n}}}

\newcommand\bX{\ensuremath{{\bm X}}}

\newcommand\ba{\ensuremath{{\bm a}}}

\newcommand\bA{\ensuremath{{\bm A}}}

\newcommand\bg{\ensuremath{{\bm g}}}

\newcommand\bd{\ensuremath{{\bm d}}}
\newcommand\bD{\ensuremath{{\bm D}}}

\newcommand\bu{\ensuremath{{\bm u}}}

\newcommand\bv{\ensuremath{{\bm v}}}

\newcommand\bY{\ensuremath{{\bm Y}}}

\newcommand\bs{\ensuremath{{\bm s}}}
\newcommand\bS{\ensuremath{{\bm S}}}

\newcommand{\Rbb}{\mathbb{R}}
\newcommand{\Cbb}{\mathbb{C}}

\newcommand{\setX}{\mathcal{X}}
\newcommand{\setR}{\mathcal{R}}
\newcommand{\setU}{\mathcal{U}}

\newcommand{\setW}{\mathcal{W}}
\newcommand{\setS}{\mathcal{S}}
\newcommand{\setC}{\mathcal{C}}

\newcommand{\setZ}{\mathcal{Z}}

\newcommand{\Exp}{\mathbb{E}}
\newcommand{\jj}{\mathfrak{j}}

\newcommand\br{\ensuremath{{\bm r}}}

\newcommand{\bI}{{\bm I}}

\newcommand\tr{\ensuremath{{\rm tr}}}

\newcommand\appxgt[1]{\dot{#1}} 

\DeclareMathOperator{\mat}{mat}

%% file: main.bbl
\begin{thebibliography}{10}
\providecommand{\url}[1]{#1}
\csname url@samestyle\endcsname
\providecommand{\newblock}{\relax}
\providecommand{\bibinfo}[2]{#2}
\providecommand{\BIBentrySTDinterwordspacing}{\spaceskip=0pt\relax}
\providecommand{\BIBentryALTinterwordstretchfactor}{4}
\providecommand{\BIBentryALTinterwordspacing}{\spaceskip=\fontdimen2\font plus
\BIBentryALTinterwordstretchfactor\fontdimen3\font minus \fontdimen4\font\relax}
\providecommand{\BIBforeignlanguage}[2]{{%
\expandafter\ifx\csname l@#1\endcsname\relax
\typeout{** WARNING: IEEEtran.bst: No hyphenation pattern has been}%
\typeout{** loaded for the language `#1'. Using the pattern for}%
\typeout{** the default language instead.}%
\else
\language=\csname l@#1\endcsname
\fi
#2}}
\providecommand{\BIBdecl}{\relax}
\BIBdecl

\bibitem{goldsmith2005wireless}
A.~Goldsmith, \emph{Wireless Communications}.\hskip 1em plus 0.5em minus 0.4em\relax Cambridge University Press, 2005.

\bibitem{love2008anoverview}
D.~J. Love, R.~W. Heath, V.~K. N.~Lau, D.~Gesbert, B.~D. Rao, and M.~Andrews, ``An overview of limited feedback in wireless communication systems,'' \emph{IEEE J. Sel. Areas Commun.}, vol.~26, no.~8, pp. 1341--1365, 2008.

\bibitem{jindal2006mimo}
N.~Jindal, ``{MIMO} broadcast channels with finite-rate feedback,'' \emph{IEEE Trans. Inf. Theory}, vol.~52, no.~11, pp. 5045--5060, 2006.

\bibitem{ngo2013energy}
H.~Q. Ngo, E.~G. Larsson, and T.~L. Marzetta, ``Energy and spectral efficiency of very large multiuser {MIMO} systems,'' \emph{IEEE Trans. Commun.}, vol.~61, no.~4, pp. 1436--1449, 2013.

\bibitem{gao2015spatially}
Z.~Gao, L.~Dai, Z.~Wang, and S.~Chen, ``Spatially common sparsity based adaptive channel estimation and feedback for {FDD} massive {MIMO},'' \emph{IEEE Trans. Signal Process.}, vol.~63, no.~23, pp. 6169--6183, 2015.

\bibitem{lu2014overview}
L.~Lu, G.~Y. Li, A.~L. Swindlehurst, A.~Ashikhmin, and R.~Zhang, ``An overview of massive {MIMO}: Benefits and challenges,'' \emph{IEEE J. Sel. Topics Signal Process.}, vol.~8, no.~5, pp. 742--758, 2014.

\bibitem{jiang2015achievable}
Z.~Jiang, A.~F. Molisch, G.~Caire, and Z.~Niu, ``Achievable rates of {FDD} massive {MIMO} systems with spatial channel correlation,'' \emph{IEEE Trans. Wireless Commun.}, vol.~14, no.~5, pp. 2868--2882, 2015.

\bibitem{hu2017channel}
D.~Hu and L.~He, ``Channel estimation for {FDD} massive {MIMO OFDM} systems,'' in \emph{Proc. IEEE 86th Veh. Tech. Conf. (VTC-Fall)}.\hskip 1em plus 0.5em minus 0.4em\relax IEEE, 2017, pp. 1--5.

\bibitem{li2021pushing}
K.~Li, Y.~Li, L.~Cheng, Q.~Shi, and Z.-Q. Luo, ``Pushing the limit of {Type I} codebook for {FDD} massive {MIMO} beamforming: A channel covariance reconstruction approach,'' in \emph{Proc. IEEE Int. Conf. Acoustics, Speech, Signal Process. (ICASSP)}, 2021, pp. 4785--4789.

\bibitem{li2023csi}
L.~Li, X.~Zeng, Y.-F. Liu, Y.~Xu, and T.-H. Chang, ``{CSI} sensing from heterogeneous user feedbacks: A constrained phase retrieval approach,'' \emph{IEEE Trans. Wireless Commun.}, vol.~22, no.~10, pp. 6930--6945, 2023.

\bibitem{auyeung2007ontheperformance}
C.~K. Au-yeung and D.~J. Love, ``On the performance of random vector quantization limited feedback beamforming in a {MISO} system,'' \emph{IEEE Trans. Wireless Commun.}, vol.~6, no.~2, pp. 458--462, 2007.

\bibitem{choi2013noncoherent}
J.~Choi, Z.~Chance, D.~J. Love, and U.~Madhow, ``Noncoherent trellis coded quantization: A practical limited feedback technique for massive {MIMO} systems,'' \emph{IEEE Trans. Commun.}, vol.~61, no.~12, pp. 5016--5029, 2013.

\bibitem{rao2014distributed}
X.~Rao and V.~K.~N. Lau, ``Distributed compressive {CSIT} estimation and feedback for {FDD} multi-user massive {MIMO} systems,'' \emph{IEEE Trans. Signal Process.}, vol.~62, no.~12, pp. 3261--3271, 2014.

\bibitem{alevizos2018limited}
P.~N. Alevizos, X.~Fu, N.~D. Sidiropoulos, Y.~Yang, and A.~Bletsas, ``Limited feedback channel estimation in massive {MIMO} with non-uniform directional dictionaries,'' \emph{IEEE Trans. Signal Process.}, vol.~66, no.~19, pp. 5127--5141, 2018.

\bibitem{zhou2017sparse}
Z.~Zhou, X.~Chen, D.~Guo, and M.~L. Honig, ``Sparse channel estimation for massive {MIMO} with 1-bit feedback per dimension,'' in \emph{Proc. 2017 IEEE Wireless Communi. Netw. Conf. (WCNC)}.\hskip 1em plus 0.5em minus 0.4em\relax IEEE, 2017, pp. 1--6.

\bibitem{qian2019algebraic}
C.~Qian, X.~Fu, and N.~D. Sidiropoulos, ``Algebraic channel estimation algorithms for {FDD} massive {MIMO} systems,'' \emph{IEEE J. Sel. Topics Signal Process.}, vol.~13, no.~5, pp. 961--973, 2019.

\bibitem{zhang2024integrated}
R.~Zhang, L.~Cheng, S.~Wang, Y.~Lou, Y.~Gao, W.~Wu, and D.~W.~K. Ng, ``Integrated sensing and communication with massive {MIMO}: A unified tensor approach for channel and target parameter estimation,'' \emph{IEEE Trans. Wireless Commun.}, vol.~23, no.~8, pp. 8571--8587, 2024.

\bibitem{kuo2012compressive}
P.-H. Kuo, H.~T. Kung, and P.-A. Ting, ``Compressive sensing based channel feedback protocols for spatially-correlated massive antenna arrays,'' in \emph{Proc. IEEE Wireless Communi. Netw. Conf. (WCNC)}.\hskip 1em plus 0.5em minus 0.4em\relax IEEE, 2012, pp. 492--497.

\bibitem{qian2018tensor}
C.~Qian, X.~Fu, N.~D. Sidiropoulos, and Y.~Yang, ``Tensor-based channel estimation for dual-polarized massive {MIMO} systems,'' \emph{IEEE Trans. Signal Process.}, vol.~66, no.~24, pp. 6390--6403, 2018.

\bibitem{bajwa2010compressed}
W.~U. Bajwa, J.~Haupt, A.~M. Sayeed, and R.~Nowak, ``Compressed channel sensing: A new approach to estimating sparse multipath channels,'' \emph{Proc. IEEE}, vol.~98, no.~6, pp. 1058--1076, 2010.

\bibitem{dai2018fdd}
J.~Dai, A.~Liu, and V.~K. Lau, ``{FDD} massive {MIMO} channel estimation with arbitrary {2D}-array geometry,'' \emph{IEEE Trans. Signal Process.}, vol.~66, no.~10, pp. 2584--2599, 2018.

\bibitem{lin2020tensor}
Y.~Lin, S.~Jin, M.~Matthaiou, and X.~You, ``Tensor-based channel estimation for millimeter wave {MIMO-OFDM} with dual-wideband effects,'' \emph{IEEE Trans. Commun.}, vol.~68, no.~7, pp. 4218--4232, 2020.

\bibitem{alkhateeb2014channel}
A.~Alkhateeb, O.~El~Ayach, G.~Leus, and R.~W. Heath, ``Channel estimation and hybrid precoding for millimeter wave cellular systems,'' \emph{IEEE J. Sel. Topics Signal Process.}, vol.~8, no.~5, pp. 831--846, 2014.

\bibitem{li1996efficient}
J.~Li and P.~Stoica, ``Efficient mixed-spectrum estimation with applications to target feature extraction,'' \emph{IEEE Trans. Signal Process.}, vol.~44, no.~2, p. 281–295, 1996.

\bibitem{shao2022massive}
M.~Shao and X.~Fu, ``Massive {MIMO} channel estimation via compressed and quantized feedback,'' in \emph{2022 56th Asilomar Conf. Signals, Syst., Comput.}, 2022, pp. 1016--1020.

\bibitem{heath2016anoverview}
R.~W. Heath, N.~González-Prelcic, S.~Rangan, W.~Roh, and A.~M. Sayeed, ``An overview of signal processing techniques for millimeter wave {MIMO} systems,'' \emph{IEEE J. Sel. Topics Signal Process.}, vol.~10, no.~3, pp. 436--453, 2016.

\bibitem{stoica2005spectral}
P.~Stoica and R.~L. Moses, \emph{Spectral analysis of signals}.\hskip 1em plus 0.5em minus 0.4em\relax Upper Saddle River, N.J.: Pearson/Prentice Hall Upper Saddle River, N.J., 2005.

\bibitem{liu2007multidimensional}
J.~Liu, X.~Liu, and X.~Ma, ``Multidimensional frequency estimation with finite snapshots in the presence of identical frequencies,'' \emph{IEEE Trans. Signal Process.}, vol.~55, no.~11, pp. 5179--5194, 2007.

\bibitem{sha2019harmonic}
Z.~Sha, Z.~Wang, and S.~Chen, ``Harmonic retrieval based baseband channel estimation for millimeter wave {OFDM} systems,'' \emph{IEEE Trans. Veh. Tech.}, vol.~68, no.~3, pp. 2668--2681, 2019.

\bibitem{nion2010tensor}
D.~Nion and N.~D. Sidiropoulos, ``Tensor algebra and multidimensional harmonic retrieval in signal processing for {MIMO} radar,'' \emph{IEEE Trans. Signal Process.}, vol.~58, no.~11, pp. 5693--5705, 2010.

\bibitem{lipshitz1992quantization}
S.~P. Lipshitz, R.~A. Wannamaker, and J.~Vanderkooy, ``Quantization and dither: A theoretical survey,'' \emph{J. Audio Engineer. Society}, vol.~40, no.~5, pp. 355--375, 1992.

\bibitem{wannamaker2000atheory}
R.~Wannamaker, S.~Lipshitz, J.~Vanderkooy, and J.~Wright, ``A theory of nonsubtractive dither,'' \emph{IEEE Trans. Signal Process.}, vol.~48, no.~2, pp. 499--516, 2000.

\bibitem{timilsina2023quantized}
S.~Timilsina, S.~Shrestha, and X.~Fu, ``Quantized radio map estimation using tensor and deep generative models,'' \emph{IEEE Trans. Signal Process.}, 2023.

\bibitem{horn1991topics}
R.~A. Horn and C.~R. Johnson, \emph{Topics in Matrix Analysis}.\hskip 1em plus 0.5em minus 0.4em\relax Cambridge University Press, 1991.

\bibitem{shalev2014understanding}
S.~Shalev-Shwartz and S.~Ben-David, \emph{Understanding Machine Learning: From Theory to Algorithms}.\hskip 1em plus 0.5em minus 0.4em\relax USA: Cambridge University Press, 2014.

\bibitem{davenport2014onebit}
M.~A. Davenport, Y.~Plan, E.~van~den Berg, and M.~Wootters, ``1-bit matrix completion,'' \emph{Inf. Inference: J. IMA}, vol.~3, no.~3, pp. 189--223, 2014.

\bibitem{ghadermarzy2018learning}
N.~Ghadermarzy, Y.~Plan, and O.~Yilmaz, ``Learning tensors from partial binary measurements,'' \emph{IEEE Trans. Signal Process.}, vol.~67, no.~1, pp. 29--40, 2018.

\bibitem{lee2020tensor}
C.~Lee and M.~Wang, ``Tensor denoising and completion based on ordinal observations,'' in \emph{Proc. Int. Conf. Mach. Learn. (ICML)}.\hskip 1em plus 0.5em minus 0.4em\relax PMLR, 2020, pp. 5778--5788.

\bibitem{zhang2020spectrum}
G.~Zhang, X.~Fu, J.~Wang, X.-L. Zhao, and M.~Hong, ``Spectrum cartography via coupled block-term tensor decomposition,'' \emph{IEEE Trans. Signal Process.}, vol.~68, pp. 3660--3675, 2020.

\bibitem{shrestha2022deep}
S.~Shrestha, X.~Fu, and M.~Hong, ``Deep spectrum cartography: Completing radio map tensors using learned neural models,'' \emph{IEEE Trans. Signal Process.}, vol.~70, pp. 1170--1184, 2022.

\bibitem{plan2013robust}
Y.~Plan and R.~Vershynin, ``Robust 1-bit compressed sensing and sparse logistic regression: A convex programming approach,'' \emph{IEEE Trans. Inf. Theor.}, vol.~59, no.~1, p. 482–494, Jan. 2013.

\bibitem{boufounos2015quantization}
P.~T. Boufounos, L.~Jacques, F.~Krahmer, and R.~Saab, ``Quantization and compressive sensing,'' in \emph{Proc. Compressed Sensing Its Appl.: MATHEON Workshop}.\hskip 1em plus 0.5em minus 0.4em\relax Springer, 2015, pp. 193--237.

\bibitem{xu2019quantized}
C.~Xu and L.~Jacques, ``{Quantized compressive sensing with {RIP} matrices: the benefit of dithering},'' \emph{Inf. Inference: J. IMA}, vol.~9, no.~3, pp. 543--586, 11 2019.

\bibitem{saab2018quantization}
R.~Saab, R.~Wang, and {\"O}.~Y{\i}lmaz, ``Quantization of compressive samples with stable and robust recovery,'' \emph{Applied and Computational Harmonic Analysis}, vol.~44, no.~1, pp. 123--143, 2018.

\bibitem{bora2017compressed}
A.~Bora, A.~Jalal, E.~Price, and A.~G. Dimakis, ``Compressed sensing using generative models,'' in \emph{Proc. Int. Conf. Mach. Learn. (ICML)}.\hskip 1em plus 0.5em minus 0.4em\relax PMLR, 2017, pp. 537--546.

\bibitem{boyd2011foundations}
S.~Boyd, N.~Parikh, E.~Chu, B.~Peleato, and J.~Eckstein, ``Distributed optimization and statistical learning via the alternating direction method of multipliers,'' \emph{Foundations Trends Mach. Learn.}, vol.~3, pp. 1--122, 01 2011.

\bibitem{zymnis2009compressed}
A.~Zymnis, S.~Boyd, and E.~Candes, ``Compressed sensing with quantized measurements,'' \emph{IEEE Signal Process. Lett.}, vol.~17, no.~2, pp. 149--152, 2009.

\bibitem{mezghani2010multiple}
A.~Mezghani, F.~Antreich, and J.~A. Nossek, ``Multiple parameter estimation with quantized channel output,'' in \emph{2010 International ITG Workshop on Smart Antennas (WSA)}, 2010, p. 143–150.

\bibitem{shao2024accelerated}
M.~Shao, W.-K. Ma, J.~Liu, and Z.~Huang, ``Accelerated and deep expectation maximization for one-bit {MIMO-OFDM} detection,'' \emph{IEEE Trans. Signal Process.}, 2024.

\bibitem{Alkhateeb2019deepmimo}
A.~Alkhateeb, ``{DeepMIMO}: A generic deep learning dataset for millimeter wave and massive {MIMO} applications,'' in \emph{Proc. of Information Theory and Applications Workshop (ITA)}, San Diego, CA, Feb 2019, pp. 1--8.

\bibitem{sun2018limited}
H.~Sun, Z.~Zhao, X.~Fu, and M.~Hong, ``Limited feedback double directional massive {MIMO} channel estimation: From low-rank modeling to deep learning,'' in \emph{Proc. IEEE 19th Int. Workshop Signal Process. Advances Wireless Commun. (SPAWC)}, 2018.

\bibitem{mo2014channel}
J.~Mo, P.~Schniter, N.~G. Prelcic, and R.~W. Heath, ``Channel estimation in millimeter wave {MIMO} systems with one-bit quantization,'' in \emph{Proc. 48th Asilomar Conf. Signals, Syst., Comput.}\hskip 1em plus 0.5em minus 0.4em\relax IEEE, 2014, pp. 957--961.

\bibitem{mo2018channel}
J.~Mo, P.~Schniter, and R.~W. Heath, ``Channel estimation in broadband millimeter wave {MIMO} systems with few-bit {ADCs},'' \emph{IEEE Trans. Signal Process.}, vol.~66, no.~5, pp. 1141--1154, 2018.

\bibitem{Remcom}
Remcom, ``{Wireless InSite},'' \url{http://www.remcom.com/wireless-insite}.

\bibitem{Baraniuk2008}
R.~Baraniuk, M.~Davenport, R.~DeVore, and M.~Wakin, ``A simple proof of the restricted isometry property for random matrices,'' \emph{Constructive Approximation}, vol.~28, no.~3, p. 253–263, 2008.

\bibitem{bartlett2017spectrally}
P.~L. Bartlett, D.~J. Foster, and M.~J. Telgarsky, ``Spectrally-normalized margin bounds for neural networks,'' \emph{Advances Neural Inf. Process. Systems (NeurIPS)}, vol.~30, 2017.

\bibitem{vershynin2011introduction}
R.~Vershynin, ``Introduction to the non-asymptotic analysis of random matrices,'' \emph{arXiv preprint arXiv:1011.3027}, 2010.

\bibitem{cao2015categorical}
Y.~Cao and Y.~Xie, ``Categorical matrix completion,'' in \emph{Proc. IEEE 6th Int. Workshop Comput. Advances Multi-Sensor Adaptive Process. (CAMSAP)}.\hskip 1em plus 0.5em minus 0.4em\relax IEEE, 2015, pp. 369--372.

\bibitem{rockafellar2009variational}
R.~T. Rockafellar and R.~J.-B. Wets, \emph{{Variational Analysis}}.\hskip 1em plus 0.5em minus 0.4em\relax Springer Science \& Business Media, 2009, vol. 317.

\bibitem{ahle2020oblivious}
T.~D. Ahle, M.~Kapralov, J.~B. Knudsen, R.~Pagh, A.~Velingker, D.~P. Woodruff, and A.~Zandieh, ``Oblivious sketching of high-degree polynomial kernels,'' in \emph{Proc. 14th Annual ACM-SIAM Symposium Discrete Algorithms}.\hskip 1em plus 0.5em minus 0.4em\relax SIAM, 2020, pp. 141--160.

\bibitem{ibrahim2020recoverability}
S.~Ibrahim, X.~Fu, and X.~Li, ``On recoverability of randomly compressed tensors with low {CP} rank,'' \emph{IEEE Signal Process. Lett.}, vol.~27, pp. 1125--1129, 2020.

\end{thebibliography}
